\documentclass[a4paper,twocolumns,11pt,accepted=2026-04-07]{quantumarticle}
\pdfoutput=1
\usepackage[numbers, sort&compress]{natbib}
\usepackage{xurl}
\usepackage{graphicx,amsmath,amsfonts,enumerate,amsthm,amssymb,mathtools,enumitem,thmtools,hyperref,subfigure,mathdots,enumitem,centernot,bm,soul,bbm,mathrsfs}
\usepackage{hyperref}
\usepackage[capitalise, noabbrev]{cleveref}
\usepackage{float}
\usepackage{physics}
\usepackage{xcolor,colortbl}
\usepackage{mathrsfs}  
\usepackage{soul}
\usepackage{comment}
\usepackage{tikz}
\usepackage[T1]{fontenc}

\usepackage{pgfplots}
\usetikzlibrary{quantikz}

\newcommand{\mybox}[4]{
    \begin{figure}[h]
        \centering
    \begin{tikzpicture}
        \node[anchor=text,text width=\columnwidth-1.2cm, draw, rounded corners, line width=1pt, fill=#3, inner sep=5mm] (big) {\\#4};
        \node[draw, rounded corners, line width=.5pt, fill=#2, anchor=west, xshift=5mm] (small) at (big.north west) {#1};
    \end{tikzpicture}
    \end{figure}
}
\graphicspath{{Figures/}}
\hypersetup{colorlinks=true,linkcolor=blue,citecolor=blue,urlcolor=blue}
\definecolor{mygreen}{rgb}{0.01, 0.31, 0.59}
\definecolor{myblue}{rgb}{0.01, 0.31, 0.59}
\definecolor{bluecyan}{rgb}{0.27, 0.66, 0.88}
\hypersetup{
  colorlinks=true,   
  linkcolor=myblue,  
  citecolor=mygreen, 
  urlcolor =myblue    
}

\def\S{ {\cal S} }
\def\O{ {\cal O} }
\def\N{ {\cal N} }
\def\M{ {\cal M} }
\def\F{ {\cal F} }

\def\>{\rangle}
\def\<{\langle}

\newcommand{\iden}{\mathbbm{1}}

\renewcommand{\v}[1]{\ensuremath{\boldsymbol #1}}

\definecolor{ppblue}{RGB}{46,117,182}
\definecolor{ppred}{RGB}{197, 90, 17}

\theoremstyle{plain}
\newtheorem{thm}{Theorem}
\newtheorem{lem}[thm]{Lemma}

\theoremstyle{definition}
\newtheorem{defn}{Definition}

\usepackage[normalem]{ulem}

\begin{document}

\title{Quantum Resource Theories beyond Convexity}

\author{Roberto Salazar}
\email{rb.salazar.vargas@gmail.com}
\affiliation{Department of Communications \& Computer Engineering, Faculty of Information \& communication technology (ICT), University of Malta, Msida, MSD 2080, Malta}
\affiliation{Faculty of Physics, Astronomy and Applied Computer Science, Jagiellonian University, 30-348 Krak\'{o}w, Poland}

\author{Jakub Czartowski}
\email{jakub.czartowski@doctoral.uj.edu.pl}
\affiliation{Faculty of Physics, Astronomy and Applied Computer Science, Jagiellonian University, 30-348 Krak\'{o}w, Poland}
\affiliation{Doctoral School of Exact and Natural Sciences, Jagiellonian University, ul.\ Lojasiewicza 11, 30-348 Kraków, Poland}
\affiliation{School of Physical and Mathematical Sciences, Nanyang Technological University, 21 Nanyang Link, 637361 Singapore, Republic of Singapore}

\author{Ricard Ravell Rodr\'iguez}
\affiliation{Institute for Cross-Disciplinary Physics and Complex Systems IFISC (UIB-CSIC),
Campus Universitat Illes Balears, E-07122 Palma de Mallorca, Spain}

\author{Grzegorz Rajchel-Mieldzio\'c} 
\affiliation{BEIT sp.\ z o.o., ul.\ Mogilska 43, 31-545 Krak{\'o}w, Poland}
\affiliation{ICFO-Institut de Ciencies Fotoniques, The Barcelona Institute of Science and Technology, Av. Carl Friedrich Gauss 3, 08860 Castelldefels (Barcelona), Spain} 

\author{Pawe{\l} Horodecki}
\affiliation{International Centre for Theory of Quantum Technologies, University of Gda\'{n}sk, Wita Stwosza 63, 80-308 Gda\'{n}sk, Poland}

\author{Karol {\.Z}yczkowski}
\affiliation{Faculty of Physics, Astronomy and Applied Computer Science, Jagiellonian University, 30-348 Krak\'{o}w, Poland}
\affiliation{Center for Theoretical Physics, Polish Academy of Sciences, 02-668 Warszawa, Poland}

\date{April 7, 2026}

\begin{abstract}

A class of quantum resource theories, based on non-convex star-shape sets, presented in this work captures the key quantum properties that cannot be studied by standard convex theories. We provide operational interpretations for a resource of this class and demonstrate its advantage to improve performance of correlated quantum discrimination tasks and testing of quantum combs. Proposed techniques provide useful tools to describe quantum discord, total correlations in composite quantum systems and to estimate the degree of non-Markovianity of an analyzed quantum dynamics. Other applications include the problem of unistochasticity of a given bistochastic matrix, with relevance for quantization of classical dynamics and studies of violation of CP-symmetry in high energy physics. In all these cases, the non-linear witnesses introduced here outperform the standard linear witnesses. Importance of our findings for quantum information theory is also emphasized.

\end{abstract}

\maketitle
\section{Introduction}

Developing quantum resource theories is a pivotal endeavor in quantum information science \cite{Ben2004,Dev2008,Horodecki2009}, with wide-ranging applications including entanglement \cite{Horodecki2003,Vidal1999}, coherence \cite{Streltsov2017}, and quantum thermodynamics \cite{LostaglioThermo2019}, among others \cite{Gour2019,Paul2019,Veitch2014,deVicente2014,PatrykPRX2021,TakagiPRL2020,TakaginonGaussianity,WojewodkaSciazko2024}. Resource theories unify the processes of identifying, quantifying and manipulating quantum resources, enabling advancements across diverse quantum technologies \cite{Gour2019}. A fundamental distinction within resource theories lies in whether resourceless devices form a convex or non-convex set \cite{Gour2019,TakagiRegula2019}. While significant progress has been made in convex resource theories, shaping innovations in quantum communication and computing, non-convex theories have seen limited exploration due to their mathematical challenges \cite{Gour2019}.  Consequently, expanding this framework to incorporate non-convex resource theories has the potential to generate new capabilities and pave the way for further advancements in quantum information science.

In fact, convex resource theories cannot capture crucial properties such as memory in stochastic processes and total correlations in quantum networks, necessitating the development of appropriate non-convex resource theories \cite{Gour2019,Vinayak2020}. These properties, present in processes with an output dependent on a long sequence of past states, are prevalent in various fields of science and applied mathematics \cite{BreuerRev2016,Vacchini2011}. Furthermore, communication limitations between nodes in large-scale networks \cite{AlsPRX2021} make total correlations a crucial resource for
future quantum network security and distributed computing functionalities~\cite{Wehner2018}.

\begin{figure}[ht]
    \centering
    \includegraphics[width=1\linewidth]{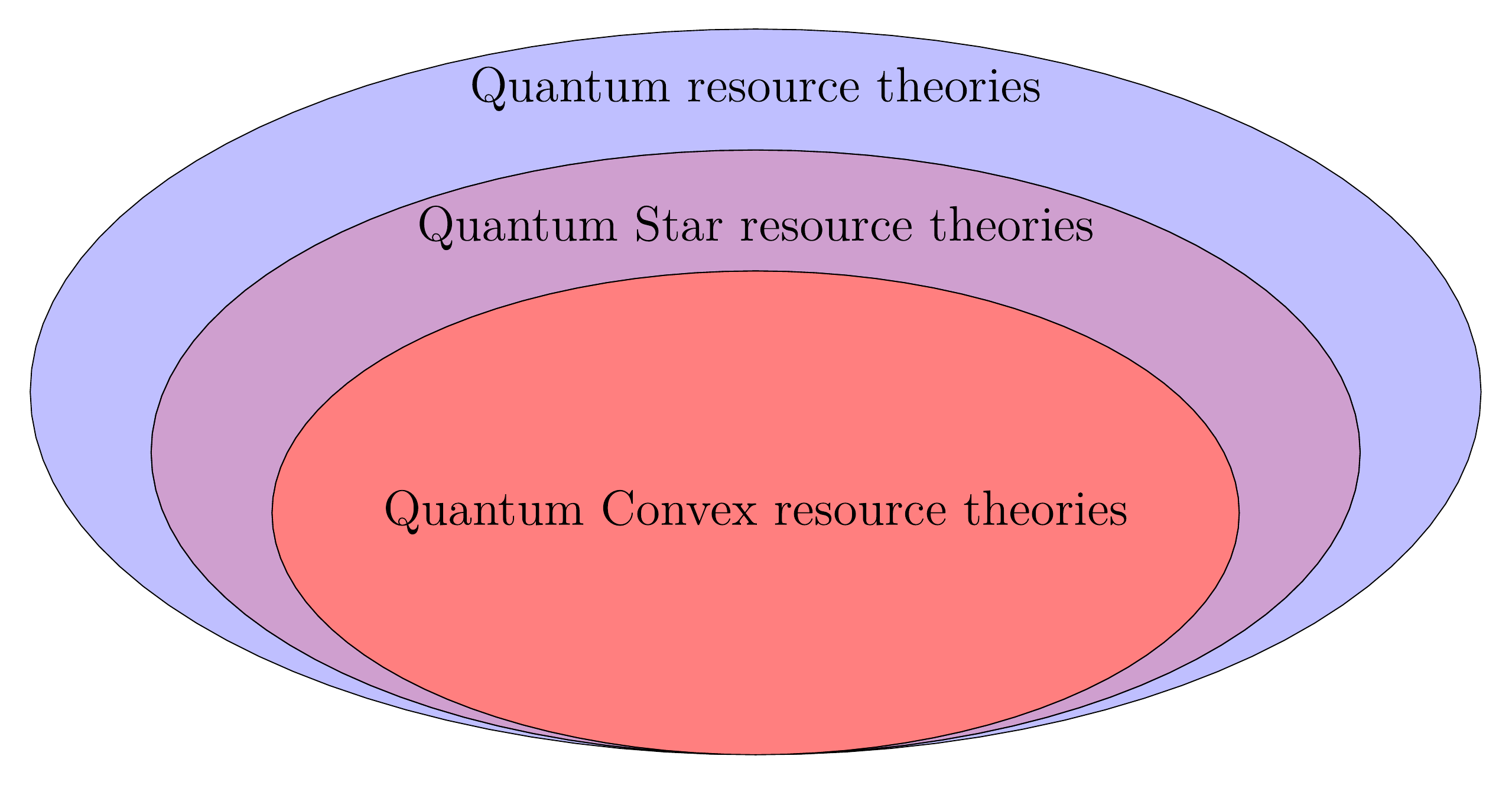}
   
    \caption{Hierarchy of Quantum Resource Theories. Since every convex set is also a star set, the well known family of convex resource theories is a subset of a more general class of star resource theories.  }
    \label{hierarchy}
\end{figure}

This article addresses the challenge of going beyond the paradigm of convexity, unveiling a novel class of resource theories: Star Resource Theories (SRTs). Unlike previous theories, SRTs pivot on a distinctive geometric attribute; the mathematical representation of  their resourceless devices form \emph{star domains} \cite{coxeter1989}. This characteristic trait delineates SRTs as a unique class, including all convex theories and crucial non-convex resource theories hitherto unexplored in their full depth and potential (see Fig.~\ref{hierarchy}). Within the SRTs class, we present two new quantifiers, showing their unprecedented operational meaning and demonstrating their enhanced capabilities for assessing and characterizing resources within the domain of non-convex theories. Notably, these quantifiers indicate advantages for \emph{correlated} discrimination tasks or the \emph{surplus} of cooperative discrimination tasks, establishing the foundations for novel quantum information applications.

Moreover, this article marks a pioneering effort, introducing a class of non-trivial free operations for the whole class of SRTs. Additionally, we study the conditions for resource non-generating operations and their connection with the internal structure of the corresponding star domain.

Intuitively, the advocated setup provides a means to exclude that a particular point belongs to a specific set, like resourceless objects. This capability is useful in various settings, even outside quantum information. Suppose that a system's evolution cannot lead out of a particular set. Then, certifying that the system's state is outside the set means that either the model or the experimental test is wrong.

 Some work along these lines can be found in \cite{Schluck_2022,adesso2023}; however, we will show that experimentally relevant theories, such as quantum-state texture \cite{Parisio2024}, are outside the scope of their approach but within the reach of our framework. Moreover, we will describe other operationally testable examples where only our approach is currently suitable for developing faithful witnesses and a complete resource theory.

Furthermore, we illustrate results obtained in well-known cases beyond the limitations of convex theories: quantum discord, total correlations, and non-Markovian processes. Additionally, we apply proposed methods to disprove the unistochasticity of a particular matrix, providing an operational test to demonstrate that, e.g., a classical walk could not have originated from a quantum one. We emphasize that traditional linear witness approaches have negligible (or even none) usefulness in the cases analyzed, underscoring the value of the methods proposed.

Beyond the previous immediate applications, the significance of this work extends to the very foundations of mathematical analysis. The introduction of SRTs hints at a broader mathematical generalization, a leap from the confines of convex analysis to the broader landscape of \emph{star-convex analysis} by pointing out the key features of star domains relevant for practical applications. Additionally, this work aligns with other recent developments in non-linear witnesses \cite{Otfried2006,Shen2020,Sen2023}, supporting a transformation in the understanding of constrained sets and offering a mathematical framework that transcends traditional boundaries.

In essence, this article not only presents a novel paradigm in resource theory but also serves as a cornerstone for future explorations. It paves the way for a foundational shift, offering a novel lens to study and manipulate resources, with potential implications ranging from quantum properties to particle physics and ushering in a new era of non-convex analysis.

The structure of this work is as follows: In Section \ref{sec:sett}, we introduce the notions of quantum mechanics
relevant to the present study 
and review  fundamental concepts of quantum resource theories. Section \ref{sec:srt} presents the star resource theories, along with their operational interpretations. Section \ref{sec:app} illustrates the practical application of our main result through specific examples of quantum discord, total correlations,  unistochasticity, and non-Markovian processes. Finally, in Section \ref{sec:dis}, we discuss implications of the results achieved
and their potential future applications.

\section{Setting the scene} \label{sec:sett}

\subsection{Quantum formalism}

A compact review of key concepts within quantum formalism becomes indispensable as we explore resource theories of quantum devices. Let us begin with the essentials: The space of pure states of a $d$-level system is described in quantum mechanics by a $d$-dimensional Hilbert space $\mathcal{H}$ and the states $\ket{\psi}\in\mathcal{H}$ are given by vectors of a unit norm, $\abs{\braket{\psi}} = 1$, equivalent up to a global phase change $\ket{\psi}\sim e^{i\phi}\ket{\psi}$.

Observables within the quantum theory are described by the bounded operators $\mathbb{B}(\mathcal{H})$ on the Hilbert space. A special subset $\mathbb{S}(\mathcal{H})\subset\mathbb{B}(\mathcal{H})$ of positive semidefinite operators $\mathbb{S}(\mathcal{H})\ni\rho\geq0$ with unit trace, $\Tr(\rho) = 1$, describe what is called density operators, which give the full state-space of quantum mechanics. More specifically, the rank-1 projectors are in one-to-one correspondence with the pure states, $\ket{\psi} \rightarrow \rho_\psi = \ketbra{\psi}$, and their convex combinations provide the mixed states. The set $\mathbb{S}$ of quantum states can be endowed with a metric relevant to the single-shot discrimination of the states, namely the trace distance induced by the trace norm, $\norm{\rho}_1 := \Tr\abs{\rho}$. With this, the optimal minimum-error probability of distinguishing between two states $\rho_1$ and $\rho_2$ by a single measurement is stated by Helstrom theorem to be $p = \frac{1}{2}\qty(1 + \norm{\rho_1 - \rho_2}_1)$.

Evolution of closed systems in quantum mechanics is described by unitary operations, which act on pure states as $\ket{\psi} \rightarrow \ket{\psi'} = U\ket{\psi}$ and by extension, density operators evolve as $\rho\rightarrow\rho' = U\rho U^\dag$. Evolution of open quantum systems can be more diverse and is described by completely positive trace preserving (CPTP) maps $\Theta: \mathbb{S}(\mathcal{H}) \rightarrow \mathbb{S}(\mathcal{H})$ which take density matrices to density matrices. It has been shown that any CPTP map can be written in terms of Kraus operators, $\Theta(\rho) = \sum_i K_i \rho K_i^\dag$, which satisfy the identity resolution condition, $\sum K_i^\dag K_i = \iden$. Unitary evolution therefore can be seen as a special case of a CPTP map defined by a single Kraus operator. A norm for the channels similar to the trace norm for the states is called the diamond norm. It is defined by the trace norm of the state resulting from acting with a channel $\Theta$, optimized over all possible states $X\in \mathbb{B}(\mathcal{H})$ on an extended space, $\norm{\Theta}_\diamondsuit := \underset{X:\norm{X}_1 \leq 1}{\max}\qty(\iden_d\otimes\Theta)\qty(X)$. It has been shown that the diamond norm can be approximated efficiently using convex programming procedures \cite{Watrous_2009}. Similarly to the Helstrom theorem for the states, the optimal probability of distinguishing between two channels $\Theta_1$ and $\Theta_2$ is expressed as $p = \frac{1}{2}\qty(1 + \norm{\Theta_1 - \Theta_2}_\diamondsuit)$. Furthermore, it has been used to prove several results concerning the complexity classes of several categories of quantum computation \cite{Watr2003,rosgen:2008}.

Quantum channels on a $d$-dimensional space and states of dimension $d^2$ are closely related by the Choi-Jamiołkowski isomorphism. Consider a maximally entangled Bell state of dimension $d^2$, $\ket{\Psi_+} = \frac{1}{\sqrt{d}} \sum_{i=1}^d \ket{i,i}$ and an arbitrary channel $\Theta$, which takes $d$-dimensional quantum states to $d$-dimensional states. The Jamiołkowski state is defined by acting with a channel $\Phi$ on the second party of the Bell state $\ket{\Psi_+}$ and leaving the first party unperturbed $J_{\Theta} := \qty(\iden\otimes\Theta)\qty(\ketbra{\Psi_+})$ \cite{choi1975completely,jamiolkowski1972linear}.

\subsection{Geometric fundamentals} \label{Geofund}

\begin{figure}[H]
    \

    \includegraphics[width=0.9\columnwidth]{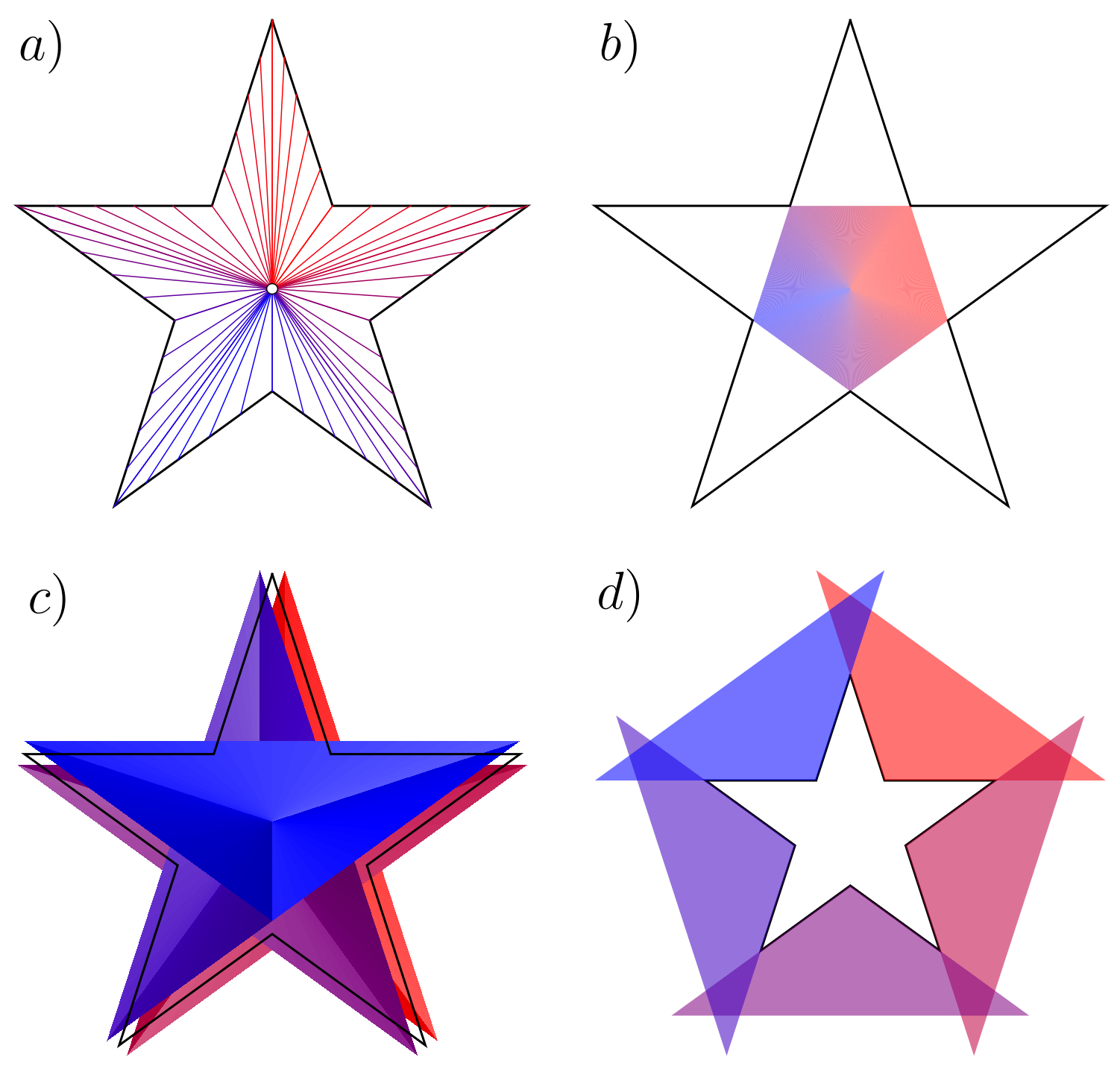}
    \caption{Examples of star domains and their natural structures. a) shows a center of the star connected to the whole set, while b) shows the set of all possible centers, known as the kernel of the star. c) Shows some of the largest convex subsets on the star under inclusion, and d) shows support cones tightly separating the star from outer space.}   
    
    \label{fig:stardom_conpol}
\end{figure}

A foundational take of essential geometric concepts becomes paramount in developing of this framework. In this section, we explore fundamental analytical and geometric definitions necessary for constructing and comprehending the
theory of star quantum resource theories -- see Fig. \ref{fig:stardom_conpol}.

In the following, given a set $A$, we use the standard notation $A^{\textbf{c}}$
and $\textrm{Int}(A)$ for the complement and interior of $A$, respectively.
Additionally, we write $\partial A$ to designate the boundary of
$A$, $\textrm{Conv}(A)$ to the convex hull of $A$ and $\textrm{Ext}(A)\equiv[\textrm{Int}(A)]^{\textbf{c}}$,
which we name the \emph{exterior} of $A$, equivalent to the complement of the interior of $A$. When $A\subset V$ of some
vector space $V$, we write $\left(A\pm x\right)$ for a particular
$x\in V$ to denote the set of all elements $a\pm x$, with $a\in A$.

The definition of  \emph{star domains} in a vector space $V$, also known as \emph{star-shaped sets}
 \cite{coxeter1989,munkres2000topology}
is central for the present work.

\begin{defn}
Let $V$ be a vector space, then a subset $\mathcal{K}\subset V$ is  \emph{star-shaped} and is called a \emph{star domain} if there exists an element $\mu_0 \in \mathcal{K}$ such that for all $\mu \in \mathcal{K}$, \mbox{$\lambda\in [0,1] $} the convex combination $\lambda\mu_0+(1-\lambda )\mu\in \mathcal{K}$. Any such element $\mu_0 $ is denoted as a \emph{center} of $ \mathcal{K}$. Additionally, the property of being a \emph{star domain} is known as \emph{star-convexity}.
\end{defn}

Conceptually,  imagine a star domain as a set where at least a single point links to every other point in the set via lines that never leave its boundaries. The definition of the star domain includes all non-empty convex sets and a selection of non-convex sets, such as stars or crosses.

 Star domains have already a long history of mathematical research, dating back to early Kepler studies of regular star polygons \cite{coxeter1989} up to other modern exceptional cases such as stellations \cite{coxeter1951} with an impact on art, culture and graphics \cite{Moreale2003,Magnenat2012}. Also, star domains recently found applications in some areas like discrete computational geometry, fixed point theory, optimization and neural networks \cite{Hansen2020,preparata2012,crespi2005,Hu2014,fang2017,Annesi2023}. 

 An important aspect of a star domain $\mathcal{K}$ is its \emph{kernel}, denoted
as $\mathrm{Ker}\left(\mathcal{K}\right)$, which comprises the union
of all the centers $\mu_{0}$ of $\mathcal{K}$ in a unique, convex set \cite{Hansen2020}. In addition to the kernel, star-convex sets have multiple internal structures, among which the \emph{convex components} stand out. Precisely, a \emph{convex component}
$\mathcal{K}_{y}$ of a star domain $\mathcal{K}$ is a maximal, with respect to
inclusion, convex subset of $\mathcal{K}$. Evidently, the convex components form a tight covering for $\mathcal{K}$:
\begin{equation}
\mathcal{K}=\bigcup_{y\in Y}\mathcal{K}_{y}.
\end{equation}

 Moreover, the full collection of convex components
$\left\{ \mathcal{K}_{y}\right\}_{y\in Y} $
of $\mathcal{K}$ could be infinite and uncountable, but always their intersection is the kernel of $\mathcal{K}$ \cite{Hansen2020}:
\begin{equation}
\textrm{Ker (\ensuremath{\mathcal{K}})}=\bigcap_{y\in Y}\mathcal{K}_{y}.
\end{equation}

Note that, for a compact star domain with polyhedral boundary, we define its \emph{convex cells} as the connected, convex, compact subsets arising from intersections of the half-hyperplanes determined by the polyhedral faces of the boundary.

\begin{figure}[ht]
    \centering
    \includegraphics[width=1\linewidth]{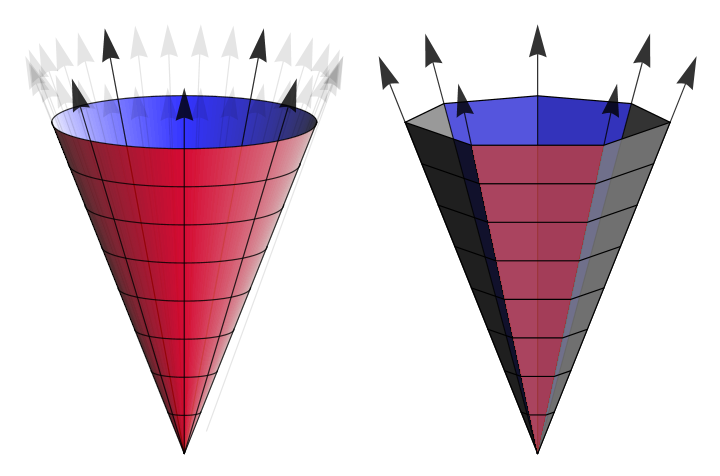}
   
    \caption{Two exemplary convex cones, with the one on the left possessing an infinite frame, and the second requiring only a finite frame of degree 7. In each case we highlighted the vectors in the frame, with the full frame for the first cone sketched, as it is infinite.}
    \label{fig:convexcones}
\end{figure}

Another key aspect of the construction proposed
is  use of \emph{cones} in a vector space $V$ \cite{schrijver1998theory,de2012algebraic}. A subset $\mathcal{C}\subset V$
is a \emph{cone} if there is an element $v_{0}\in V$ such that for
each $v\in V,\,\lambda\in]0,\infty[$ we have:
\begin{equation}
v\in\mathcal{C}\Rightarrow v_{0}+\lambda\left(v-v_{0}\right)\in\mathcal{C},
\end{equation}
which often is writen as $\lambda\left(\mathcal{C}-v_{0}\right)\subset\left(\mathcal{C}-v_{0}\right)$.
If $\mathcal{C}$ is also a convex set, we call it a \emph{convex cone}. A set $\mathcal{T} \subset \mathcal{C}$ is called a \emph{frame} of $\mathcal{C}$,
if $\mathcal{T}$, but no proper subset of~$\mathcal{T}$, spans~$\mathcal{C}$ positively. Particularly relevant to us is the specific case of \emph{convex polyhedral cones}:

\begin{defn}
Let $V$ be a vector space; then a set $\mathcal{C}$ is a \emph{convex polyhedral cone} if it is generated by the following convex combination of finitely many vectors $\left\{ v_{0},v_{1},...,v_{n}\right\} $ in $V$,	
	\begin{equation}\label{eq:polycone}
	\mathcal{C}:\left\{ v_{0}+\sum_{i=1}^{n}\alpha_{i}\left(v_{i}-v_{0}\right)\mid\alpha_{i}\in\mathbb{R}_{\geq0},v_{i}\in V\right\},
	\end{equation}
where we denote $v_{0}$ as an \emph{apex} of $\mathcal{C}$. Additionally
if $\left\{ v_{0},v_{1},...,v_{n}\right\} $ is the minimal number
of vectors to write every $v\in\mathcal{C}$ as in (\ref{eq:polycone})
we denote them as \emph{principal vectors}, and $n$ is the \emph{degree}  of the cone. In fact, the difference between principal vectors and the apex form the finite frame of the convex polyhedral cone $\mathcal{C}$,  See Fig \ref{fig:convexcones}. 
\end{defn}

Notably, when $\mathcal{C}$ has a unique apex $v_{0}$, the differences between the principal
vectors and  $v_{0}$ are conically independent,
hence $\alpha_{i}\!\in\!\mathbb{R}_{\geq0}$,\!\! $\sum_{i=1}^{n}\!\alpha_{i}\!\left(\!v_{i}\!-\!v_{0}\!\right)\!=\!0$
implies $\alpha_{1}\!=\!\alpha_{2}\!=\!...\!=\!\alpha_{n}\!=\!0$ \cite{Marshall1967}.

A way to visualize a convex polyhedral cone $\mathcal{C}$ is as a cone made up of flat faces, like a pyramid. Additionally, the flat faces that make up the cone must be polyhedra, meaning they have a finite number of flat faces and vertices. A fundamental property of every $\mathcal{C}$ is its equivalence to an intersection of a finite collection  of halfspaces in $V$ with a common vector $v_{0}$  \cite{lovasz1986matching}.

\begin{figure}[ht]
    \centering
    \includegraphics[width = .75\linewidth]{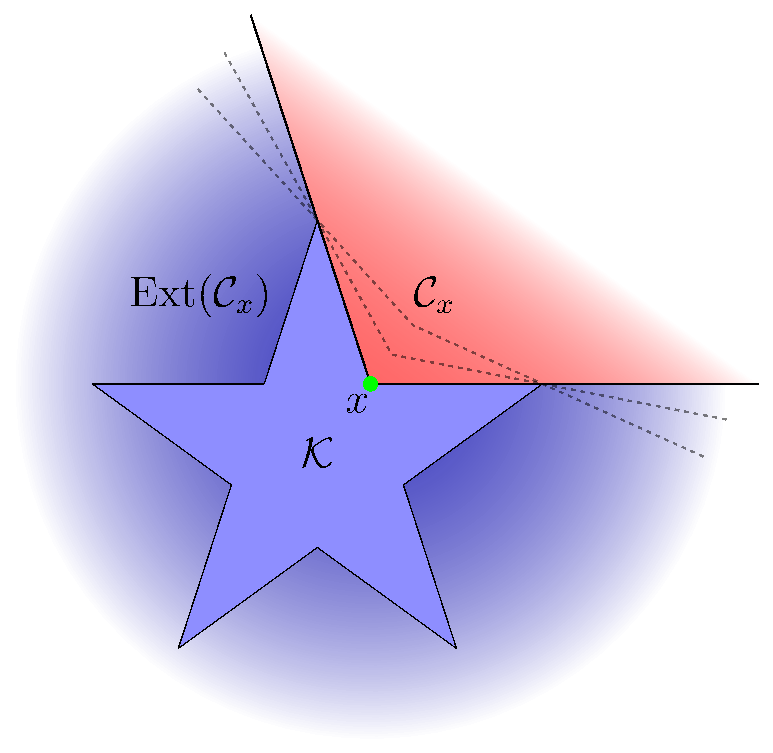}
    \caption{An example of a support cone $\mathcal{C}_x$ for the star set $\mathcal{K}$ with the apex $x\in\partial\mathcal{K}$ (red gradient with solid boundary), with the entire star set in the exterior of the cone, $\mathcal{K}\subseteq\operatorname{Ext}(\mathcal{C}_x)$.}
    \label{fig:suppcone}
\end{figure}

Likewise, other key types of cones worth to mention are \emph{reflected} and \emph{support cones}. For a given convex cone $\mathcal{C}$ with frame $\mathcal{T}$, the reflected cone  $\mathrm{Refl}\left(\mathcal{C}\right)$ has the same apex vectors as $\mathcal{C}$, but with frame $-\mathcal{T}$. In particular if $\mathcal{C}$ is a polyhedral cone as defined in (\ref{eq:polycone}), then any $v_r\in\mathrm{{Refl}}\left(\mathcal{C}\right)$ it is of the form: \begin{equation}\label{eq:refpolycone}
	v_r= v_{0}-\sum_{i=1}^{n}\alpha_{i}\left(v_{i}-v_{0}\right), \textrm{with  }\alpha_{i}\in\mathbb{R}_{\geq0}.
	\end{equation}
A convex cone $\mathcal{C}_{x}$ with apex $x$ and
non-empty interior is a \emph{support cone} of $A$ at $x$ if $x\in V$,
$\textrm{Ext}(\mathcal{C}_{x})\supseteq A$ and $\mathcal{C}_{x}$
is a maximal (with respect to inclusion) convex cone with these properties.
Support cones play the same role for star domains as support half-spaces
for convex sets, hence in \cite{Hansen2010} is shown that for a closed
and bounded star domain $\mathcal{K}\subset V=\mathbb{R}^{d}$, there
exist a support cone $\mathcal{C}_{x}$ at every $x\in\partial\mathcal{K}$, See Fig. \ref{fig:suppcone}.

Additionally, to eliminate redundancy in 
the construction presented, the following mathematical tool will be helpful: Let $\mathfrak{A}=\left\{ \mathcal{A}_{x}\right\} _{x\in X}$ be a
collection of subsets of a set $V$ and let $\mathcal{K}\subseteq V$
be fixed. We define the \emph{redundancy deletion} of $\mathfrak{A}$
through $\mathcal{K}$ as the sub-collection:
\begin{eqnarray}
\mathrm{Red}_{\mathcal{K}}\left(\mathfrak{A}\right) & : & =\left\{ \mathcal{A}_{x}\in\mathfrak{A}\mid\mathcal{A}_{x}\cap\mathcal{K}\:\textrm{is \ensuremath{\subseteq}-maximal }\right.\nonumber \\
 &  & \left.\;\;\quad\qquad\textrm{\textrm{ in \ensuremath{\left\{  \mathcal{A}_{x^{\prime}}\cap\mathcal{K}\mid x^{\prime}\in X\right\} } } }\right\}. \nonumber
\end{eqnarray}
\noindent
In addition, we will denote $\mathfrak{A}\succ_{\mathcal{K}}\mathfrak{B}$
to indicate that all members of $\mathfrak{A}$ whose intersection
with $\mathcal{K}$ is strictly contained in that of another member
are removed, i.e. $\mathfrak{B}=\mathrm{Red}_{\mathcal{K}}\left(\mathfrak{A}\right).$

Note, that for given $\mathfrak{A}$ and $\mathcal{K}$, the redundancy
deletion is unique because the set of $\ensuremath{\subseteq}$-maximal
elements of a poset is uniquely determined. Also, we remark that while
for the general case, we could show the existence of $\mathrm{Red}_{\mathcal{K}}\left(\mathfrak{A}\right)$
by invoking Zorn's lemma, in our work we do not need this
lemma to guarantee existence of such maximal intersections. Indeed,
in our work we only consider redundancy deletions with $\mathfrak{A}$
and $\mathcal{K}$ such that each $\mathcal{A}_{x}\cap\mathcal{K}$
is a non-empty compact subset of a compact set, and
that the collection is closed under limits of increasing chains. Under
the above conditions, every increasing chain of intersections has
a supremum (given by the closure of the union), which remains compact.
Hence, the poset of intersections is chain-complete and therefore contains
maximal elements.

Having covered the previous concepts, we are ready for the final crucial geometrical concept:

\begin{defn}\label{fortress}
We will say that a \emph{fortress} $\mathfrak{T}_{\mathcal{K}}$
for the set $\mathcal{K}$ is a collection of convex cones
$\left\{ \mathcal{C}_{x}\right\}_{x\in X} $
such that: 

\begin{enumerate}
    \item[i)] $ \left\{ \mathcal{K},\mathcal{C}_{x}\right\}_{x\in X} \:\textrm{is a covering for \ensuremath{V}}$,
    \item[ii)]  $\textrm{Ext}(\mathcal{C}_{x})\supseteq\mathcal{K}\:\textrm{for every \ensuremath{\mathcal{C}_{x}} in \ensuremath{\mathfrak{T}_{\mathcal{K}}}}$,
    \item[iii)]  $\textrm{Refl}(\mathcal{C}_{x})\supseteq \textrm{Ker}(\mathcal{K})\:\textrm{for every \ensuremath{\mathcal{C}_{x}} in \ensuremath{\mathfrak{T}_{\mathcal{K}}}}$,
    \item[iv)] $\partial\mathcal{K}\!=\!\bigcup_{\ensuremath{\mathcal{C}_{x}}\in\ensuremath{\mathfrak{T}_{\mathcal{K}}}}\!\left(\mathcal{C}_{x}\cap\mathcal{K}\right)\:\textrm{with \ensuremath{\mathcal{C}_{x}\!\cap\!\mathcal{K}\!\neq\!\emptyset,}  \ensuremath{\!\forall\: \mathcal{C}_{x}}}$.
\end{enumerate}

\end{defn}

A fortress $\mathfrak{T}_{\mathcal{K}}$ could be infinite and uncountable, thus  when all cones $\mathcal{C}_{x}$ are convex polyhedral, that is, of finite frame, we will denote it as a \emph{polyhedral fortress}. Even more, 
note that in the case of  $V=\mathbb{R}^{d}$ the support cones of $\mathcal{K}$ always form a fortress $\mathfrak{T}_{\mathcal{K}}$. Precisely, condition ii) follows from the definition of support cone, while iv) is a simple consequence that a support cone with apex at $x$ exists for every  $x\in\partial\mathcal{K}$.

Now, we show the collection of support cones to satisfy conditions i) and iii) by means of the following lemmas:

\begin{lem}\label{lemma1}
Let $\mathcal{K}\subset V$ be a compact star-shaped set with kernel
$\textrm{Ker (\ensuremath{\mathcal{K}})}$. For each $x\in\partial\mathcal{K}$,
let $\mathcal{C}_{x}$ be any support cone of $\mathcal{K}$ with
apex at $x$, and $\mathfrak{T}_{x}$ the collection of all $\mathcal{C}_{x}$.
Then
\begin{equation*}
V=\mathcal{K}\cup\qty(\bigcup_{x\in\partial\mathcal{K}}\bigcup_{\mathcal{C}_{x}\in\mathfrak{T}_{x}}\mathcal{C}_{x}).
\end{equation*}

\end{lem}
\emph{Proof: }Let $z\notin\mathcal{K}$ be arbitrary and choose any
$w\in\textrm{Ker (\ensuremath{\mathcal{K}})}$. For a compact star
domain $\mathcal{K}$ the segment connecting $z$ and $w$ intersects the boundary
$\partial\mathcal{K}$ at a single point $z_{w}=\lambda z+(1-\lambda)w$,
for some $\lambda\in[0,1)$, and consequently: 
\begin{equation*}
z-z_{w}=(1-\lambda)(z-w),
\end{equation*}
i.e., the ray with apex $z_{w}$ and frame $z-z_{w}$ points outward
$\mathcal{K}$, and therefore belongs to every support cone $\mathcal{C}_{z_{w}}$.
Since the previous ray contains $z$, we obtain $z\in \mathcal{C}_{z_{w}}$
for every support cone $\mathcal{C}_{z_{w}}$. Thus, every point is
in $\mathcal{K}$ or in some $\mathcal{C}_{z_{w}}$, proving the claim.

\begin{lem}\label{lemma2}
Under the  same assumptions of Lemma \ref{lemma1}, for each $x\in\partial\mathcal{K}$,
let $\mathcal{C}_{x}$ be any support cone of $\mathcal{K}$ with
apex at $x$. Then
\begin{equation*}
\textrm{Ker (\ensuremath{\mathcal{K}})}\subseteq\textrm{Refl (\ensuremath{\mathcal{C}_{x}})},
\end{equation*}
with $\textrm{Refl (\ensuremath{\mathcal{C}_{x}})}$ the reflected
cone of $\mathcal{C}_{x}$.

\end{lem}

\emph{Proof: }From the construction in Lemma 1 it follows that for
the boundary point $x=\lambda x_{w}+(1-\lambda)w$ for some point
$x_{w}\in\mathcal{C}_{x}$ and an arbitrary $w\in\textrm{Ker (\ensuremath{\mathcal{K}})}$.
Hence, 
\begin{equation*}
w=x-\left(\frac{\lambda}{1-\lambda}\right)\left(x_{w}-x\right),
\end{equation*}
is a vector in $\textrm{Refl (\ensuremath{\mathcal{C}_{x}})}$, but
since $x$ and $w$ are arbitrary, every kernel point lies in the reflection cone of every support cone, i.e., the claim follows.

From the above, it follows that Lemma \ref{lemma1} shows property i) and Lemma \ref{lemma2} the property iii), ensuring that the full collection of support cones forms a fortress of $\mathcal{K}$.

Intuitively, a fortress is a collection of cones whose outer surfaces are glued together to sections of the set, defining the maximal visibility region from a point to the outer space, analogously to the shooting visibility area for artillery at a defense wall of an 18th-century fortress. Indeed,  $\mathfrak{T}_{\mathcal{K}}$  must allow each part of the set's outer space to be distinguished, including every point outside it in at least one of the collection's cones. In the 18th-century defensive structures, the above condition recalls the need for exterior points to serve as clear targets for the defender's cannons while  excluding the fortress's interior from the potential targets.

\mybox{Summary 1}{blue!40}{blue!10}{\textbf{Key geometrical concepts:} 
\begin{enumerate}
    \item[1)] \textbf{Star domain:} Abstract geometrical definition of sets with a star shape.
    \item[2)] \textbf{Kernel:} Set of all centers of a star domain.
    \item[3)]  \textbf{Convex components:} Largest convex sets inside a star domain.
    \item[4)] \textbf{Convex polyhedral cones:} Cones generated by a finite collection of vectors.
    \item[5)] \textbf{Fortress of a set:} A collection of convex cones tightly separating the set from the outer vector space.
\end{enumerate}
}
   
\subsection{Quantum Resource Theories}

Quantum resource theories (QRTs) offer a way to analyze the employment of quantum properties as resources for practical tasks. It's common to manipulate fundamental quantum objects like states $\rho$, measurements $\mathbf{M}$, and quantum channels $\Theta$ in order to exploit the desired quantum properties \cite{Gour2019}. In present investigation, we describe objects as quantum channels for convenience, but the results obtained can also be extended to states and measurements, which will be covered later. 

A resource theory starts by defining the free set $\F $, a specific subset of the objects under study $\mathcal{S}$. The free set must consist solely of objects lacking the desired property and serves as a benchmark for measuring the advantage of objects with the property. Then, we should include the theory operations $\O$, known as free operations, that formally preserve the free set $\F $ and are naturally suitable for their intended use. At last, we quantify the resourcefulness of $\S$ by  a  monotone measure $\M$ defined as a non-negative valued function, $\M: \S \mapsto [0, +\infty[$, satisfying $\M(\v{f}) = 0$ for all $\v{f} \in \F$ and being non-increasing under free operations: $\M(\psi(\v{s})) \leq \M(\v{s})$ for all $\psi \in \O$ and $\v{s} \in \S$ \cite{Gour2019}. 

The mathematical structure of a resource theory is comprised of the triple $ \left\{ \mathcal{F},\mathcal{O},\M \right\} $, but for it to be considered physically meaningful, it needs an operational interpretation. This interpretation links $\M(\v{s})$ to the advantage of $\v{s}$ in performing a practical task, as described in \cite{Gour2019}. Typically, the operational interpretation is specific to each case and based on a prior understanding of the particular task at hand \cite{Gour2019}. However, there are also examples of universal interpretations, where the mathematical structure guarantees it to hold for a wide range of resource theories.

An essential class of QRTs with universal interpretation are those in which $\mathcal{F}$ is a convex set and the limit of any sequence $\{\psi_k\in \O \}_{k\in\mathbb{N}} $ is also a free operation, designated as convex resource theories. This class encompasses several well-known QRTs, including entanglement, coherence, asymmetry, and athermality \cite{Gour2019}. Convex QRTs, with their inherent mathematical structure, gain from the rich outcomes of convex analysis. The hyperplane separation theorem \cite{rock1970} plays a crucial role in explicitly providing a universal interpretation for these theories. The study in \cite{TakagiRegula2019} revealed a crucial insight: hyperplane separations determine discrimination tasks for each class of quantum mechanics objects. This breakthrough led to a universal interpretation of convex QRTs.

Takagi and Regula's work in \cite{TakagiRegula2019} revealed the significance of separation hyperplanes in determining universal interpretations for convex QRTs and opened up a new avenue in convex QRT research. However, this raises the question: \emph{Can other hypersurfaces achieve the same outcome?} The following section answers in the affirmative and presents a unified class of theories with applications in quantum discord, total correlations, unistochasticity and markovianity.

\mybox{Summary 2}{blue!40}{blue!10}{ 
\textbf{Quantum resource theory:}

It comprises a triple $ \left\{ \mathcal{F},\mathcal{O},\M \right\} $. Here, $\F$ represents the set of free objects, encompassing entities like quantum states or operations, within a larger set $\S$, encompassing the same kind of quantum objects. The set $\mathcal{O}$ comprises free operations, consisting of functions $\psi :  \S \mapsto \S$ that leave the set of free objects invariant. A function $\M: \S \mapsto [0, +\infty[$, exhibiting monotonicity under free operations and satisfying $\M(\v{f}) = 0$ for all $\v{f} \in \F$, is crucial. This function $\M$ is designated to quantify the resourcefulness inherent to objects within the set $\S$.

\vspace{0.2 cm}

 \textbf{Quantum convex resource theory:}

 An important class of resource theories, characterized by a free set $\F$ which is convex. Remarkably, there are monotones with universal operational interpretation for this class.
}

\section{Star Resource Theories} \label{sec:srt}

The captivating allure of star-shaped geometrical objects transcends the boundaries between art, mathematics, and the exploration of natural phenomena. Throughout history, artists have woven these intricate shapes into a fabric of expression,  ancient symbols from natives of Latin America, adorning Islamic architectural wonders, and finding their place in the meticulous knotwork of Celtic art. Beyond their aesthetic appeal, star-shaped figures have also fascinated mathematicians, particularly in studying star domains and intricate star polygons. As we delve into the depths of aesthetic mathematics, where beauty and precision converge, the historical significance of these shapes beckons us to unravel their secrets for practical applications. Indeed, envisioning the free set as a star domain in the theory of resources opens new possibilities.

Embarking on the journey to construct a star resource theory necessitates solely a compact star
domain as the free set $\mathcal{F}$, coupled with a fortress $\mathfrak{T}_{\mathcal{F}}$. Remarkably, we demonstrate the
consistent construction of an appropriate fortress $\mathfrak{T}_{\mathcal{F}}$
for any compact star domain $\mathcal{F}$ encompassing quantum states,
channels, or measurements with suitable representations. Building upon this foundation, we introduce
two novel quantifiers $\mathcal{G}_{\mathcal{L}}\left(\cdot\mid\mathcal{F}\right)$ and $\mathcal{G}_{\mathcal{R}}\left(\cdot\mid\mathcal{F}\right)$, depicting their
practical significance in correlated discrimination tasks and the
surplus of cooperative discrimination tasks, respectively. 

Later, in the quest for non-trivial
free operations $\mathcal{O}$, we unveil a universally valid class
applicable to any compact star domain $\mathcal{F}$. Furthermore,
we delve into expanded classes imposing additional suitable constraints on the
star structure of $\mathcal{F}$ and its relation with the corresponding
fortress $\mathfrak{T}_{\mathcal{F}}$, uncovering their meaningful
implications across multiple concrete applications.

 Aside from the fascinating geometry of star shapes, we will see how to exploit their features to construct resource theories applicable in quantum technologies, deepen our understanding of physical models and motivate the development of new mathematical tools. Then, we formally denote a star resource theory in the following: 

\begin{defn}{\textbf{(SRT)}:} 
We designate a resource theory  $\left\{ \mathcal{F},\mathcal{O},\M \right\} $ as a \emph{star resource theory}  if: \\ \\
$1)\quad    $The free set $\mathcal{F}$ is a compact star domain. \\ \\
   $2)\quad    $There exists a fortress $\mathfrak{T}_{\mathcal{F}}$ for the free set. \\ \\
   $3)\quad    $The limit of any sequence $\{\widehat{\Lambda}_k\in \O \}_{k\in\mathbb{N}} $ is  also a free operation in $\O$.

 \end{defn}

 Note that for any type of quantum object there exist linear representations
where the set of valid objects $\mathcal{S}$ is bounded \cite{Ritter2005,Bertlmann2008}, which implies that also $\mathcal{F}$ is bounded in such representation, making natural the choice of a compact $\mathcal{F}$.

Examples of geometrical shapes satisfying the conditions for a free set in an SRT are common in some ancient cultures as well as artistic designs (See Fig. \ref{fig:my_label}). Exploiting these geometrical wonders becomes a conduit for understanding and harnessing the inherent properties of nature, revealing a harmonious intersection where artistry, mathematics, and resource theory coalesce to unlock the mysteries of the world around us.

\mybox{Summary 3}{blue!40}{blue!10}{ \textbf{Key advantages of Quantum SRTs:}

\begin{enumerate}
    \item[1)] \textbf{Free set:} The simple condition of being a star domain applies to all known connected free sets of the literature, and provides a rich geometrical structure with potential physical meaning.
    \item[2)] \textbf{Quantifier:} Faithful, convex function, well-behaved continuity, reduces ambiguity and suppresses the relative errors in contrast to previous approaches. We later show it has novel operational interpretations. 
    \item[3)]  \textbf{Free operations:} We introduce  class of free operations existing in every star resource theory. Additionally, we provide two simple, sufficient conditions to certify resource non-generating operations. Furthermore, we describe a third class of operations rich in instances and relevant for applications.

    \end{enumerate}
    
    }

\subsection{Free set and fortress property}\label{freesetfortress}

A crucial feature of the free set in the theories proposed
stems from being a \emph{star domain} compact subset $\mathcal{F}\subset\mathcal{S}$ of the objects under study, where $\mathcal{S}$ also possesses a vector space structure with an inner product $\left\langle \cdot,\cdot\right\rangle $. However, it is worth noting that linear real representations, like the generalized
Bloch ball \cite{Ritter2005,Bertlmann2008}, exhibit a property where convex combinations
of two quantum objects transform into convex combinations of the same
objects. Consequently, this property ensures that convex sets of quantum
objects retain their convexity within the linear real representation.
Furthermore, the above phenomenon extends to star domains, transforming them into star domains as well, as they retain their characteristic structure by preserving convex combinations between any free quantum object and those within the kernel.

The previous property allows
us to consistently choose $V=\mathbb{R}^{d}$ in the geometric definitions
of subsection \ref{Geofund} when restricted to quantum objects. Indeed, any quantum star domain within the original set translates to a star domain within a linear real representation. Even more, appropriate linear representations exist in the larger landscape of causal general probabilistic theories~\cite{Chiribella2010,Scandolo2021}, and quantifiers analogous to the ones we use in this work \cite{TakagiRegula2019}. In summary, the methods presented are extensive to essentially all meaningful physical theories.

Utilizing the above observation, we could regard a quantum
$\mathcal{F}$ as a star domain within $\mathbb{R}^{d}$, and select the collection $\mathfrak{I}_{\mathcal{F}}=\left\{ \mathcal{C}_{x}^{\prime}\right\}_{x\in\partial\mathcal{F}} $
composed of support cones, as the fortress in the linear representation. While this fact already proves the existence of a fortress for quantum SRTs, employing the entire set of support cones might prove impractical due to its demand for a cone at each boundary point of $\mathcal{F}$. Since, later, we will evaluate the quantifier separately at each cone in the fortress,  selecting $\mathfrak{I}_{\mathcal{F}} $ would often necessitate redundant evaluations. Nonetheless, we provide a method for substantial enhancement in this selection, offering the prospect of a notable reduction in the required support cones, limiting the evaluation of quantifiers to a finite number in several scenarios.

Addressing the previous difficulty, we employ the process of redundancy deletion through $\mathcal{F}$ on
the collection~$\mathfrak{I}_{\mathcal{F}}$. This streamlining method eliminates unnecessary cones while retaining vital information about boundary~$\partial\mathcal{F}$, resulting in a more practical fortress. This produces a sub-collection
\begin{equation*}
\mathfrak{T}_{\mathcal{F}}^{\prime}=\mathrm{Red}_{\mathcal{F}}\left(\mathfrak{I}_{\mathcal{F}}\right),
\end{equation*}
which retains only those cones whose intersections with $\mathcal{F}$
are maximal. The resulting family $\mathfrak{T}_{\mathcal{F}}^{\prime}$
significantly reduces the number of cones while still encoding all
the boundary information required from $\mathfrak{I}_{\mathcal{F}}$.
Note, that as we mentioned in Section \ref{Geofund} the redundancy deletion
for a given $\mathfrak{I}_{\mathcal{F}}$ and $\mathcal{F}$ is unique,
and since $\mathcal{F}$ is compact, the intersections $\mathcal{C}_{x}\cap\mathcal{F}$
form a chain-complete poset under inclusion, making $\mathfrak{T}_{\mathcal{F}}^{\prime}$
well-defined. Next, to ensure that this reduction preserves the structural
properties we need, we verify that each cone $\mathcal{C}_{x}\in\mathfrak{T}_{\mathcal{F}}^{\prime}$
satisfies the fortress conditions i)--iv). Precisely, since each
$\mathcal{C}_{x}$ is a support cone of $\mathcal{F}$, conditions
ii) and iii) follow directly. Condition iv) is preserved by redundancy
deletion, which removes only cones with strictly smaller intersections
with $\mathcal{F}$ and therefore cannot eliminate maximal ones. The
proof of the covering property i) is more delicate, and for clarity
is postponed to  Appendix \ref{app:covering}.  

We now exploit the geometric structure that a fortress induces inside
a compact star domain $\mathcal{F}$. As we mentioned in section 2.2,
such $\mathcal{F}$ admits a unique decomposition into convex components:
\begin{equation*}
\mathfrak{C}_{\mathcal{F}}=\left\{ \mathcal{F}_{y}^{\prime}\right\} _{y\in Y}.
\end{equation*}
Given a fortress $\mathfrak{T}_{\mathcal{F}}^{\prime}$, each cone
$\mathcal{C}_{x}\in\mathfrak{T}_{\mathcal{F}}^{\prime}$ selects a
sub-collection of these convex components by redundancy deletion,
\begin{equation*}
\mathfrak{C}_{\mathcal{F}}\succ_{\mathcal{C}_{x}}\{\mathcal{F}_{y}^{\prime}\}_{y\in Y_{x}}.
\end{equation*}
Note, here that every $\mathcal{C}_{x}\cap\mathcal{F}_{y}^{\prime}$
is a non-empty convex compact subset of the compact set $\mathcal{C}_{x}\cap\mathcal{F}$,
and similarly as in the previous case form a chain-complete poset
under inclusion, such that every sub-collection $\{\mathcal{F}_{y}^{\prime}\}_{y\in Y_{x}}$
is unique and well-defined. However, several convex components may
still share the same maximal intersection with $\mathcal{C}_{x}$,
a remaining redundancy that we also wish to eliminate. To solve the
previous inconvenience, first consider a sub-collection
\begin{equation*}
\{\mathcal{F}_{z}^{\prime}\}_{z\in Z_{x}^{\eta}}\subset\{\mathcal{F}_{y}^{\prime}\}_{y\in Y_{x}},
\end{equation*}
consisting of all components satisfying:
\begin{equation*}
\mathcal{F}_{z}^{\prime}\cap\mathcal{C}_{x}=\mathcal{J}_{\eta}^{x},
\end{equation*}
where $\mathcal{J}_{\eta}^{x}$ is one of the maximal convex and compact
intersections $\mathcal{C}_{x}\cap\mathcal{F}_{y}^{\prime}$ under
inclusion.

Second, we define a \emph{free section} $\mathcal{F}_{\eta}^{(x)}$
of $\mathcal{F}$ as the intersection
of all such convex components $\{\mathcal{F}_{z}^{\prime}\}_{z\in Z_{x}^{\eta}}$:
\begin{equation}
\mathcal{F}_{\eta}^{(x)}=\bigcap_{z\in Z_{x}^{\eta}}\mathcal{F}_{z}^{\prime},\label{eq:freesection}
\end{equation}
which are all convex subsets of $\mathcal{F}$, and \begin{equation*}
    \textrm{Conv}(\textrm{Ker}\ensuremath{(\mathcal{F})}\cup\mathcal{J}_{\eta}^{x})\subseteq\mathcal{F}_{\eta}^{(x)}.
\end{equation*} We remark here, that our streamlining method produces a unique and well-defined set at every step, hence generating a unique set of free sections, for a given initial fortress $\mathfrak{T}_{\mathcal{F}}^{\prime}$ of a compact star domain $\mathcal{F}$. 

Furthermore, from the definition of fortress:
\begin{equation*}
    \partial\mathcal{F}=\bigcup_{x,\eta}\mathcal{J}_{\eta}^{x}.
\end{equation*}Then, since for a star domain $\mathcal{F}$ every point is a convex combination
between a point at the boundary and the kernel, we have: 
\begin{equation}
\bigcup_{x,\eta}\mathcal{F}_{\eta}^{(x)}\supseteq\bigcup_{x,\eta}\textrm{Conv}(\textrm{Ker}\ensuremath{(\mathcal{F})}\cup\mathcal{J}_{\eta}^{x})=\mathcal{F}.
\end{equation}
Even more, since each $\mathcal{F}_{\eta}^{(x)}$ is a subset of $\mathcal{F}$
follows: 
\begin{equation}
\mathcal{F}=\bigcup_{x,\eta}\mathcal{F}_{\eta}^{(x)},
\end{equation}
and in conclusion the free sections generated by our method form a covering of $\mathcal{F}$. Later, we select the collection

\begin{equation*}
    \mathfrak{C}_{\mathcal{F}}^{(x)}=\{ \mathcal{F}_{\eta}^{(x)}\}_{\eta\in \Sigma_x},
\end{equation*}   containing all the information about the intersection
between $\mathcal{F}$ and $\mathcal{C}_{x}$, using a minimal number of convex subsets. 

If at this point every number $\Sigma_x$ of maximal
sections $\mathcal{J}_{\eta}^{x}$ is finite we are done and we could
select $\mathfrak{T}_{\mathcal{F}}=\{ \mathcal{C}_{x}\} _{x\in X}$, denoting it as the \emph{canonical fortress}.

It is important to clarify at this point how the computational complexity of constructing a 
canonical fortress depends on the geometry of the free set. In general, both 
the number and the local visibility structure of the boundary faces of 
$\mathcal{F}$ contribute to the complexity: constructing each support cone 
amounts to computing the set of directions that immediately exit 
$\mathcal{F}$, which is a visibility computation in the ambient space 
$\mathbb{R}^d$. For a polyhedral free set with $m$ facets, the worst-case 
combinatorial complexity of such a visibility region scales as 
$m^{\lfloor d/2 \rfloor}$, following classical bounds on polyhedral visibility 
and facet enumeration \cite{McMullen1970,Ziegler1995}. Consequently, constructing all support cones 
requires at most $O(m^{\,1+\lfloor d/2 \rfloor})$ operations. 
The subsequent redundancy-deletion step, which tests inclusion between 
intersections of cones and boundary facets, hence in dimension $d-1$, takes 
$O(m^{2+\lfloor (d-1)/2 \rfloor})$ time in the worst case, dominating over the construction of support cones.

While these dimension-dependent bounds describe the worst-case scenario, the 
practical computational cost is usually governed not by the ambient dimension 
but by the \emph{geometric complexity} of $\partial\mathcal{F}$: the number of 
facets, their arrangement, and the presence (or absence) of irregular nonconvex features. In many physically relevant settings, such as free 
sets with symmetric, smooth, or low-complexity boundaries, the number of 
distinct support cones is small, and the fortress can be computed efficiently 
even in high dimensions.

A simple example shows that worst-case complexity bounds can be overly pessimistic 
for structured free sets. Let $P$ be a $(d\!-\!1)$-dimensional polytope with $m$ 
facets, embedded in $\mathbb{R}^d$, and form the lateral boundary of a 
$d$-dimensional pyramid by connecting every boundary point of $P$ to a single 
apex $a$. Although general visibility algorithms in dimension $d$ would have a 
worst-case complexity scaling as $O(m^{1+\lfloor d/2 \rfloor})$ to 
identify all relevant support cones, the canonical fortress in this case is 
immediate: it consists of the $m$ supporting half-spaces induced by the facets 
of the pyramid, together with a single internal cone with apex $a$ whose 
generators are the vectors from $a$ to the vertices of $P$. If the facets of $P$ 
are given explicitly, this fortress can be constructed in linear time in the 
number of facets, i.e.\ $O(m)$. Similarly, constructing the free  sections amounts to taking the convex hull between $a$ and each facet of $P$ which is 
therefore  also linear in the number of facets. This example illustrates that 
the practical complexity of fortress construction is often governed by geometric 
structure rather than by worst-case asymptotic bounds.

Furthermore, for settings where $\partial\mathcal{F}$ is extremely irregular—e.g., 
high-dimensional many-body systems—one need not compute the exact fortress. 
The free set can be approximated from inside and outside by star-shaped sets 
with simpler, more regular boundaries, yielding efficiently computable upper 
and lower bounds on the SRT monotone. This approach parallels standard techniques in computational geometry \cite{ComGeo1,ComGeo2,ComGeo3}. A systematic development of scalable approximation schemes for such complex free sets is an interesting direction for future work.

 Moving forward, it is known
that each closed subset of the reals is the disjoint union of a perfect
set and a countable set \cite{engelking1989general}. While a perfect set could still be
pathological, such as a Cantor set and the countable set could be infinite,
in practice quantum objects have always a finite fidelity $\epsilon$
which determines a neighborhood of the object where each one is indistinguishable
from others. Assuming a finite distinguishability $\delta_{\epsilon}$ --two points whose separation is less than $\delta_\epsilon$ cannot be operationally distinguished-- we could
use a minimal finite union of disjoint compact intervals in
the reals, $\delta_{\epsilon}$-indistinguishable from both perfect and countable sets composing the
set of intersections $\mathcal{J}_{\eta}^{x}$, to replace it. More formally, we can replace $\mathcal{J}^x_\eta$ by its $\delta_\epsilon$–approximation $(\mathcal{J}^x_\eta)_{\delta_\epsilon}$ such that
\begin{equation}
d_{\delta_\epsilon}\bigl(\mathcal{J}^x_\eta,(\mathcal{J}^x_\eta)_{\delta_\epsilon}\bigr)=0,
\end{equation}
where $d_{\delta_\epsilon}$ is the measure-based symmetric-difference pseudometric \cite{folland1999real}.

In other words, distinctions finer than $\delta_\epsilon$ are operationally meaningless. Therefore, the index set is finite, $|\Sigma_x|<\infty$. Similarly, a finite $\delta_{\epsilon}$-approximation of every frame $\mathcal{T}_x$ corresponding to a $\mathcal{C}_x$ generates a polyhedral fortress.
Hence, for quantum objects with finite distinguishability $\delta_{\epsilon}$,
we always can find a fortress $\mathfrak{T}_{\mathcal{F}}=\left\{ (\mathcal{C}_{x})_{\delta_\epsilon}\right\} _{x\in X}$
leading to sections $\mathfrak{C}_{\mathcal{F}}^{(x)}=\{ (\mathcal{F}_{\eta}^{(x)})_{\delta_\epsilon}\} _{\eta\in \Sigma_x}$
where $\Sigma_x$ is finite, and any further refinement is physically
irrelevant. 

In what follows, we assume that the free set admits finitely many free sections. We emphasize that this is an operational assumption, reflecting the finite resolution with which any physical experiment can distinguish different boundary features of the free set. Consequently, only finitely many sections are operationally relevant. This finiteness is not required for our mathematical results involving only the geometric quantifier and free operations, which continue to hold when the free set possesses infinitely many sections. However, the assumption streamlines the operational interpretations and simplifies the notation.

\mybox{Summary 4}{blue!40}{blue!10}{ \textbf{Key construction remarks of Quantum SRTs:}
\begin{enumerate}
    \item[1)] A star domain of quantum objects is also a star domain in any linear real representation.
    \item[2)] For any star domain in a linear real representation we can select the set of support cones as a fortress.
    \item[3)]  The full set of support cones is usually an impractical choice of fortress. However, we show that under redundancy deletion, we obtain a more economic fortress. 
    \item[4)] We denote as free section the intersection of all convex components overlapping at the same maximal intersection surface with a cone of the fortress.
    
    \end{enumerate}
    
    }

\subsection{Monotones}
\label{subsec:mon}

For every $\mathcal{F}$ which is a closed star domain of quantum
objects we can exploit the geometrical considerations of the previous
section, to always obtain a fortress $\mathfrak{T}_{\mathcal{F}}=\left\{ \mathcal{C}_{x}\right\} _{x\in X}$,
with every cone $\mathcal{C}_{x}$ leading to a finite collection
of free sections  $\mathfrak{C}_{\mathcal{F}}^{(x)}=\{ \mathcal{F}_{j}^{(x)}\} _{j=1}^{M_{x}}$.
Then, provided a quantifier $\mathcal{M}(\cdot\!\!\mid\!\!\mathcal{X}):\mathcal{S}\rightarrow\mathbb{R}_{\geq0}$,
such that $\mathcal{M}(\Phi\mid\mathcal{X})=0$ if $\Phi\in\mathcal{X}$
we define for every $\mathcal{C}_{x}\in\mathfrak{T}_{\mathcal{F}}$
a domain $\mathcal{D}^{(x)}\!=\mathcal{C}_{x}\!\cup_{j}\!\mathcal{F}_{j}^{(x)}$
and a monotone $\mathcal{G}_{\mathcal{M}}\!\!:\!\mathcal{D}^{(x)}\!\!\rightarrow\mathbb\!{R}_{\geq0}$:
\begin{equation}\label{eq:Conemeasure}
\mathcal{G}_{\mathcal{M}}\left(\Theta\mid\mathfrak{C}_{\mathcal{F}}^{(x)}\right)=\mathbb{G}_{j}\left[\mathcal{M}\left(\Theta\mid\mathcal{F}_{j}^{(x)}\right)\right],
\end{equation}
where $\mathbb{G}_{j}\left[x_{j}\right]:=\sqrt[M]{\prod_{j=1}^{M}x_{j}}$
is the geometric mean of the $M$ variables $x_{j}$, which in
the case analyzed are the $M_{x}$ convex subsets $\mathcal{F}_{j}^{(x)}$
when we evaluate an object $\Theta\in\mathcal{D}^{(x)}$. Some examples
of $\mathcal{M}(\cdot\mid\mathcal{F}_{j}^{(x)})$ used in QRTs are the generalized robustness, the minimum trace distance for states, the diamond norm for channels and the operator norm for measurements \cite{Gour2019}. The geometric intuition of (\ref{eq:Conemeasure}) becomes evident in the particular case that $\mathcal{M}$ is a distance quantifier $\mathcal{L}$; in such a case, the condition \mbox{$\mathcal{G}_{\mathcal{L}}(\Theta\mid\mathfrak{C}_{\mathcal{F}}^{(x)})=\mathrm{const.}$} determines a hyperbolic surface asymptotically approaching the free sections $\mathcal{F}_{j}^{(x)}$. This observation reveals that $\mathcal{G}_{\mathcal{L}}$ corresponds to the hyperbolic hypersurface --asymptotic to the free sections of $\mathcal{F}$-- that best distinguishes the resource from free objects. In this way, proposed quantifier extends previous methods based on separation hyperplanes \cite{Takagi2019,TakagiPRL2020,TakagiRegula2019} to hyperbolic separation hypersurfaces.

 Because the collection $\{ \mathcal{D}^{(x)}\}_{x\in X} $
is a covering of  $\mathcal{S}$ for all  $\Theta\in\mathcal{S}$ there exists at least one $x^{*}$ such that $\Theta\in\mathcal{D}^{(x^{*})}$, allowing us to define:
\begin{equation}
\mathcal{G}_{\mathcal{M}}\left(\Theta\mid\mathcal{F}\right)=\sup_{\mathcal{D}^{(x)}:\Theta\in\mathcal{D}^{(x)}}\left\{ \mathcal{G}_{\mathcal{M}}\left(\Theta\mid\mathfrak{C}_{\mathcal{F}}^{(x)}\right)\right\}. \label{eq:monotoneF}
\end{equation}
By construction 
when $\Phi\in\mathcal{F}$ we obtain $\mathcal{G}_{\mathcal{M}}(\Phi\mid\mathcal{F})=0$ since in that case $\Phi\in\mathcal{F}_{j}^{(x)}$
for some $j$ at the domains evaluated in (\ref{eq:monotoneF}). Even more, it is direct to verify that a quantifier $\mathcal{G}_{\mathcal{M}}$ is faithful if  $\mathcal{M}$ is faithful. Additionally, if \mbox{$\mathcal{M}(\cdot\mid\mathcal{X})$} is a convex function relative to a convex set $\mathcal{X}$, then every $\mathcal{G}_{\mathcal{M}}(\cdot\mid\mathfrak{C}_{\mathcal{F}}^{(x)})$ is also convex. Hence, for any convex combination of two resources $\Theta_1, \Theta_2$ with the same domain $\mathcal{D}^{(x)}$ of evaluation in (\ref{eq:monotoneF}), the quantifier  $\mathcal{G}_{\mathcal{M}}$ is also convex. Consequently, the virtues of $\mathcal{M}(\cdot\mid\mathcal{X})$ when $\mathcal{X}$ is convex are inherited by $\mathcal{G}_{\mathcal{M}}$ when $\mathcal{F}$ is a non-convex star domain, which does not happen when applying  $\mathcal{M}$ directly to $\mathcal{F}$ \cite{Schluck_2022}.

A general case in which the quantifier $\mathcal{M}(\cdot\mid\mathcal{X})$ is directly applicable to  $\mathcal{F}$, we consider monotones 
based on a decomposition into sub-convex sets \cite{adesso2023,Kuroiwa2023}:
\begin{equation}
\mathcal{M}\left(\Theta\mid\mathcal{F}\right)=\inf_{\mathcal{F}_{\zeta}\subseteq\mathcal{F}}\mathcal{M}\left(\Theta\mid\mathcal{F}_{\zeta}\right),\label{infdec}
\end{equation}
with $\{\mathcal{F}_{\zeta}\}$ any collection of convex subsets, satisfying $\mathcal{F}=\bigcup_{\zeta}\mathcal{F}_{\zeta}$. Such quantifiers rely on identifying a minimally distinguishable subset from the resource.

Then, it is straightforward to show the following simple upper and lower bounds:
\begin{equation}
\mathcal{M}(\Theta\mid\mathcal{F}) \leq \mathcal{G}_{\mathcal{M}}\left(\Theta\mid\mathcal{F}\right) \leq \mathcal{M}(\Theta\mid\mathrm{Ker}\left[\mathcal{F}\right]).
\label{eq:monotoneBounds}
\end{equation}
Remarkably, Lipschitz-continuity of several $\mathcal{M}(\cdot\mid\mathcal{F})$ is enforced by star-convexity of $\mathcal{F}$, for instance in the case of generalized robustness \cite{Schluck_2022}. The former remark implies that the assumptions made contribute to showing the existence of well-behaved lower bounds and underscore its relevance  for the mathematical well-behavior of quantifiers of the form~(\ref{infdec}).

Additionally, an upgrade of quantifier (\ref{eq:monotoneF}) helpful to deal with general resource non-generating operations is the following:
\begin{equation}
\overline{\mathcal{G}}_{\mathcal{M}}\left(\Theta\mid\mathcal{F},\overline{\mathcal{O}}\right)=\sup_{\widehat{\Lambda}\in\overline{\mathcal{O}}}\mathcal{G}_{\mathcal{M}}\left(\widehat{\Lambda}\left[\Theta\right]\mid\mathcal{F}\right),\label{eq:GupgradeRNG}
\end{equation}
 where $\overline{\mathcal{O}}$ is a subset of valid operations mapping $\mathcal{F}$
into itself, and such that any sequence $\{ \widehat{\Lambda}_{n}\in\overline{\mathcal{O}}\} _{n\in\mathbb{N}}$
converges to an operation in $\overline{\mathcal{O}}$. With these definitions
the supremum in (\ref{eq:GupgradeRNG}) is reached by a concrete
$\widehat{\Lambda}^{*}\in\overline{\mathcal{O}}$ and the quantifier $\overline{\mathcal{G}}_{\mathcal{M}}$ is non increasing under the full set $\overline{\mathcal{O}}$.

\begin{figure}[h]
	\begin{tikzpicture}
	\node at (0,0) {\includegraphics[width=.475\linewidth]{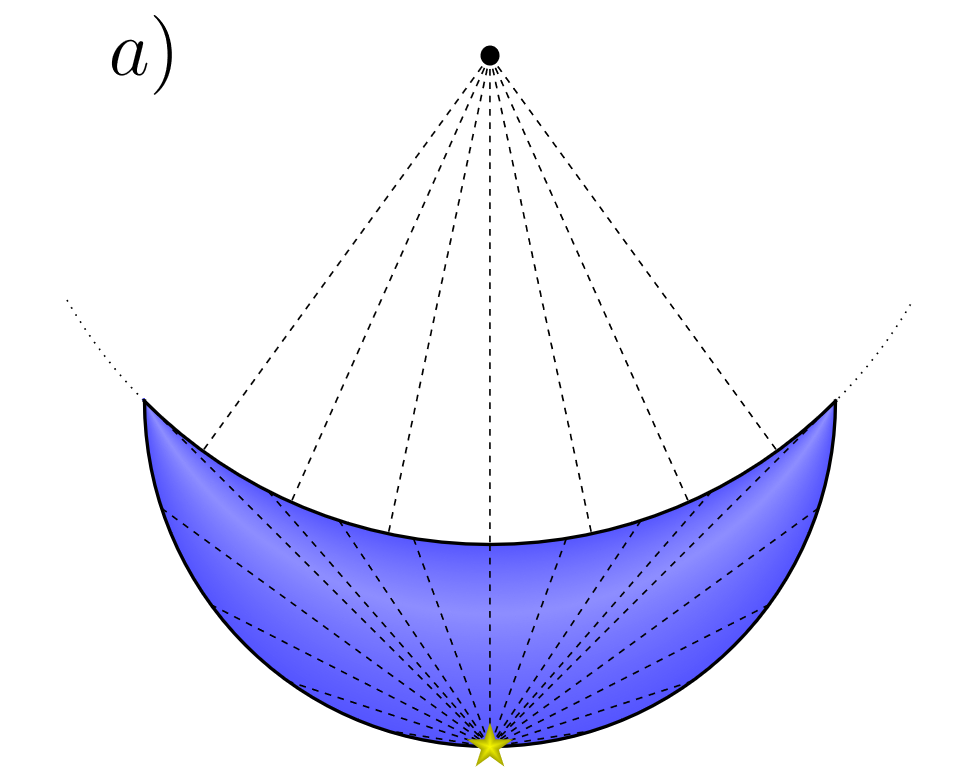}
    \includegraphics[width=.475\linewidth]{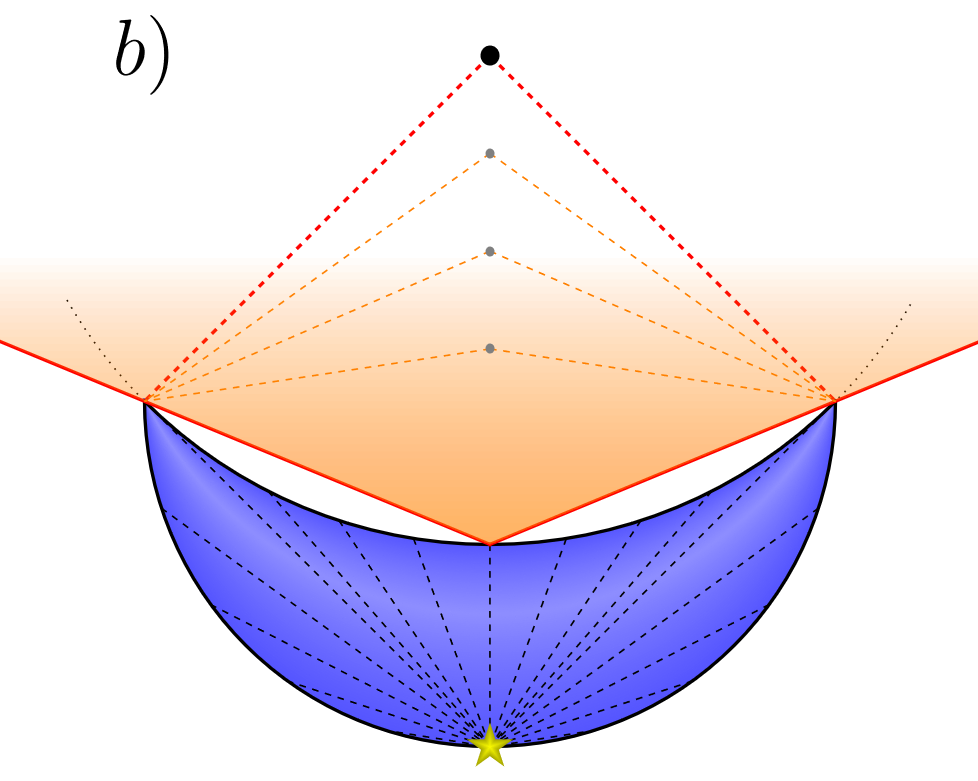}};		
	\node at (-0.4\columnwidth,-0.15\columnwidth) {$\mathcal{F}$};
        \node at (0.09\columnwidth,-0.15\columnwidth) {$\mathcal{F}$};
	\end{tikzpicture}
	\caption{Comparison between different approaches. a)  Previous methods associate the monotone $\mathcal{M}$ with the least distinguishable convex subset of the free set $\mathcal{F}$. b) Our approach selects only the convex subsections of $\mathcal{F}$ intersecting the support cone that best separate the resource, to construct quantifier $\mathcal{G}_\mathcal{M}$.
    In this way we reduce ambiguity and make the quantifier more robust against measurement or computational errors.}
    \label{fig:ambiguity}
\end{figure}

Contrary to the claim of previous works \cite{Schluck_2022,adesso2023,turnerAdesso2024}, here we show there exist quantum properties where their quantifiers are inapplicable. Concretely, the quantum-state texture \cite{Parisio2024} is an experimentally valuable resource where the free set consists of a single point on the \emph{boundary} of quantum states. However, the generalized robustness quantifier (See \cite{Vidal1999} and definition (\ref{eq:genrobust}) in this article) for any free set lying entirely on the boundary of the set of objects, results in an infinite value, trivializing the resource theory. Since the approach in \cite{Schluck_2022,adesso2023,turnerAdesso2024} involves minimizations of the form (\ref{infdec}), solely over the generalized robustness, it became inapplicable in its current form to any resource theory with a free set entirely inside the boundary of the space of objects, particularly quantum-state texture. Our approach does not rely solely on generalized robustness because we include quantifiers based on the trace distance, for which the above mentioned issue does not apply. Furthermore, in Appendix \ref{app:counterexample}, we provide an example of a case in which the free set lies entirely in the boundary of the set of objects and forms a star domain, a situation that, at present, only our framework can treat.

Additionally, while our quantifiers are tailored specifically for star domains, this distinction does not pose a significant restriction. Notably, all currently known \emph{connected} free sets associated with quantum properties exhibit the characteristic of being star domains. Therefore, our geometric condition aligns seamlessly with the prevailing structure of relevant free sets.

We remark that all established quantum resource theories—such as entanglement, coherence, asymmetry, thermodynamics based on Gibbs-preserving operations, quantum correlations, and non-Markovianity—feature connected free sets. The only example we are aware of that proposes a disconnected free set is Example 3 in \cite{Kuroiwa2023}, but this construction is non-standard and raises questionable conceptual issues (e.g., depolarising noise can create a resource). It is also not representative of any experimentally relevant scenario. Furthermore, the work of Wojewódka et al. \cite{WojewodkaSciazko2024}, cited in \cite{Kuroiwa2023}, does not involve such a structure. In that framework, each Gibbs state at a fixed temperature constitutes its own free set, so no disconnected free set appears.  These considerations suggest that disconnected free sets do not arise in physically meaningful or operationally justified quantum resource theories, even though they can be artificially constructed.

Moreover, earlier quantifiers often are prone to non-uniqueness, thereby introducing ambiguity in the task for which the resource is advantageous (Fig. \ref{fig:ambiguity}). In contrast, our approach adopts a distinctive strategy by simultaneously considering all pertinent convex sections of the complete set. This simultaneous consideration significantly mitigates ambiguity in resource assessment, offering a distinct understanding of the advantage provided. 

Furthermore, with simple error propagation arguments, we  show in Appendix \ref{relerrosupr} that the relative error of the quantifier (\ref{eq:monotoneF}) is smaller than the corresponding relative error of monotone~(\ref{infdec}):
\begin{equation}
\frac{1}{\sqrt{\Sigma}}\frac{\delta\mathcal{M}(\Theta\mid\mathcal{F})}{\mathcal{M}(\Theta\mid\mathcal{F})} \geq \frac{\delta\mathcal{G}_{\mathcal{M}}\left(\Theta\mid\mathcal{F}\right)}{\mathcal{G}_{\mathcal{M}}\left(\Theta\mid\mathcal{F}\right)},
\label{eq:monotoneRelError}
\end{equation}
where $\Sigma\geq 1$ depends of $\Theta$ and $\mathcal{F}$.  We remark that errors here mean deterministic Lipschitz-type worst-case sensitivity bounds for the functional considered. In fact, we consider errors as arbitrary, possibly even correlated, and do not need to follow any noise model. The bound we derive is valid for all possible realisations of such errors, and therefore does not depend on any probability distribution.

As a result, the approach presented suppresses the effect of relative errors in contrast to previous works. Hence, by encompassing multiple convex sections simultaneously, 
the method introduced provides a robust and involved framework for quantifying resources in non-convex sets, promising advancements in precision and reliability compared to previous methodologies.

\subsection{Free operations}
Delineating the fundamental classes of free operations within star resource theories is crucial for defining how one can manipulate, interconvert, and model resources in the face of environmental degradation; we categorize them into distinct classes, each imbued with unique attributes. Foremost is the class of \emph{section preserving operations}, which ensures the preservation of free sections within the star domain. Our exploration presents a notable non-trivial case within this category, demonstrating its relevance in real-world scenarios. The second class, Resource Non-Generating (RNG) operations, introduces natural sufficient conditions with the potential to provide profound physical insights. The third class, \emph{hyperbolic contraction operations}, emerges as a beacon of practical significance, albeit demanding additional conditions. Hyperbolic contractions consistently manifest across diverse applications, underscoring their inherent relevance in resource theories. Beyond these established classes, our exploration extends to potential free operations such as hyperbolic rotations and squeezing maps, and offers promising avenues for further refining of the framework.

\subsubsection{Section preserving operations}

Given a free star-convex set of free objects $\mathcal{F}$ with a suitable fortress $\mathfrak{T}_{\mathcal{F}}=\left\{ \mathcal{C}_{x}\right\} _{x\in X}$, the most natural class of free operations $\mathcal{O}$ would be those non-increasing, for any  $\mathcal{M}(\cdot\mid\mathcal{F}_{j}^{(x)})$ composing  the quantifier (\ref{eq:monotoneF}), under mild conditions. The simplest way to achieve the above is by demanding every operation  to leave invariant each free section $\mathcal{F}_{j}^{(x)}$, hence the name \emph{section preserving operations}. Formally this requires every  $\widehat{\psi}\in \mathcal{O}$ to satisfy:
\begin{eqnarray}
\forall\,\Phi\in\!\mathcal{F}_{j}^{(x)}\Rightarrow &  &\,\,\, \widehat{\psi}\left(\Phi\right)\in\mathcal{F}_{j}^{(x)}.\label{eq:isomorphism}
\end{eqnarray}
Depending on how complex the free sections $\mathcal{F}_{j}^{(x)}$ are, a complete description of \eqref{eq:isomorphism} could be a formidable task. However, $\mathcal{F}$'s star domain structure helps us narrow down the possibilities. Note, for
instance, that from (\ref{eq:isomorphism}), $\mathcal{O}$ must leave invariant $\mathrm{Ker}\left(\mathcal{F}\right)$ since it is a subset of each free section. Hence, the set of kernel isomorphisms is a super-set of section preserving operations. The above fact is generally beneficial when the kernel has a simple structure, which is the usual case in most applications.

Nevertheless, we will
see that we can always define section preserving  free operations which are nontrivial and meaningful for the universal interpretations introduced later in this article. From the definition of a free section $\mathcal{F}^{(x)}_j$ it directly follows  that any operation \mbox{$\widehat{\Delta}_{\lambda}(\Phi)= \lambda \Phi + (1-\lambda) \Delta$} with $\lambda \in [0,1]$, $\Delta \in \mathrm{Ker}\left(\mathcal{F}\right)$ for $\Phi\in\mathcal{F}^{(x)}_j$   satisfies (\ref{eq:isomorphism}). Moreover, in Appendix A, we show that each operation $\widehat{\Delta}_{\lambda}$  is non-increasing for  (\ref{eq:monotoneF})  if the  $\mathcal{M}(\cdot\mid\mathcal{F}_{j}^{(x)})$ are convex functions like in the case of the generalized robustness or satisfies the triangle inequality as in the case of the trace distance, or diamond norm.

Noteworthy, the previous result is not trivial as it exploits the star-convexity of $\mathcal{F}$ in several ways. Formerly, the very existence of $\mathrm{Ker}\left(\mathcal{F}\right)$ and the convexity of every free section are a natural consequence of the star domain structure. Precisely faithful monotones
$\mathcal{M}(\cdot\mid\mathcal{X})$, such as the generalized robustness, are convex functions only if the reference set $\mathcal{X}$ is convex, as noted in \cite{Schluck_2022}. The above remark points out an advantage of the proposed framework for identifying free operations, in contrast to the previous approaches 
considering arbitrary reference sets $\mathcal{X}$ for simple faithful quantifiers $\mathcal{M}(\cdot\mid\mathcal{X})$. Later, in Appendix \ref{app:Monproofs}, the star properties, as well the fortress definition, play an explicit role in deriving that if $\widehat{\Delta}_{\lambda}\left[\Theta\right]\in\mathcal{C}_x$
then also $\Theta\in\mathcal{C}_x$, a crucial step in demonstrating the monotonicity of (\ref{eq:monotoneF}) under $\widehat{\Delta}_{\lambda}$.

Furthermore, we show this operations inherit a physical meaning from the special role of  $\Delta \in \mathrm{Ker}\left(\mathcal{F}\right)$ for the concrete examples of non-convex SRTs investigated.

\subsubsection{Resource non-generating operations}

As in the traditional definitions, in framework itroduced, we identify resource non-increasing operations (RNG) mapping the free set into itself, comprising the broadest set of potentially free operations. In addition, the structure determined by  $\mathcal{F}$ and $\mathfrak{T}_{\mathcal{F}}=\left\{ \mathcal{C}_{x}\right\} _{x\in X}$ within discussed framework enables us to articulate two distinct sufficient conditions with potential physical implications. Let us denote by $\widehat{\psi}(\mathcal{F}_{j}^{(x)})$  the sets
of objects $\widehat{\psi}\left(\Phi\right)$, where $\Phi\in\mathcal{F}_{j}^{(x)}$.
Then the first suficient condition requires that for every $j$ in $\Sigma_{x}$
\begin{equation}
\widehat{\psi}\left(\mathcal{F}_{j}^{(x)}\right)\subseteq\mathcal{F}_{k}^{(x)},\label{eq:RNGcond1}
\end{equation}
for some $k$ in $\Sigma_{x}$. A direct consequence of (\ref{eq:RNGcond1})
is that for every two free objects $\Phi_{1},\Phi_{2}\in\mathcal{F}_{j}^{(x)}$,
each of their convex combinations,
get mapped by a linear $\widehat{\psi}$ into convex combinations of $\widehat{\psi}(\Phi_{1}),\widehat{\psi}(\Phi_{2})\in\mathcal{F}_{k}^{(x)}$, hence preserving convex combinations inside the free sections.
In this way, a linear $\widehat{\psi}$ satisfying (\ref{eq:RNGcond1}), not
only is RNG, but in some sense evokes the property of preserving commutations, where free sections play the role of a common basis.
Similarly, if the collection $\{ \overline{\mathcal{F}}_{y}\} _{y\in Y}$
denote the convex components of $\mathcal{F}$, the second sufficient
condition states that for every $y$ in $Y$:
\begin{equation}
\widehat{\psi}\left(\overline{\mathcal{F}}_{y}\right)\subseteq\overline{\mathcal{F}}_{y^{\prime}},\label{eq:RNGcond2}
\end{equation}
for some $y^{\prime}$ in $Y$. In this case a linear $\widehat{\psi}$ satisfying
(\ref{eq:RNGcond2}) also possess the property of preserving convexity
from one $\overline{\mathcal{F}}_{y}$ into another $\overline{\mathcal{F}}_{y^{\prime}}$,
and later we will see that in concrete cases, such as quantum discord,
this property translates exactly into preservation of commutativity. Evidently,
operations $\widehat{\Delta}_{\lambda}$ with $\Delta\in\textrm{Ker\ensuremath{(\mathcal{F})}}$
 satisfy conditions (\ref{eq:RNGcond1}) and (\ref{eq:RNGcond2}),
however we can find many others by simply studying the isometries
of $\mathcal{F}$, which are rather simple to characterize in several
applications. Once we collect the operations $\widehat{\psi}$ satisfying conditions
(\ref{eq:RNGcond1}) or (\ref{eq:RNGcond2}), to determine a subset
$\overline{\mathcal{O}}$ of RNGs, we can check if $\mathcal{G}_{\mathcal{M}}$
is non-increasing under $\overline{\mathcal{O}}$ or arternatively,
we can update it and obtain a quantifier $\overline{\mathcal{G}}_{\mathcal{M}}$ as described in Eq. (\ref{eq:GupgradeRNG}).

\subsubsection{Hyperbolic contraction operations}

We introduce a distinctive class of potential free operations denoted
\emph{hyperbolic contraction operations}. While demanding additional
assumptions, this class proves particularly relevant in numerous practical
scenarios where these assumptions find fulfillment. This class consist in operations
applied cone-wise, that is $\widehat{\psi}^{x}:\mathcal{C}_{x}\rightarrow\mathcal{C}_{x}$,
forming a collection of operation sets $\{\mathcal{O}_{x}\}_{x\in X}$
induced by the fortress $\mathfrak{T}_{\mathcal{F}}$. Consequently,
for a cone $\mathcal{C}_{x}$ with apex $v_{0}^{(x)}$ and frame $\mathcal{T}^{(x)}=\{t_{y}^{(x)}\}_{y\in Y_{x}}$
we can describe them in terms of their action over the frame positive
span. For instance if $\Theta\in\mathcal{C}_{x}$ with positive span:
\begin{equation}
\Theta=v_{0}^{(x)}+\sum_{y\in Y_{x}}\alpha_{y}t_{y}^{(x)},\label{eq:spanT}
\end{equation}
with $\alpha_{y}\geq0$. A hyperbolic contraction $\widehat{\psi}^{x}$ is the following kind of unequal scaling over
the frame elements:
\begin{eqnarray}
\widehat{\psi}^{x}\left(\Theta\right)=v_{0}^{(x)}\!+\!\!\sum_{y\in Y_{x}}\!\alpha_{y}\lambda_{y}t_{y}^{(x)},\!\!\;\label{eq:HCO} \\ \textrm{with}\;\lambda_{y}\geq0,\; \textrm{and}\;1\!\geq\!\!\prod_{\lambda_{y}>0}\lambda_{y}.\nonumber
\end{eqnarray}
These operations include several interesting cases, such as squeeze
maps when the product of all $\lambda_{y}$ equals 1, conic non-increasing
order maps when every $\lambda_{y}\leq1$, or hyperbolic rotations
when for every $\lambda_{y_{+}}=e^{\varphi}$ there is a $\lambda_{y_{-}}=e^{-\varphi}$.
However, for such operations to be authentic free operations in our
framework, it is necessary to ensure the monotonically decrease of
quantifier (\ref{eq:monotoneF}).

We can guarantee the previous property through multiple geometric arguments, which vary according to the geometry of the free set $\mathcal{F}$ and the associated fortress $\mathfrak{T}_{\mathcal{F}}$. Since each argument cannot be applied to the most general case but covers a broad spectrum of possibilities, we will include them as observations when discussing 
specific cases.
It is enough to establish two essential conditions that will aim to show such arguments: I) when applying a $\widehat{\psi}^{x}$ operation, the evaluation of the domain in (\ref{eq:monotoneF}) is precisely $\mathcal{D}^{(x)}$, and II) the existence of a correspondence between the free sections and the boundary $\partial \mathcal{C}_x$ within $\mathcal{S}$, which allows showing that each geometrical mean (\ref{eq:Conemeasure}) is non-increasing under $\widehat{\psi}^{x}$.

\subsection{Universal operational interpretation}\label{optinter}

\subsubsection{Interpretation for distance based monotone}

Interactive proofs are a popular way of solving computational problems in computer science \cite{Gold1989}. They involve two types of players: \emph{provers} and a \emph{verifier}. The provers send messages to the verifier to help solve a problem, and the verifier uses these messages and their operations to determine the solution accurately. The problem's difficulty depends on the operations the verifier can perform and the resources used in the messages. If players use quantum resources like states, channels and measurements, the proof is called a \emph{Quantum Interactive Proof} (QIP) \cite{Watr2003}. In the following, we will follow the tradition in quantum information to simplify the nomenclature by presenting the QIP problems as a game between agents.

The promise game  \emph{Close-Images}, which we denote as $\mathtt{G_{I}}$, is known to be complete for QIP \cite{rosgen:2008} and it played a major role in proving QIP=PSPACE \cite{Watr2003}.  In the game $\mathfrak{G_{I}}$ a verifier (Bob) should distinguish between two convex sets of devices, by using the information and devices provided by a yes-prover (Alice) and a no-prover (Eve), which attempt to convince him that both sets are always distinguishable or not, respectively.

We will use a particular instance of the $\mathtt{G_{I}}$ game to provide a universal operational interpretation to all SRT of channels, when the base  monotones $\mathcal{M}(\cdot\mid\mathcal{F}_{j}^{(x)})$ are given by the diamond norm distance of $\Theta $ to the respective free section: 

\begin{equation}
\mathcal{L}(\Theta\mid\mathcal{F}_{j}^{(x)})=\inf_{\Phi\in\mathcal{F}_{j}^{(x)}}\frac{1}{2}\left\Vert \Theta-\Phi\right\Vert _{\diamondsuit}.\label{eq:diamnormgamma}
\end{equation}

We develop the operational interpretation for channels and diamond norm distances, since this later reduce naturally to the special cases of state, trace norm and measurement, operator norm respectively.

\begin{figure}[H]
    \centering
    \includegraphics[width=\columnwidth]{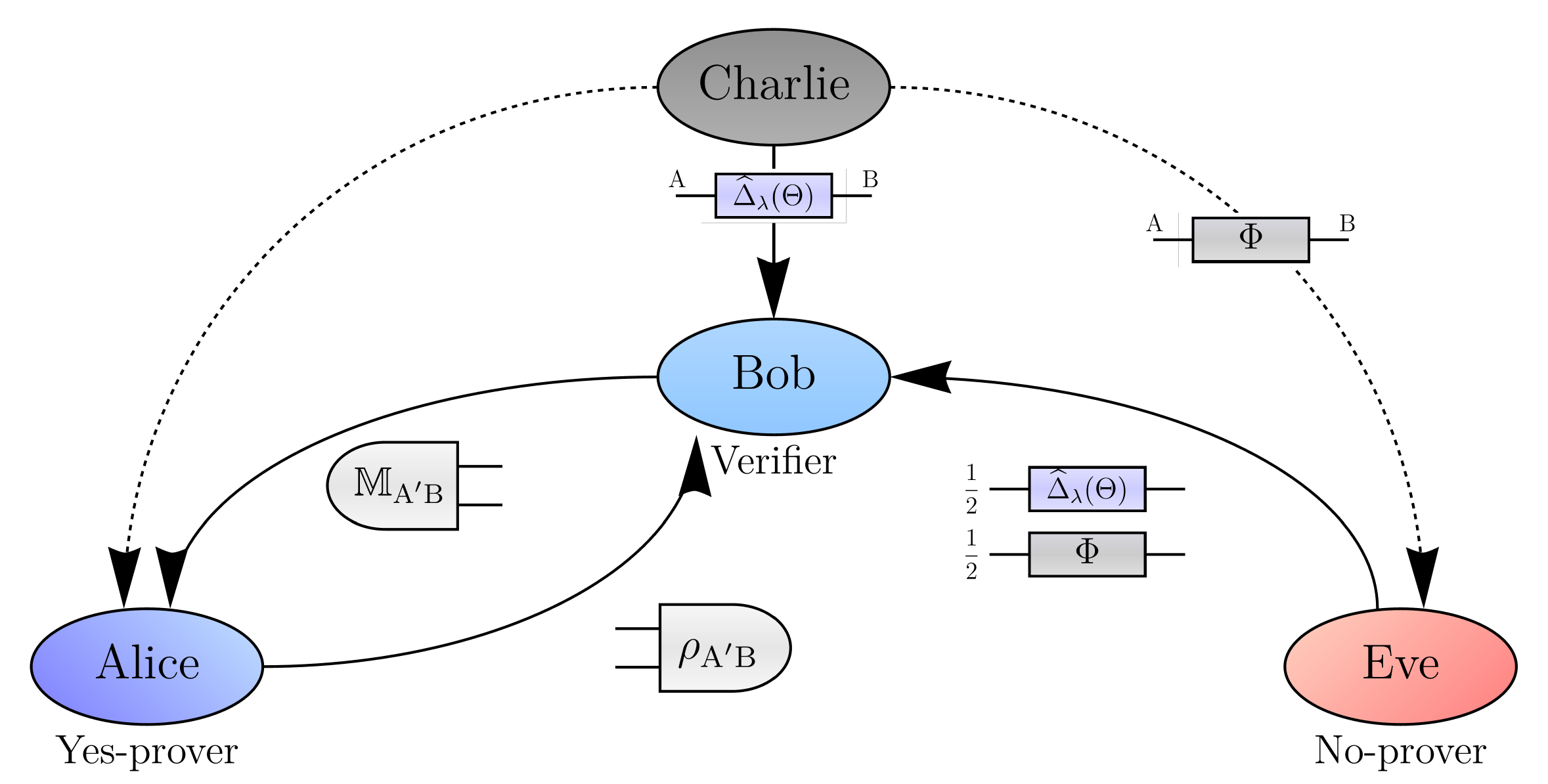}
    \caption{In the game $\mathtt{G_{I}}$, Alice and Eve compete to convince Bob of the answer to a problem posed by Charlie. Bob must determine if a black box from Charlie is different from a test box from Eve. The process starts with Eve sending her box to Bob, then Alice and Bob exchange messages, and finally Bob decides if the boxes are different.}
    
    \label{fig:singleplayer_CI}
    
\end{figure}

The game $\mathtt{G_{I}}$ we consider is the distinction  between the convex sets $\{ \Theta \}$  and $\mathcal{F}_{1}^{(x)} $, for a $\Theta\in\mathcal{D}^{(x)}$. The setting is presented in Figure \ref{fig:singleplayer_CI} where a referee (Charlie) provides a black box to Bob  with the promise of originally having the channel $\{ \Theta \}$, but potentially affected by noise. 

Simultaneously, Charlie informs Alice and Eve of the precise model of noise, represented by one of the free operations $\mathcal{O}$, say for instance $\widehat{\Delta}_{\lambda}$ or any operation satisfying \eqref{eq:isomorphism}.
In this setting, Eve sends to Bob a device with a channel $\Phi\in \mathcal{F}_{1}^{(x)}$. Since Eve is a no-prover, her best strategy is to choose a channel $\Phi$ as close to  $\widehat{\Delta}_{\lambda}(\Theta)$ as possible.

\begin{figure}[H]
    \centering
    \includegraphics[width=\columnwidth]{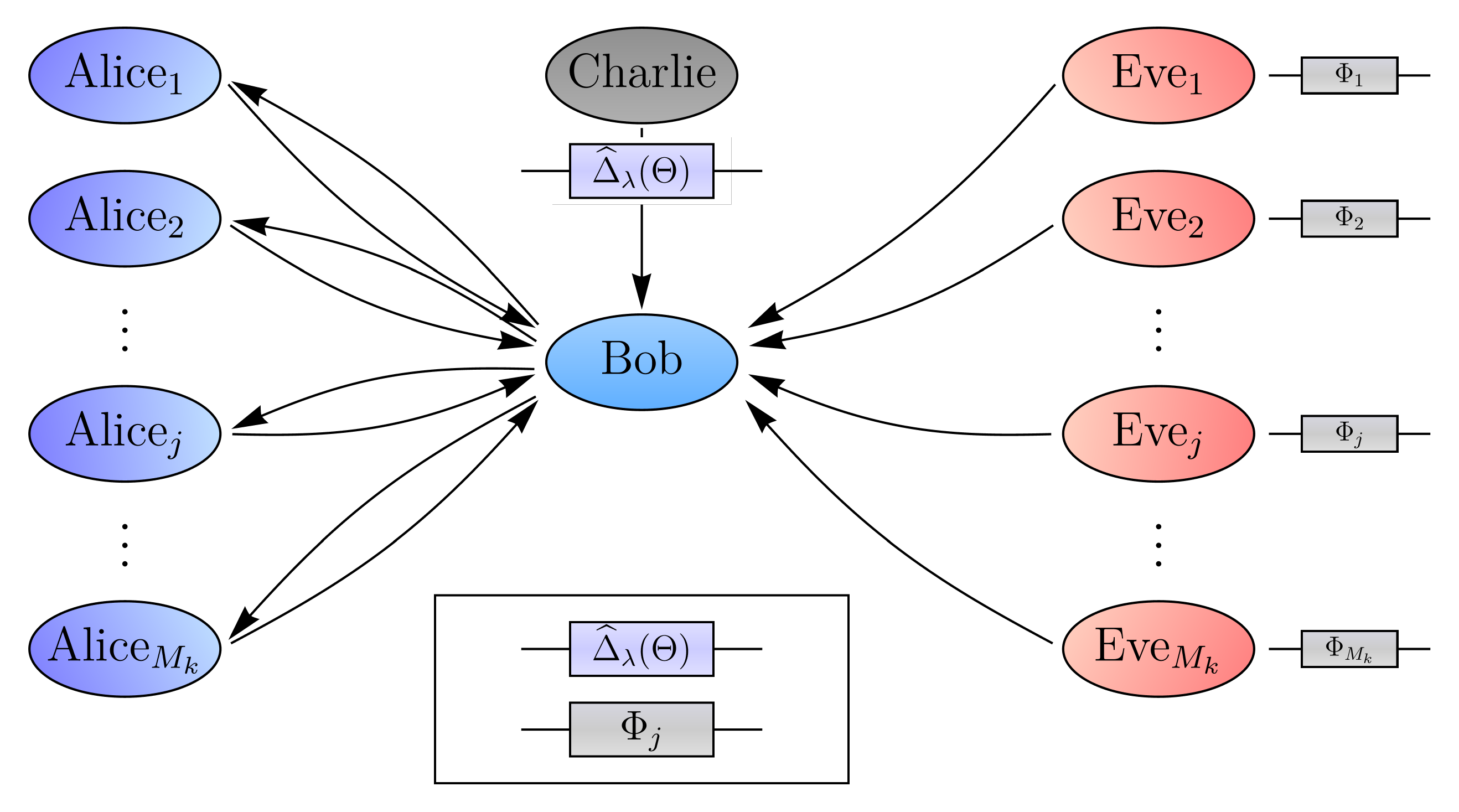}
    \caption{ In the extended version of the $\mathtt{G_{I}}$ game, there is a pair of players, Alice$(j)$ and Eve$(j)$, for each section $\mathcal{F}_j$ of a domain $\mathcal{D}$. They try to convince a verifier player, Bob, whether a black box is different from a $j$-th test box sent by Eve$(j)$.  
    }
    \label{fig:multiplayer_CI}
\end{figure}

After this, Alice sends to Bob a subsystem $A$ of a state $\rho_{A A^{\prime}}$, Bob labels $0$ the black box provided by Charlie and $1$ the test box provided by Eve, selecting one of them with uniform probability and applying it to the system send by Alice. After this, Bob sends the output system $B$ back to Alice, where Alice applies a measurement $\mathbb{M}_{B A^{\prime}}$ and sends a binary output back to him. Bob,  accepts if Alice's output matches the label of the box applied and reject otherwise. Since, Alice is a yes-prover she will choose the best  state $\rho_{A A^{\prime}}$ and measurements $\mathbb{M}_{B A^{\prime}}$ to distinguish between $\widehat{\Delta}_{\lambda}(\Theta)$ and the closest $\Phi\in \mathcal{F}_{1}^{(x)}$. If the binary variable $a$ takes value $1$ when Alice succeed to convince Bob to accept and $0$ if it fails, the optimal success probability for Alice is \cite{Watr2003}: 

\begin{equation}
    P\left(a\!=\!1\!\mid\widehat{\Delta}_{\lambda}(\Theta)\right)=\frac{1+\mathcal{L}\left(\widehat{\Delta}_{\lambda}(\Theta)\mid\mathcal{F}_{1}^{(x)}\right)}{2},
\end{equation}
with $\mathcal{L}(\cdot\mid\mathcal{F}_{1}^{(x)})$ given by (\ref{eq:diamnormgamma}).  It is known that the above protocol is optimal for Bob's certification \cite{Watr2003}. Now, assume that for each set $\mathcal{F}_j^{(x)}$ there is a pair of Alice($j$) and Eve($j$) performing independently with Bob an instance of $\mathtt{G_{I}}$ with the goal to distinguish between $\{ \Theta \}$  and $\mathcal{F}_{j}^{(x)} $, for a $\Theta\in\mathcal{D}^{(x)}$ exactly as before (See Figure \ref{fig:multiplayer_CI}). Suppose for simplicity that there is no noise for channel $\{ \Theta \}$, when Charlie send it to Bob, i.e. $\widehat{\Delta}_{\lambda}$ is such that $\lambda=1$. In this case, the success probability of every Alice($j$) is:

\begin{equation}
P\left(a_{j}=1\!\mid\Theta\right)=\frac{1+\mathcal{L}\left(\Theta\mid\mathcal{F}_{j}^{(x)}\right)}{2}.\label{succalj}
\end{equation}
Note that knowing the success probability (\ref{succalj}), is it possible to compute the correlation of success or failures in convincing Bob to accept. In Appendix \ref{OperTask1}, we demonstrate the following relation among the correlation between the  success $e_j=a_{j}\oplus1$ of every  Eve($j$) and the monotone $\mathcal{G}_{\mathcal{L}}(\cdot\mid\mathfrak{C}_{\mathcal{F}}^{(x)})$ as defined in (\ref{eq:Conemeasure}):

\begin{thm}\label{Theorem1}
For a given domain $\mathcal{D}^{(x)}$ of a star resource theory with free sections $\mathfrak{C}_{\mathcal{F}}^{(x)}=\{ \mathcal{F}_{j}^{(x)}\} _{j=1}^{M_x}$,
a channel $\Theta\in\mathcal{D}^{(x)}$ and game $\mathtt{G_{I}}$
we have:
\begin{equation}
P\left(\bigoplus_{j=1}^{M_x}e_{j}=0\mid\Theta\right)=\frac{1}{2}\left[1+\left[\mathcal{G}_{\mathcal{L}}(\Theta\mid\mathfrak{C}_{\mathcal{F}}^{(x)})\right]^{M_x}\right],
\end{equation}
where $\oplus$ is the sum mod(2) and $\mathcal{G}_{\mathcal{L}}(\cdot\mid\mathfrak{C}_{\mathcal{F}}^{(x)})$
is the geometric mean of the diamond norm distances $\mathcal{L}\left(\Theta\mid\mathcal{F}_{j}^{(x)}\right)$
of $\Theta$ to the free sections in  $\mathfrak{C}_{\mathcal{F}}^{(x)}$.
    
\end{thm}

Then, playing the same kind of game for every domain $\mathcal{D}^{(x)}$ including $\Theta$ and finding the domain with the maximum correlation $\bigoplus_{j=1}^{M_{x}}e_{j}=0 $ we have the statement involving the conditional probability $P(\cdot|\cdot)$,
 \begin{equation}\label{operquant}
\mathcal{G}_{\mathcal{L}}\left(\Theta\mid\mathcal{F}\right)=\!\!\!\max_{\mathcal{D}^{(x)}:\Theta\in\mathcal{D}^{(x)}}\!\!\!\sqrt[M_{x}]{2P\!\!\left(\bigoplus_{j=1}^{M_{x}}e_{j}=0\!\mid\!\Theta\right)\!\!-\!1}.
\end{equation}

We can effortlessly adapt the operational task described in Theorem 1  to the case in which the study set involves states or measurements. If the theory is about states, then each box is replaced by a channel that, for every input, generates a single output state; then, in the $\mathtt{G_{I}}$ game, each Alice($j$) does not send any bipartite state to Bob but performs the optimal measurement to discriminate the resource output state from the output state sent by Eve($j$). Likewise, if the theory is about POVMs, then each channel is replaced by a measure-and-prepare channel where the measurement is the POVM, and the output is a classic pointer state labeling the result of the measurement. Alice($j$) selects the best simple system input state to discriminate the resource POVM from Eve($j$)'s POVM and performs a projective measurement to distinguish the output pointer state. In the case of state theories, the diamond norm turns into a trace norm, while in the case of measurement theories, it turns into the operator norm. Additionally, we treat the particular case of a theory about probability distributions as a theory of diagonal states.

Moreover, Theorem 1 provides a universal operational interpretation of SRT
theories, also enabling the recovery of convex RTs as a limit case. If $\mathcal{F}$ is a convex
set, the cones $\mathcal{C}^{(x)}$ transform into half-planes, implying
$M_{x}=1$, while due to convexity of $\mathcal{F}$ we have, $\mathcal{F}_{1}^{(x)}=  \mathrm{Ker}\left(\mathcal{F}\right)=\mathcal{F}$  and thus $\mathcal{L}(\Theta\mid\mathcal{F}_{1}^{(x)})=\mathcal{L}\left(\Theta\mid\mathcal{F}\right)$.
Furthermore, the additional restrictions \eqref{eq:isomorphism}  used in $\mathtt{G_{I}}$,
where the noises affecting $\Theta$ must keep each $\mathcal{F}_{j}^{(x)}$
invariant, reduce to invariance of a single $\mathcal{F}$, exactly as in convex resource theories. These facts show how SRTs allow for the recovery of resource
quantification and manipulation from convex resource theories.

To sum up, we remark how the necessity for domains $\mathcal{D}^{(x)}$ with multiple free sections reveals the novelty of SRTs for scenarios with a non-convex set $\mathcal{F}$. In such cases, the resource influences Eve's advantage over Alice in each game and simultaneously determines the correlation between game results, with the global correlation being the factor quantified by the measure used.

\subsubsection{Interpretation for robustness based monotone} \label{robmonotenesection}

For the quantifier $\mathcal{G}_{\mathcal{R}}(\cdot\mid\mathfrak{C}_{\mathcal{F}}^{(x)})$ based on,
the generalised robustness quantifier~\cite{Vidal1999},

\begin{equation}
\mathcal{R}\left(\Theta\mid\mathcal{X}\right)=\min_{\Lambda\in\mathcal{S}}\left\{ r\geq0\mid\frac{\Theta+r\Lambda}{1+r}\in\mathcal{X}\right\}. \label{eq:genrobust}
\end{equation}

We demonstrate that it relates to the boost of a coalition
of $M_x$ players holding each, a different free set section $\mathcal{F}_{j}^{(x)}\in\mathcal{D}^{(x)}$ in a discrimination game. The game's goal is to  distinguishing between
different evolution branches $\Xi_{i}$  induced by a so-called \emph{quantum comb} $\Upsilon=\sum_{i}p_{i}\Xi_{i}$  \cite{Chiribella2008}, a class of maps taking $M$ quantum operations as input and outputs a
target new operation. Quantum combs are extensively used to address process transformation challenges and maximize performance potential. These problems encompass various transformations of unitary operations, such as inversion, complex conjugation, control-U analysis, and machine learning tasks \cite{Tulio2019,Murao2023,Chiribella2016op,Perinotti2010,Sedlak2019}. Moreover, they serve as a tool for examining broader process analyses \cite{zhu2024,mo2024} and resource dependence in algorithm success probability \cite{Ahnefeld2022}. To simplify notation, we consider quantum combs with branches $\Xi_{j}$ providing a classical output, i.e. $\Xi_{j}:\textrm{CPTP}^{M}\longrightarrow q_{j}$, for
all $j$  a probability $q_{j}\in\left[0,1\right]$, $\sum_{j}q_{j}=1$. The previous quantum combs have quantum circuit implementations of the form: \begin{equation}\label{combwithmeasurement}
\Upsilon\left(\cdot\right)=\textrm{Tr}\left\{ \mathbb{N} \,\mathcal{I\otimes}(\cdot)\circ\mathcal{N}_{M-1}\circ...\circ\mathcal{N}_{1}\circ\mathcal{I\otimes}(\cdot)\left[\zeta_{0}\right]\right\}, 
\end{equation}
 as illustrated in diagrams \eqref{eq:diagram1} and \eqref{eq:diagram2}. In the diagrams the gate $\mathcal{N}_{\zeta_0}$ is a channel always preparing the bipartite state $\zeta_0$, subsequent gates $\mathcal{N}_{j}$ are general bipartite channels and finally gate $\mathcal{N}_{\mathbb{N}}$ performs a POVM $\mathbb{N}$ with outputs $j$ identifying the branch $\Xi_{j}$. The reason to choose the structure \eqref{combwithmeasurement}, is to integrate into the ensemble of branches $\left\{ p_{j},\Xi_{j}\right\} _{j=1}^{N}$ the preparation $\zeta_0$ and measurement $\mathbb{N}$ of the discrimination strategy. We can assume the condition \eqref{combwithmeasurement} without loss of generality, because later in the discrimination task, both $\zeta_0$, and $\mathbb{N}$ are  optimized together with the quantum comb.

 \begin{widetext}
 \begin{equation}
 		\label{eq:diagram1}
 \begin{quantikz}
     &\gate[2]{\mathcal{N}_{\zeta_0}}\gategroup[2,steps=9,style={dashed,rounded corners,fill=blue!20,inner sep=2pt},background,label style={label
 position=below,anchor=north,yshift=-0.2cm}]{
 In blue an implementation \eqref{combwithmeasurement} of quantum comb $\Upsilon$ tested by channels $\{\Phi_1, ... , \Phi_M\}$ at the red slots. }& \qw &\gate[2]{\N_{1}}& \qw &\gate[2]{\N_{2}}& \qw &\gate[2]{\N_{M-1}}& \qw &  \gate[2]{\mathcal{N}_{\mathbb{N}}} \\
     & & \gate{\Phi_1}\gategroup[1,steps=1,style={dashed,rounded corners,fill=red!20,inner sep=2pt},background,label style={label
 position=below,anchor=north,yshift=-0.2cm}]{{\sc
 }}  & & \gate{\Phi_2}\gategroup[1,steps=1,style={dashed,rounded corners,fill=red!20,inner sep=2pt},background,label style={label
 position=below,anchor=north,yshift=-0.2cm}]{{\sc
 }} & & \gate{\ldots}\gategroup[1,steps=1,style={dashed,rounded corners,fill=red!20,inner sep=2pt},background,label style={label
 position=below,anchor=north,yshift=-0.2cm}]{{\sc
 }} & &  \gate{\Phi_M}\gategroup[1,steps=1,style={dashed,rounded corners,fill=red!20,inner sep=2pt},background,label style={label
 position=below,anchor=north,yshift=-0.2cm}]{{\sc
 }} &  & 
 \end{quantikz}
 \end{equation}

 \begin{equation}
 		\label{eq:diagram2}
 \begin{quantikz}
     &\gate[2]{\mathcal{N}_{\zeta_0}}\gategroup[2,steps=9,style={dashed,rounded corners,fill=blue!20,inner sep=2pt},background,label style={label
 position=below,anchor=north,yshift=-0.2cm}]{
 Here the slots are all filled with the resource channel $\{\Theta\}$ to distinguish the evolution branches of  $\Upsilon$.}& \qw &\gate[2]{\N_{1}}& \qw &\gate[2]{\N_{2}}& \qw &\gate[2]{\N_{M-1}}& \qw &  \gate[2]{\mathcal{N}_{\mathbb{N}}} \\
     & & \gate{\Theta}\gategroup[1,steps=1,style={dashed,rounded corners,fill=red!20,inner sep=2pt},background,label style={label
 position=below,anchor=north,yshift=-0.2cm}]{{\sc
 }}  & & \gate{\Theta}\gategroup[1,steps=1,style={dashed,rounded corners,fill=red!20,inner sep=2pt},background,label style={label
 position=below,anchor=north,yshift=-0.2cm}]{{\sc
 }} & & \gate{\ldots}\gategroup[1,steps=1,style={dashed,rounded corners,fill=red!20,inner sep=2pt},background,label style={label
 position=below,anchor=north,yshift=-0.2cm}]{{\sc
 }} & &  \gate{\Theta}\gategroup[1,steps=1,style={dashed,rounded corners,fill=red!20,inner sep=2pt},background,label style={label
 position=below,anchor=north,yshift=-0.2cm}]{{\sc
 }} &  & 
 \end{quantikz}
 \end{equation}
 
 \end{widetext}

In games $\mathtt{G_{D}}\left[M\right]$,  each player $\textrm{Alice}(j)$ holding channels $\mathcal{F}_{j}\ni\Phi_{j}$
is allowed to test the causal network accessing only one comb slot of the quantum comb depict as red slots in diagram \eqref{eq:diagram1}. Then, without loss of generality
in each game $\left\{ p_{i},\Xi_{i}\right\} \in\mathtt{G_{D}}\left[M\right]$
we compare the best strategy of the coalition $\left\{ \textrm{Alice}(j)\right\} _{j=1}^{M}$
against the coalition $\left\{ \textrm{Bob}(j)\right\} _{j=1}^{M}$
using $\Theta$ in every comb slot as portrayed in diagram \eqref{eq:diagram2}. In Appendix \ref{OperTask2}, we prove a general statement, applicable in multi-resource \cite{Sparaciari2020} and multi-object \cite{Brunner2021,DucuaraMO2020,salazar2022} scenarios, which, under the previous particular conditions, reduces to: 
 \begin{widetext}
\begin{thm}\label{Theorem2}
For a given domain $\mathcal{D}$ of an SRT
with free sections $\mathfrak{C}_\mathcal{F}=\left\{ \mathcal{F}_{j}\right\} _{j=1}^{M}$,
a channel $\Theta\in\mathcal{D}$ and discrimination games $\left\{ p_{i},\Xi_{i}\right\} \in\mathtt{G_{D}}\left[M\right]$
we have:
\begin{equation}
\max_{\left\{ p_{i},\Xi_{i}\right\} \in\mathtt{G_{D}}\left[M\right]}\!\!\!\!\!\frac{P_{\textrm{s}}\!\left(\!\left\{ p_{i},\Xi_{i}\right\} \!,\{\Theta\}\!\right)}{\underset{\left\{ \Phi_{k}\in\mathcal{F}_{k}\right\} _{k=1}^{M}}{\max}\!\!\!\!\!P_{\textrm{s}}\!\left(\!\left\{ p_{i},\Xi_{i}\right\} \!,\left\{ \Phi_{k}\right\}\!\right)}\!=\!\prod_{k=1}^{M}\!\left[1\!+\!\mathcal{R}\left(\Theta\mid\mathcal{F}_k\right)\right],\label{eq:Th2main}
\end{equation}
with $P_{\textrm{s}}$ the optimal discrimination probability,  $\left\{ \Phi_{k}\right\}$ stands for $\{\Phi_{1},\ldots,\Phi_{M}\}$ and $\{\Theta\}$ for $\{\Theta,\ldots,\Theta\}$, $M$ times. Besides,
each $\mathcal{R}\left(\Theta\mid\mathcal{F}_k\right)$ is the generalised
robustness of $\Theta$ with respect to the free section $\mathcal{F}_{k}$. 
\end{thm}
\end{widetext}
Next, we will adopt the approach of cooperative games \cite{peleg2003,muros2019}
to evaluate the increase in a coalition's advantage brought by resource
$\Theta$ when playing games $\mathtt{G_{D}}\left[M\right]$
as a \emph{dividend}. Cooperative games, a branch of game theory,
have relevance in areas such as economics \cite{friedman1986}, machine learning
\cite{curiel1997,SUN2012} and the voting power of coalitions \cite{chakravarty2015}. A pair $(S,\mathbf{u})$
formally defines a cooperative game, where $S$ is the complete set
of players, and $\mathbf{u}:2^{\left|S\right|}\rightarrow\mathbb{R}$
is a function that maps each possible coalition $T\subseteq S$ of
players within set $S$ to their respective profit or utility and
satisfies $\mathbf{u}\left(\emptyset\right)=0$.

In  scenario analyzed,
set $S$ contains players $\left\{ \textrm{Bob}_{j}\right\} _{j=1}^{M}$,
and $\mathbf{u}\left(\Theta\mid T\right)$ evaluates the relative
advantage each coalition $T$ of Bobs gains over a coalition of Alice's
with the same labels $j$ as Bob's in $T$ when playing games $\mathtt{G_{D}}\left[\left|T\right|\right]$,
i.e:

    \begin{equation}
{\scriptsize
\mathbf{u}\left(\Theta\!\mid\!T\right)=\!\!\!\!\!\!\max_{\left\{ p_{i},\Xi_{i}\right\} \in\mathtt{G_{D}}\!\left[\left|T\right|\right]}\!\!\!\!\!\!\!\!\!\!\!\frac{P_{\textrm{s}}\left(\!\left\{ p_{i},\Xi_{i}\right\} \!,\!\{\Theta\}\!\right)\!-\!P_{\textrm{s}}\left(\!\left\{ p_{i},\Xi_{i}\right\} \!,\!\{\Phi_{k}^{*}\}\!\right)}{\!P_{\textrm{s}}\left(\!\left\{ p_{i},\Xi_{i}\right\} \!,\!\{\Phi_{k}^{*}\}\!\right)}},\label{eq:reladv}
\end{equation}
where we set $\mathbf{u}\left(\Theta\mid\emptyset\right)=0$ by default and $\{\Phi_{k}^{*}\}$ represents the best strategy  of Alice's  coalition for the discrimination of $\left\{ p_{i},\Xi_{i}\right\} $.
Afterwards, we use the well-known Harsanyi method~\cite{harsanyi2010} to calculate the
surplus or \emph{dividend} for each coalition $T$ in $S$.
This method identifies a coalition's surplus as the difference between
its utility and the surplus of its smaller sub-coalitions. To this
end, the Harsanyi dividend $\mathbf{d}_{\mathbf{u}}\left(\Theta\!\mid\! T\right)$
of coalition $T$ with utility $\mathbf{u}\left(\Theta\!\mid\! T\right)$
is recursively determined by:
\begin{equation}
\mathbf{d}_{\mathbf{u}}\left(\Theta\!\mid\! T\right)=\!\!\begin{cases}
\mathbf{u}\left(\Theta\!\mid\! T\right)-\!\!\sum_{T^{\prime}\subset T}\mathbf{d}_{\mathbf{u}}\left(\Theta\!\mid\! T^{\prime}\right)  &\!\!\!\!\!;\left|T\right|>1 \\
\mathbf{u}\left(\Theta\!\mid\! T\right)  &\!\!\!\!\!;\left|T\right|=1. 
\end{cases}\label{eq:harsdiv}
\end{equation}
By using Harsanyi dividends, we can effectively study games and associated
solution concepts (i.e. individual profits). One such concept is the \emph{Shapley value}, calculated
by dividing the coalition surplus evenly among its members \cite{peleg2003,chakravarty2015}.
In fact, employing Theorem 4 in Appendix \ref{OperTask2}, we demonstrate that the
dividend determined by utility (\ref{eq:reladv}) completely describes the quantifier based on robustness, allowing a simple computation of every Shapley value. Concretely, we show the following operational relation:
\begin{thm}
\label{Theorem3} For a given domain $\mathcal{D}$ of an SRT\
with free sections $\mathfrak{C}_\mathcal{F}=\left\{ \mathcal{F}_{j}\right\} _{j=1}^{M}$,
a channel $\Theta\in\mathcal{D}$,  and a coalition $S=\left\{ \textrm{\emph{Bob}}(j)\right\} _{j=1}^{M}$ and sub-coalitions $T\subseteq S$, playing games $\mathtt{G_{D}}\left[\left|T\right|\right]$
with utility function $\mathbf{u}$ given by (\ref{eq:reladv}), we have:
\begin{equation}
\mathbf{d}_{\mathbf{u}}\left(\Theta\mid S\right)=\prod_{k=1}^{M}\mathcal{R}\left(\Theta\mid\mathcal{F}_k\right)=\left[\mathcal{G}_{\mathcal{R}}(\Theta\mid\mathfrak{C}_{\mathcal{F}})\right]^{M},\label{eq:Th2main-1}
\end{equation}
where $\mathcal{G}_{\mathcal{R}}(\Theta\mid\mathfrak{C}_{\mathcal{F}})$ is the
geometric mean of the generalised robustness $\mathcal{R}\left(\Theta\mid\mathcal{F}_k\right)$
of $\Theta$ with respect to the free section $\mathcal{F}_{k}$. 
\end{thm}

Similarly as in the previous task,  by playing the same kind of game for every domain $\mathcal{D}^{(x)}$ including $\Theta$ and finding the domain with the maximum dividend for the players, we have:
\begin{equation}
\mathcal{G}_{\mathcal{R}}(\Theta\mid\mathcal{F})=\!\!\max_{\mathcal{D}^{(x)}:\Theta\in\mathcal{D}^{(x)}}\!\left[\mathbf{d}_{\mathbf{u}}\left(\Theta\mid S_{x}\right)\right]^{1/M_{x}},
\end{equation}
where $S_{x}$ is now the corresponding coalition of Bob's using the resource $\Theta$ to play against Alice's holding
free sections $\{ \mathcal{F}_{j}^{(x)}\} _{j=1}^{M_{x}}$. Additionally, as in the distance-based monotone interpretation, we include state or measurement theories by replacing the general channels for channels with a constant output state or a measure-and-prepare channel, measuring on a POVM and preparing the classical pointer state as output, respectively.

Theorem 5 introduces another universal interpretation for SRTs, recovering
the results for convex RTs in the same limit as Theorem 3
, $M_{x}=1$ and $\mathcal{F}_{1}^{(x)}=\mathcal{F}$ for all $x$,
using the same previous arguments. Nevertheless, $\mathcal{G}_{\mathcal{R}}(\Theta\mid\mathcal{F})$'s
interpretation implies a different aspect of the resource, specifically,
its superadditivity when used consecutively in testing the ensemble $\left\{ p_{j},\Xi_{j}\right\} _{j=1}^{N}$. Theorem 5's insight into the advantage provided by $\Theta$ also
opens new avenues for connecting resource theory and game theory,
echoing the path established in \cite{Ducuara2022}.

Previous works on non-convex resource theories obtain operational interpretations by decomposing the free set into its convex components and applying the generalized robustness to each component separately \cite{adesso2023,Kuroiwa2023}. In those settings, the operational meaning follows directly from established results in convex resource theories, since the monotone is defined as the minimum robustness over the components, as described in equation (\ref{infdec}).

The present work differs substantially from that approach. Here we introduce and analyse products of distance-type monotones and robustness-type, which do not follow directly from the convex case and therefore do not admit operational interpretations by direct appeal to known results. Establishing their operational meaning requires new arguments, as developed in the proofs contained in Appendix \ref{Genproofs}. We emphasize this distinction in the boxed summary below, which highlights the genuinely new contributions of our framework.

\mybox{Summary 5}{red!40}{red!10}{ \textbf{Novel operational interpretations of SRT monotones:}

\begin{enumerate}
    \item[1)] \textbf{Distance-based monotone}: Demonstrates an advantage in a multiparty variant of the  \textit{Close-Images} game \textit{(Theorem \ref{Theorem1})}. This constitutes a \textit{genuinely new development}, with no prior analogue in the literature on non-convex resource theories.
    \item[2)] \textbf{Robustness-based monotone}: Yields an advantage in multiple quantum-comb discrimination tasks \textit{(Theorem \ref{Theorem2})}. Although the proof employs known semidefinite programming techniques, the appearance of products of robustness values means the proof does \textit{not} follow directly from existing results and requires new and more involved arguments.
    \item[3)] \textbf{Game-theoretic interpretation:} We obtain additional operational insight by using tools from cooperative game theory, specifically the Harsanyi dividend \cite{harsanyi2010} and Shapley values \cite{chakravarty2015, peleg2003} \textit{(Theorem \ref{Theorem3})}.
    This represents a \textit{completely new perspective}, without any previous analogue in the resource-theoretic literature.
    \end{enumerate}
    }

\section{Applications} \label{sec:app} 
Assessing the quantum resourcefulness through operational witnesses is a crucial step in quantum information processing \cite{Gour2019}, as it provides a reliable method for quantifying and comparing the resource content of different physical systems. Section \ref{optinter} develops general, theory-independent quantitative results valid for any star resource theory , while section \ref{freesetfortress} also details the construction of canonical fortresses, together with redundancy-deletion procedures, providing an explicit route to a constructive evaluation. These results extend naturally to broad classes of free sets with structured geometry, including star domains with polyhedral boundaries with $m$ faces in dimension $d$: in this case, its convex components with maximal intersection with the boundary are precisely its convex cells. These cells can be enumerated with worst-case complexity $O(m^{d})$, as shown in \cite{Zaslavsky1975}. By combining these convex components with the canonical fortress and then performing the redundancy-deletion steps described in section \ref{freesetfortress}, we can obtain the full star resource theory univocally. As shown in section \ref{freesetfortress}, redundancy deletion complexity scales as $O(m^{\,2+\lfloor (d-1)/2 \rfloor})$ and consequently, the full construction of a resource theory for a polyhedral star domain is dominated by the initial cell enumeration and therefore also scales as $O(m^{d})$ in the worst case.

The present section complements these general results by illustrating their application in four representative scenarios. We begin with the characterization of quantum discord and total correlations, proceed to the detection of non-unistochastic maps, and conclude with the identification of non-Markovianity in mixtures of Pauli channels. These examples demonstrate how the general framework and fortress construction operate across diverse quantum settings.

\mybox{Summary 6}{blue!40}{blue!10}{ \textbf{Applications of Quantum SRTs:}
\begin{enumerate}
    \item[1)] \textbf{Quantum Discord:} We provide a novel operational interpretation, extending the set of two-qubit states with known analytical close formulas to quantify discord.   
    \item[2)] \textbf{Total correlations:}  We completely characterize the bipartite case,  introduce an easy to compute witness of total correlations, and outline further generalizations to networks. 
    \item[3)]  \textbf{Non-Unistochasticity:} We obtain a fully operational test for quantum to classical transition theories in 
    high energy physics.
    \item[4)]  \textbf{Non-Markovianity:} We illustrate how to use the results obtained to develop operational tests to detect and quantify non-Markovianity in a simple case. 
   
    \end{enumerate}
    }

\subsection{Quantum discord as a star resource theory}

Quantum phenomena are unequivocally distinguished from classical physics by the powerful correlations they generate, most prominently quantum entanglement. However, there are other quantum properties with the capacity to generate non-classical correlations, such as quantum discord~\cite{Zurek_2000}, widely exploited in computation and communication scenarios \cite{Paterek2012,Adesso2016,Borivoje2010,Quesada2012}. The quantum discord corresponds to the difference between fully quantum and measurement-induced
mutual information being ubiquitous among quantum states.
Nevertheless, discord is a property that proves to be elusive, as computing it, involves minimization over all possible measurements of one out of two subsystems. Furthermore, it does not admit a linear witnesses similar to the celebrated Bell inequalities, to detect entanglement, an absence that stems from the non-convex structure of states with zero quantum discord.

In the face of this limitation, introduced approach proves fruitful, revealing that the set of states with zero discord constitutes a star domain. Precisely, the geometry of the set enables us to assign a simple polyhedric fortress, facilitating the application of the 
framework proposed   
to quantum discord. This results in the derivation of a non-linear operational quantifier and the identification of nontrivial free operations. Furthermore, we demonstrate how to analytically calculate the discord quantifier for relevant cases of the literature.

Consequently, to develop
a resource theory of discord, it is crucial to identify the set of
zero-discord states. Fortunately, in the literature 
the following form for two-qubit zero-discord states (on Alice's subsystem) is well characterized
\cite{Borivoje2010}:
\begin{equation}
\chi=\sum_{i\in\left\{ 1,2\right\} }p_{i}\left|\vartheta_{i}\right\rangle \!\left\langle \vartheta_{i}\right|\otimes\varrho_{i}\label{eq:zero-discord}
\end{equation}
where $\left\{ \left|\vartheta_{1}\right\rangle ,\left|\vartheta_{2}\right\rangle \right\} $
is a single-qubit orthonormal basis, $\varrho_{1},\varrho_{2}$ are
qubit states and $p_{1},p_{2}$ are non-negative numbers such that
$p_{1}+p_{2}=1$. At this point, we can already distinguish crucial
geometric features in the structure of the states with zero-discord
$\Omega_{0}$. First, the convex combination of any state $\chi$
with the maximally mixed state $I\otimes I/4$ preserves the form
(\ref{eq:zero-discord}), and therefore, $\Omega_{0}$ is a star domain with $I\otimes I/4$
in $\mathrm{Ker}\left(\Omega_{0}\right)$. Second, it is immediately
possible to distinguish that the convex components of $\Omega_{0}$
are precisely the states described in (\ref{eq:zero-discord}) but associated with the
same basis, $\left\{ \left|\vartheta_{1}\right\rangle ,\left|\vartheta_{2}\right\rangle \right\} $
since these are all the convex and maximal subsets under inclusion.
From this, it turns out that operations \eqref{eq:isomorphism} preserve the base
$\left\{ \left|\vartheta_{1}\right\rangle ,\left|\vartheta_{2}\right\rangle \right\} $
in (\ref{eq:zero-discord}), like $\widehat{\Delta}_{\lambda}$, which represents a mixture
with $I\otimes I/4$, or physically, a white noise. Likewise, the
sufficient condition (\ref{eq:RNGcond2}) for RNGs is equivalent to preserving commutativity
since if two $\chi_{1},\chi_{2}\in\Omega_{0}$ are described in (\ref{eq:zero-discord})
by the same basis, the same must be true after an operation which satisfies  (\ref{eq:RNGcond2}) and the converse follows from the fact that commuting
states span by the same basis in (\ref{eq:zero-discord}). Thus, the general framework
allows us to recover the sufficiency part of previous results directly \cite{Borivoje2010}.

 However, to continue exploiting the techniques proposed
 we need to characterize the zero-discord states in a bounded linear real representation of two-qubit states. Along the above direction, we choose the so-called Fano representation  \cite{Fano_1957,OmarGamel2016}:
\begin{equation}
\rho\!=\!\frac{1}{4}\!\!\left\{ \!\!I\!\otimes\! I\!+\!\!\!\!\!\!\!\sum_{i\in\left\{ 1,2,3\right\} }\!\!\!\!\!\{x_{i}\sigma_{i}\otimes I\!+y_{i}I\!\otimes\!\sigma_{i}\}\!+\!\!\!\!\!\!\!\!\sum_{i,j\in\left\{ 1,2,3\right\} }\!\!\!\!\!\!\!T_{ij}\sigma_{i}\!\otimes\!\sigma_{j}\!\right\} \label{eq:Bloch matrix}
\end{equation}
where $\left\{ \sigma_{i}\right\} _{i=1}^{3}$ represent Pauli matrices,
$x_i$ and $y_i$ denote components of the local Bloch vectors of the first and second reduced states respectively,
while the matrix $T_{ij}$ is
known as the correlation matrix. We apply the Fano parametrization to characterize analytically
the zero-discord set for a large class of two-qubit states, some of
them beyond the currently known in the literature.

Firstly, to characterize the zero-discord set we show the following theorem (Appendix \ref{Qdiscord}):

\begin{thm}
\label{Theorem4}

For any two-qubit state, parametrized with $(x_i, T_{ij})$ as in (\ref{eq:Bloch matrix}) and satisfying the positivity constraints from section IV.D in~\cite{OmarGamel2016}, if its correlation matrix is diagonal
\begin{equation}\label{symmetricT}
T_{ij}=t_i \delta_{ij},  
\end{equation}
then the state has zero discord on the first reduced state if and only if all but one of the coordinate pairs $\left\{ \left(x_{i},t_{i}\right)\right\} _{i=1}^{3}$ equal $\left(0,0\right)$ .
\end{thm}

The above condition includes the case when the correlation matrix is symmetric, since in this situation the Pauli matrices could be re-defined using the orthogonal matrix $O_{ij}$ that diagonalizes $T_{ij}$,   with $t_i$ being the corresponding eigenvalues. Similarly, in that case the $x_i$ and $y_i$ are the re-defined parameters after a rotation by $O_{ij}$.

\begin{figure} 
    \centering
    \includegraphics[width = \columnwidth]{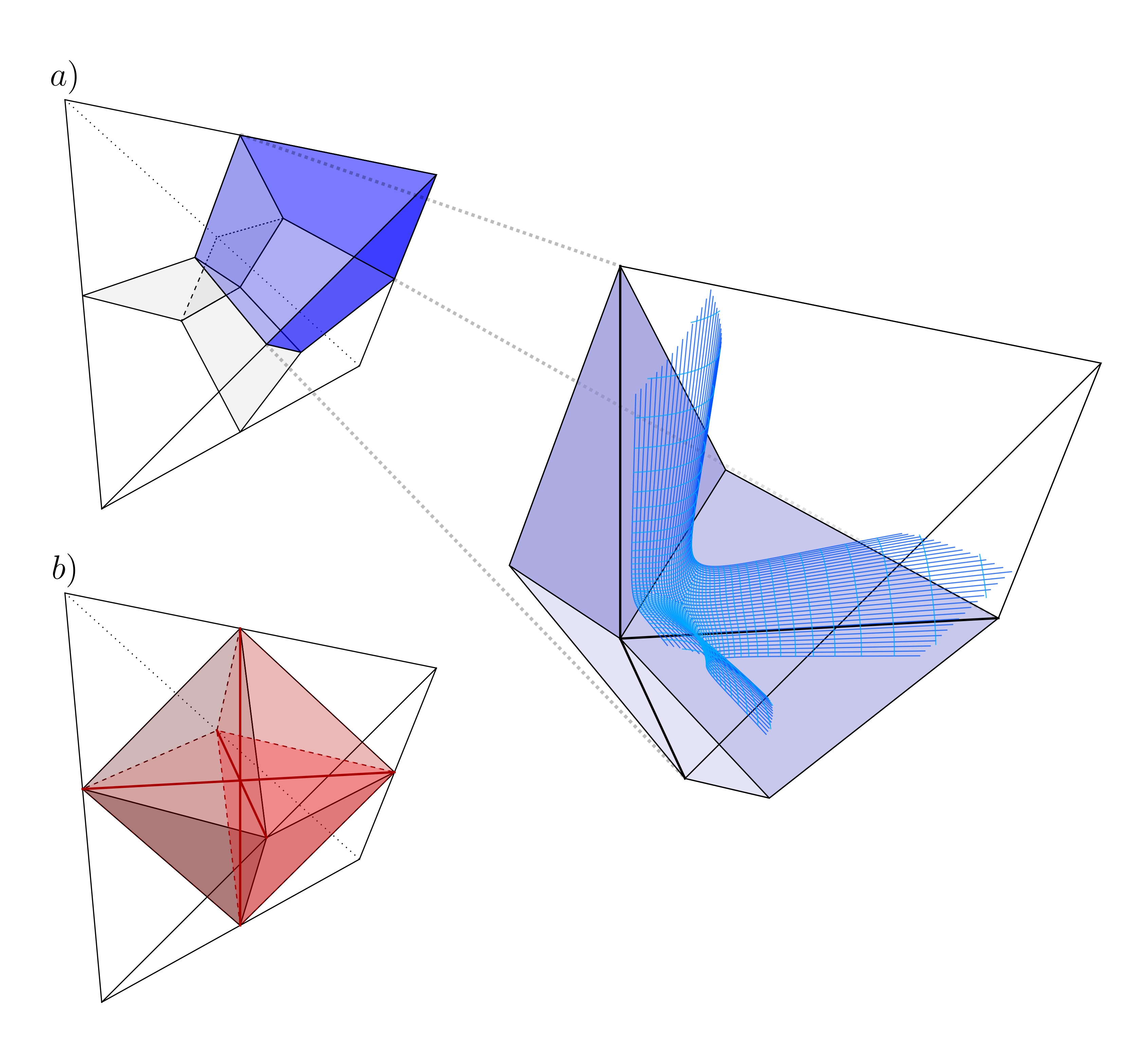}
    \caption{The figures show a tetrahedron representing two-qubit states with maximally mixed marginals \eqref{eq:studystates1}, in coordinates $\mathbf{t}=\left(t_{1},t_{2},t_{3}\right)$. The upper figure presents in light blue a surface of states with equal resource as quantified by \eqref{bellstatemonotone}. The lower figure highlights in light red the octahedron formed by the subset of separable states, and in solid red the zero discord set $\Omega_{0}$ comprised by the axes in $t$-coordinates.}
    \label{fig:discord}
\end{figure}

Later, to further characterize the geometry of $\Omega_0$ we include the constraint that $\mathbf{y}=(y_{1},y_{2},y_{3})$ must have the same
support as $\mathbf{x}=(x_{1},x_{2},x_{3})$, i.e. if a coordinate $x_{i}$ is zero this
implies that a coordinate $y_{i}$ is also zero, but the converse is not necessarily true. Then also in Appendix \ref{Qdiscord} we show,

\begin{thm}
\label{regions}
 Consider the set of two-qubit states  $\Omega$
with the Bloch matrix parametrization  (\ref{eq:Bloch matrix}), correlation matrix (\ref{symmetricT})  and $\mathbf{y}$ of equal support as $\mathbf{x}$. Then, the set  of zero-discord $\Omega_{0}$
states  has only one triple $\left\{ (x_{i},y_{i},t_{i})\right\} $  which could be different from $\left\{ (0,0,0)\right\} $ and must satisfy:
\begin{equation}
\left|x_{i}+t_{i}\right|\leq1+y_{i}\:\textrm{and}\:\left|x_{i}-t_{i}\right|\leq1-y_{i}
\end{equation} which for a given $y_i$ determines the region of possible $\left\{ (x_{i},t_{i})\right\} $  of a state in $\Omega_{0}$. 
\end{thm} While the constraint on supports, may seem arbitrary, it actually contributes to include important cases such as  the well
known X states ( $x_{1}=x_{2}=y_{1}=y_{3}=0$)
\cite{Quesada2012} or the symmetric states ($\mathbf{y}=\mathbf{x}$) and expand the set of states studied analytically.

Taking advantage of the previous results we can develop a star resource theory of Alice’s discord (for Bob’s discord, exchange $\mathbf{x}$ and $\mathbf{y}$). Precisely, Theorem 5 shows that $\Omega_{0}$ is constituted by three rectangular regions in the planes of coordinates $\left\{ (x_{i},t_{i})\right\} $, with axes of symmetry $ x_{i}=t_{i} $ and $ x_{i}=-t_{i} $. This suggests that rotating the coordinates by $\pi/4$ in each plane provides a simpler description, therefore we define: 
\begin{equation}
  u_i  =  \frac{x_i+t_i}{\sqrt{2}}\quad v_i  =  \frac{x_i-t_i}{\sqrt{2}}.  
\end{equation}

Coordinates $\left\{ (u_{i},v_{i})\right\}$ align with the symmetry axes of $\Omega_{0}$ providing a suitable reference frame for a geometric description. Indeed, in the corresponding coordinate system $\left\{ \hat{\mathbf{u}}_{i},\hat{\mathbf{v}}_{i}\right\} _{i=1}^{3}$ for   $\Omega\subset\mathbb{R}^{6}$ the set of quadrants $\mathcal{C}_{\mathring{u}\mathring{v}}$ form an appropriate fortress $\mathfrak{T}_{\Omega_{0}}$:

\begin{equation}
\mathcal{C}_{\mathring{u}\mathring{v}}\!:\!\!\left\{ \!\mathbf{r}\!\mid\!\mathbf{r}=\!\!\sum_{i=1}^{3}\!\left\{ \!\alpha_{i}\left(-1\right)^{\mathring{u}_{i}}\!\hat{\mathbf{u}}_{i}\!+\!\beta_{i}\left(-1\right)^{\mathring{v}_{i}}\!\hat{\mathbf{v}}_{i}\right\} \!,\alpha_{i},\beta_{i}\geq0\!\right\}, \label{eq:quadrants}
\end{equation}
where $\mathring{u}$ and $\mathring{v}$ are the bit strings $(\mathring{u}_{1}\mathring{u}_{2}\mathring{u}_{3})$,
and $(\mathring{v}_{1}\mathring{v}_{2}\mathring{v}_{3})$ respectively.
Despite $\mathcal{C}_{\mathring{u}\mathring{v}}$ not being support
cones, it is straightforward to verify they satisfy the i)-iv) fortress
conditions. Moreover, every $\mathcal{C}_{\mathring{u}\mathring{v}}$
defines exactly three free sections: 
\begin{eqnarray}
\Omega_{0,i}^{(\mathring{u}_{i}\mathring{v}_{i})} & \!:\! & \left\{ \mathbf{r}\mid\mathbf{r}=\!\alpha_{i}\left(-1\right)^{\mathring{u}_{i}}\hat{\mathbf{u}}_{i}+\beta_{i}\left(-1\right)^{\mathring{v}_{i}}\hat{\mathbf{v}}_{i},\right.\nonumber \\
 &  & \left.0\leq\alpha_{i}\leq\frac{\left(1+y_{i}\right)}{\sqrt{2}},0\leq\beta_{i}\leq\frac{\left(1-y_{i}\right)}{\sqrt{2}}\right\} .\nonumber \\
 &  & \label{eq:freesectiondiscord}
\end{eqnarray}
with $i\in\{1,2,3\}$. Hence, it is very simple to compute the distance based quantifier
for any two-qubit state $\rho$ with representation:
\begin{equation}\label{eq:framespandisc}
\mathbf{r}_{\rho}=\!\!\sum_{j=1}^{3}\left\{ \left|u_{\rho,j}\right|\left(-1\right)^{\mathring{u}_{\rho,j}}\!\hat{\mathbf{u}}_{j}\!+\!\left|v_{\rho,j}\right|\left(-1\right)^{\mathring{v}_{\rho,j}}\!\hat{\mathbf{v}}_{j}\right\}, 
\end{equation}
where $\left|u_{\rho,j}\right|\leq\frac{\left(1+y_{\rho,j}\right)}{\sqrt{2}},\left|v_{\rho,j}\right|\leq\frac{\left(1-y_{\rho,j}\right)}{\sqrt{2}}$, and $\left(-1\right)^{\mathring{u}_{\rho,j}}=\textrm{sgn}(u_{\rho,j})$, $\left(-1\right)^{\mathring{v}_{\rho,j}}=\textrm{sgn}(v_{\rho,j})$ 
for every $j\in\{1,2,3\}$.

Evidently, $\mathbf{r}_{\rho}\in\mathcal{C}_{\mathring{u}_{\rho}\mathring{v}_{\rho}}$
and consequently we need to compute (\ref{eq:Conemeasure})  the trace distance $\mathcal{L}$ to the free sections
$\mathfrak{C}_{\Omega_{0}}^{(\mathring{u}_{\rho}\mathring{v}_{\rho})}=\{\Omega_{0,i}^{(\mathring{u}_{\rho,i}\mathring{v}_{\rho,i})}\}_{i=1}^{3}$:
\begin{equation}
\mathcal{L}\left(\rho\mid\Omega_{0,i}^{(\mathring{u}_{\rho,i}\mathring{v}_{\rho,i})}\right)=\!\sqrt{\sum_{j\neq i}\left\{\left|u_{\rho,j}\right|^{2}+\left|v_{\rho,j}\right|^{2}\right\}},
\end{equation}
and the corresponding geometrical average (\ref{eq:Conemeasure}):
\begin{equation}\label{eq:discordquant}
\mathcal{G}_{\mathcal{L}}\left(\rho\mid\mathfrak{C}_{\Omega_{0}}^{(\mathring{u}_{\rho}\mathring{v}_{\rho})}\right)=\sqrt[3]{\prod_{i=1}^{3}\left(\!\sqrt{\sum_{j\neq i}\left\{\left|u_{\rho,j}\right|^{2}+\left|v_{\rho,j}\right|^{2}\right\}}\right)}.
\end{equation}

Note, that for every intersection $\mathcal{C}_{\mathring{u}_{1}\mathring{v}_{1}}\cap\mathcal{C}_{\mathring{u}_{2}\mathring{v}_{2}}=\partial\mathcal{C}_{\mathring{u}_{1}\mathring{v}_{1}}\cap\partial\mathcal{C}_{\mathring{u}_{2}\mathring{v}_{2}}$, by construction corresponds to a symmetry hyperplane of $\Omega_{0}$,
thus for any $\mathbf{r}_{\rho}\in\mathcal{C}_{\mathring{u}_{1}\mathring{v}_{1}}\cap\mathcal{C}_{\mathring{u}_{2}\mathring{v}_{2}}$we
have: 
\begin{equation}
\mathcal{G}_{\mathcal{L}}\left(\rho\mid\mathfrak{C}_{\Omega_{0}}^{(\mathring{u}_{1}\mathring{v}_{1})}\right)=\mathcal{G}_{\mathcal{L}}\left(\rho\mid\mathfrak{C}_{\Omega_{0}}^{(\mathring{u}_{2}\mathring{v}_{2})}\right)=\mathcal{G}_{\mathcal{L}}\left(\rho\mid\Omega_{0}\right),
\end{equation}
leading to a simple analytical computation of the quantifier proposed. To illustrate further, consider the important case of two-qubit states
with local marginals, \mbox{$\mathrm{Tr}_A \rho = \mathrm{Tr}_B \rho = I/2$} which are,
up to local unitaries $U_{1}\otimes U_{2}$, equivalent to states
of the form:
\begin{equation}
\rho\left(\mathbf{t}\right)=\frac{1}{4}\left\{ I\otimes I+\sum_{i}t_{i}\sigma_{i}\otimes\sigma_{i}\right\}, \label{eq:studystates1}
\end{equation}
where $\mathbf{t}=\left(t_{1},t_{2},t_{3}\right)$. The state  $\rho\left(\mathbf{t}\right)$
has a positive matrix  if $\mathbf{t}$ belongs to the tetrahedron $\Omega$
defined by the set of vertices $\left(-1,-1,-1\right)$, $\left(-1,1,1\right)$,
$\left(1,-1,1\right)$ and $\left(1,1,-1\right)$ as shown in
Fig.~\ref{fig:discord}. In this particular case  $u_{\rho,j}=t_{j}/\sqrt{2}$ and $v_{\rho,j}=-t_{j}/\sqrt{2}$.
Then, from (\ref{eq:discordquant}) we obtain:
\begin{equation}\label{bellstatemonotone}
\mathcal{G}_{\mathcal{L}}\left(\rho\mid\Omega_{0}\right)=\sqrt[3]{\prod_{i=1}^{3}\left(\!\sqrt{\sum_{j\neq i}\left|t_{j}\right|^{2}}\right)}.
\end{equation}

Moving on to the free operations, we notice that section preserving operations are unital
since $\textrm{Ker (\ensuremath{\Omega_{0}})}=\{\mathbf{0}\}$ in the basis $\left\{ \hat{\mathbf{u}}_{i},\hat{\mathbf{v}}_{i}\right\} _{i=1}^{3}$,
thus in particular operations $\widehat{\Delta}_{\lambda}$ act over
$\mathbf{r}_{\rho}$ as full contractions $\widehat{\Delta}_{\lambda}(\mathbf{r}_{\rho})=\lambda\mathbf{r}_{\rho}$ constituting white noise.
RNGs satisfying (\ref{eq:RNGcond1}) include reflections with respect to the origin,
axes and planes spanned by a pair $\left\{ \hat{\mathbf{u}}_{i},\hat{\mathbf{v}}_{i}\right\} $
or perpendicular to it, which are positive but not completely positive operators \cite{Altafi2006}. Also, RNGs include rotations $U\otimes U$
swapping simultaneously every pair $\left\{ \hat{\mathbf{u}}_{i},\hat{\mathbf{v}}_{i}\right\} $
into another $\left\{ \hat{\mathbf{u}}_{j},\hat{\mathbf{v}}_{j}\right\} $.
Because of (\ref{eq:framespandisc}) we can immediately write $\mathbf{r}_{\rho}\in\mathcal{C}_{\mathring{u}\mathring{v}}$
in terms of the frames:
\begin{equation*}
\mathcal{T}^{(\mathring{u}\mathring{v})}=\{(-1)^{\mathring{u}_{j}}\hat{\mathbf{u}}_{j},(-1)^{\mathring{v}_{j}}\hat{\mathbf{v}}_{j}\}_{j=1}^{3},    
\end{equation*}
and denote by  $\lambda_{j}^{v,\mathring{v}_{j}}$  and $\lambda_{j}^{u,\mathring{u}_{j}}$ the constants of an operation \eqref{eq:HCO} associated with the elements of their frame
$(-1)^{\mathring{u}_{j}}\hat{\mathbf{u}}_{j}$ and $(-1)^{\mathring{v}_{j}}\hat{\mathbf{v}}_{j}$
respectively.

Therefore, by studying Eq.~(\ref{eq:discordquant}) we identify three cone non-increasing
order operations  \eqref{eq:HCO}, first $\lambda_{3}^{u,\mathring{u}_{3}}=\lambda_{3}^{v,\mathring{v}_{3}}=\lambda_{3}\leq1$
determining a class of stochastic transition between populations,
second $\lambda_{1}^{u,\mathring{u}_{1}}=\lambda_{1}^{v,\mathring{v}_{1}}=\lambda_{2}^{u,\mathring{u}_{2}}=\lambda_{2}^{v,\mathring{v}_{2}}=\lambda_{1,2}\leq1$
representing a local qubit dephasing channel on Alice's side \cite{li1997special,kye1995positive,DephSuper2021},
and third a depolarising noise $\lambda_{j}^{u,\mathring{u}_{j}}=\lambda_{j}^{v,\mathring{v}_{j}}=\lambda\leq1$
for all $j$, equivalent to $\widehat{\Delta}_{\lambda}$. In fact
all previous hyperbolic contractions constitute  operations satisfying \eqref{eq:isomorphism}, precisely because the free sections
are spanned by a pair of vectors in some frame $\mathcal{T}^{(\mathring{u}\mathring{v})}$,
being part of $\partial\mathcal{C}_{\mathring{u}\mathring{v}}$ and
reaching the boundary of the two-qubit states $\Omega$.

\subsection{Witness of  total correlations}

Consider two parties, Alice and Bob, located far apart. They draw outcomes from their respective alphabets, which have $n_A$ and $n_B$ letters, based on the local probability distributions, $p_A$ and $p_B$, respectively. Sharing a common source of randomness $\lambda$ with a probability distribution $q(\lambda)$ allows two remote parties to establish classical correlations between their local devices, which they could exploit to solve distributed computational problems \cite{BuhrmanNonlocality2010}. In such setting a conditional probability $p_{\text{SR}}(a,b|x,y)$ of getting outcome $a$ from Alice and $b$ from Bob given inputs $x$ and $y$, respectively, is given by
\begin{equation}
    p_{\text{SR}}(a,b|x,y) = \int \dd{\xi} q(\xi) p_A(a|x;\xi) p_B(b|y;\xi),\label{SharRand}
\end{equation}
where $q(\xi)$ is the probability distribution of the shared random variable, and $p_A(a|x;\xi),\,p_B(b|y;\xi)$ are local conditional probability distributions for Alice and Bob, respectively, conditioned on the separate inputs $x,\,y$ and the shared variable $\xi$. 
It is well known that the classical correlations achievable in this way give what is called the Bell polytope \cite{brunner2014bell}, faces of which define the Bell inequalities,  allowing to distinguish classical correlations from the quantum correlations stemming from shared nonlocal quantum states.

When disconnected from their classical or quantum source of shared randomness, Alice and Bob's accessible, but uncorrelated distributions are limited to the product of local distributions, $p_{\text{NC}}(a,b) = p_A(a)p_B(b)$, which represent a limited subset of the more general bipartite distribution. Investigating classical and quantum sources of correlation relative to this reference distribution provides a theory of \emph{total correlations} \cite{Gour2019}. However,  such a theory cannot be convex, hindering its development until now.
A simple example would be two deterministic scenarios, with the outcomes being 0 for Alice and Bob in the first scenario and 1 for both in the second scenario. A convex combination between the two scenarios clearly cannot be written as a product of two local distributions. 

To demonstrate a constructive appication of the 
framework introduced, we analyze
witnesses of total correlations, which certify that a given distribution could not have been achieved without the use of shared random variables.

In the context of the total correlation model between Alice and Bob, we incorporate a third party, Charlie, who receives the messages sent by Alice and Bob. Assuming the communication channel is affected by depolarising noise of unknown level $\epsilon$, the probability distribution for messages received by Charlie is:
    \begin{equation}\label{eq:TC_free_set}
        p_C(a,b) = (1-\epsilon) p_{A}(a)p_{B}(b) + \epsilon \eta,
    \end{equation}
where $\eta \equiv 1/n_An_B$ is the flat distribution. Furthermore, for later convenience we introduce notation for deterministic probability distributions $ p_{ij}(a,b)\equiv \delta_{ij}^{ab}= \delta_{ia}\delta_{jb}$.
   Already at this stage we can identify two simple families of free operations, namely shrinking towards the kernel of the free set, the flat distribution, $\phi_\delta(p(a,b)) = (1-\delta)p(a,b) + \delta\eta$, and independent local output relabelings by  arbitrary permutations, $i_A \leftrightarrow \sigma_A(i_A),\,j_B\leftrightarrow\sigma_B(j_B)$. 
   
    Although $\mathcal{F}$ is a star domain with the flat distribution $\eta$ as its kernel, its  canonical fortress is neither polyhedric nor finite, requiring extra effort to apply the SRT framework being more costly to compute quantifiers and describe free operations.

     \begin{figure}[H]
	\centering
	\includegraphics[width=\linewidth]{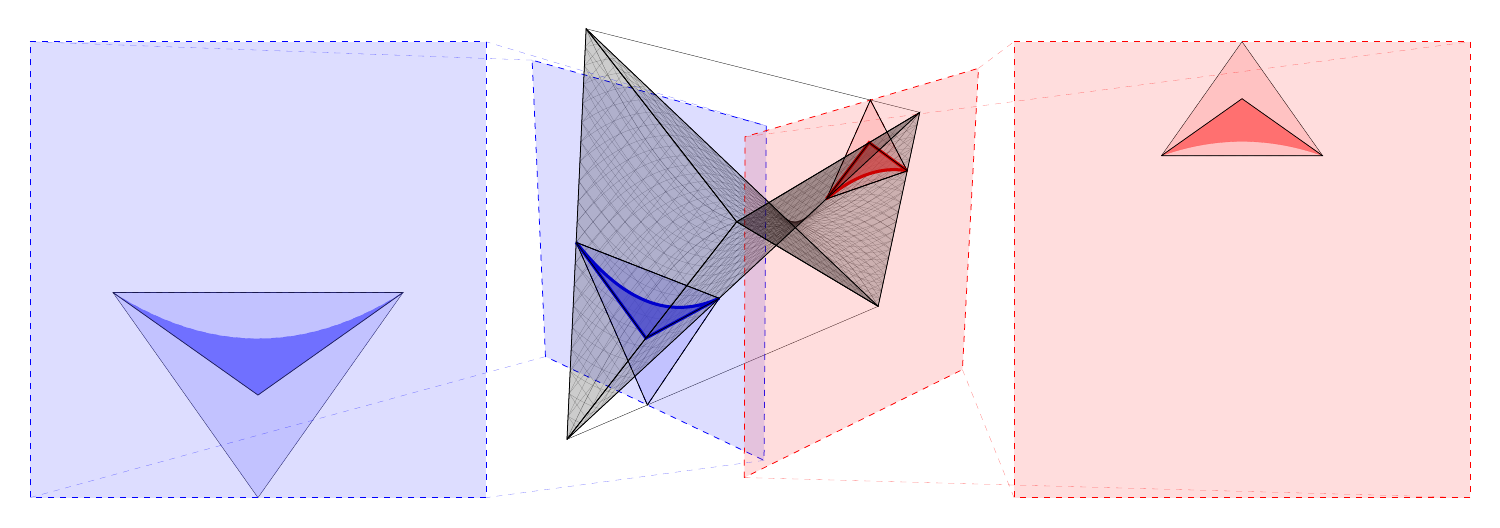}
	\caption{Free set for the total correlations, described by Eq.~\eqref{eq:TC_free_set} with $n_A = n_B = 2$, depicted in the simplex $\Delta_3$ of 4-dimensional probabilites, with the vertices defined by the deterministic distributions, top two corresponding to $00$ and $11$ outcomes, and bottom ones to $01$ and $10$. Note that the shaded gray set is indeed a star domain with center the orthocenter of the simplex. As can be observed in the two selected cross-sections (blue and red), the set is not convex.}
	\label{fig:TC_1}
    \end{figure}

However, a strategy to make the problem tractable is to identify a subset $\mathcal{F}'\subset\mathcal{F}$  which is a  star domain, but with a finite polyhedric fortress. The auxiliary free set $\mathcal{F}'$ should intuitively depict more restrictive physical situations than those accounted for by $\mathcal{F}$ to serve as a resourceless reference for $\mathcal{F}$ itself. To fulfil these requirements, a natural suggestion is to build an $\mathcal{F}'$ made up of device pairs for Alice and Bob, in which one device can access the entire set of local probabilities, and the other can only generate a deterministic output. 
A natural choice is to fix Alice's (Bob's) output, making the distribution fully deterministic, while allowing Bob's (Alice's) local distribution to vary arbitrarily. This produces two families of subsets of the probability simplex $\Delta_{n_An_B}$,
    \begin{equation}\label{eq:TC_faces}
        \begin{aligned}
            \mathcal{F}'_{Ai} & = \qty{p_C(a, b) = (1-\epsilon) p_{B}(b)\delta_{ai} + \epsilon \eta}, \\
            \mathcal{F}'_{Bj} & = \qty{p_C(a, b) = (1-\epsilon) p_{A}(a)\delta_{bj} + \epsilon \eta}.
        \end{aligned}        
    \end{equation}
        Union of the above families gives us the subset:
    \begin{equation}\label{eq:auxfree}
        \mathcal{F}' = \qty(\bigcup_{i=1}^{n_A} \mathcal{F}'_{Ai}) \cup \qty(\bigcup_{j=1}^{n_B} \mathcal{F}'_{Bj} ).
    \end{equation}
    The auxiliary free set $\mathcal{F}'\subset \mathcal{F}$ forms, by construction, a polyhedral star domain with kernel equal to the uniform probability $ \eta $, and free sections equal to the hyperfaces (\ref{eq:TC_faces}).  Furthermore, $\mathcal{F}'$ provides a natural polyhedral fortress  $\mathfrak{T}_{\mathcal{F}'} = \qty{\mathcal{C}_{ij}}$, comprised of cones with adjacent pairs of hyperfaces $\qty{\mathcal{F}'_{Ai},\mathcal{F}'_{Bj}}$: 
 
\begin{eqnarray*}
\mathcal{C}_{ij} & : & \left\{ p_{C}\left(a,b\right)=\eta +\gamma\left[ \delta^{ab}_{ij}-\eta \right]\!\!+\!\!\!\sum_{k=0}^{n_{B}-1}\!\!\alpha_{ik}\left[\delta^{ab}_{ik}\!-\!\eta\right]\right.\\
... &  & \left.+\!\sum_{l=0}^{n_{A}-1}\!\!\beta_{lj}\left[\delta^{ab}_{lj}\!-\!\eta\right]\!\mid\!\alpha_{ik},\beta_{lj},\gamma\in\mathbb{R}_{\geq0}\right\} .
\end{eqnarray*}

\begin{figure}[H]
	\centering
	\includegraphics[width=\linewidth]{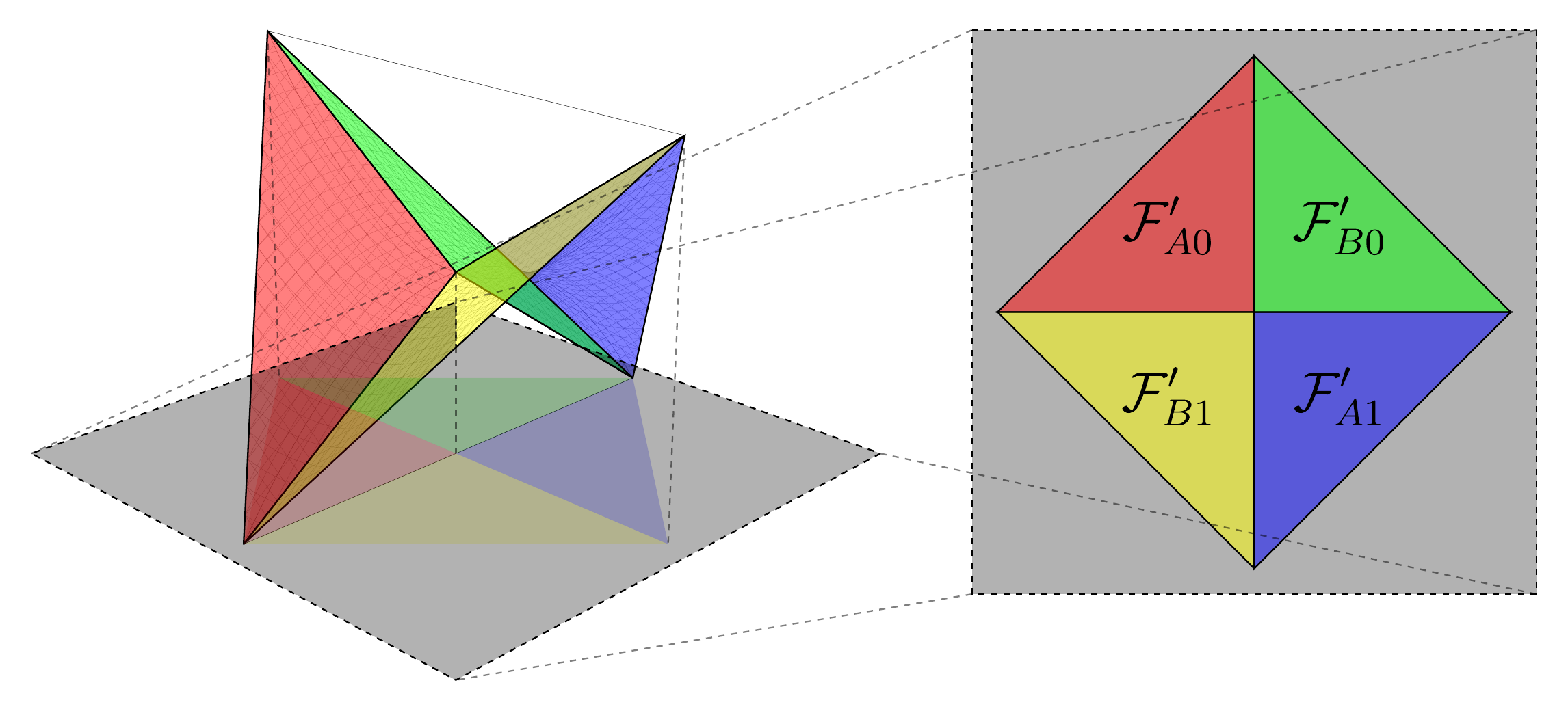}
	\caption{Four faces $\mathcal{F}'_{Ai}$ (red and blue) and $\mathcal{F}'_{Bj}$ (green and yellow) of the auxiliary polyhedral free set $\mathcal{F}'$, as defined in \eqref{eq:TC_faces}. 
 Note that $\mathcal{F}'$ is a lower-dimensional subset of the proper free set $\mathcal{F}\subset\Delta_3$ of the theory of total correlations.  }
	\label{fig:TC_2}
    \end{figure}

 To visualize a concrete case, we fix $n_A = n_B = 2$, and represent all joint probabilities $p(a,b)$ in a 3-dimensional simplex $\Delta_3$, where each vertex represents a deterministic distribution, 
  with the set $\mathcal{F}\subset\Delta_3$ taking approximately $30\%$ of the full simplex, as depicted in Fig.~\ref{fig:TC_1}. In this case there are just four faces, as shown in Fig.~\ref{fig:TC_2}, with each cone lying between two adjacent faces, eg. red-yellow or blue-green, but not yellow-green. Consequently, in the above case   the domains $\mathcal{D}^{(ij)}$, each corresponding to the cone $\mathcal{C}_{ij}$, have a simple form:   
  \begin{equation}\label{DomainsTot}
\mathcal{D}^{(ij)}  = \operatorname{Conv}\qty[\delta^{ab}_{ij},\delta^{ab}_{i(j\oplus1)},\delta^{ab}_{(i\oplus1)j},\eta]. 
       \end{equation}

In the exemplary case analyzed, it is easy to identify the regions where two cones meet,
    \begin{align*}        \mathcal{C}_{00}\cap\mathcal{C}_{11}=\partial\mathcal{C}_{00}\cap\partial\mathcal{C}_{11}= \operatorname{Conv}\qty[\delta^{ab}_{00},\delta^{ab}_{11},
        \eta], \\
        \mathcal{C}_{01}\cap\mathcal{C}_{10}=\partial\mathcal{C}_{01}\cap\partial\mathcal{C}_{10} = \operatorname{Conv}\qty[\delta^{ab}_{01},\delta^{ab}_{10},\eta],
    \end{align*}
   with similar boundary regions for arbitrary numbers of outcomes $n_A$ and $n_B$. Since, the overlapping regions consist in  symmetry planes of $\mathcal{F}'$, the value of the quantifier (\ref{eq:Conemeasure})  for a probability distribution inside the plane, will be the same for  both overlapping domains. Noteworthy is that the supporting cones of $\mathcal{F}$ intersect in multiple ways, requiring a more involved analysis to evaluate \eqref{eq:monotoneF} compared to $\mathcal{F}'$-based quantifiers.

   Furthermore, every support cone of $\mathcal{F}$ with an apex in one of the domains $\mathcal{D}^{(ij)}$ described in \eqref{DomainsTot} intersects the edges $\operatorname{Conv}[\delta^{ab}_{ij},\delta^{ab}_{i(j\oplus1)}]$ and $ \operatorname{Conv}[\delta^{ab}_{ij},\delta^{ab}_{(i\oplus1)j}] $, therefore involving three free sections: one that intersects the apex, and the other two are $\qty{\mathcal{F}'_{Ai},\mathcal{F}'_{Bj}}$. This attribute reveals that $\mathcal{F}'$-based quantifiers include two out of the three free sections involved in $\mathcal{F}$-based quantifiers.

    Besides, using the free sections and domains defined above we find a set of hyperbolic contracting  (\ref{eq:HCO}) type of free operations for the theory based on the auxiliary free set $\mathcal{F}'$. Concretely, the frames $\mathcal{T}^{(ij)}$ of  cones $\mathcal{C}_{ij}$ are:
    
    \begin{equation}\label{frameTotCorr}
         \mathcal{T}^{(ij)}\equiv \{\delta^{ab}_{ij}-\eta, \delta^{ab}_{ik}-\eta,\delta^{ab}_{lj}-\eta\}_{k\neq j, l\neq i},
    \end{equation}
then, we present the set of free operations \eqref{eq:HCO} with constants $\lambda_{rq}^{(ij)}$ associated with vector $\delta^{ab}_{rq} - \eta\in \mathcal{T}^{(ij)}$  in Appendix \ref{app:total_correlations_HCO}. Nevertheless, here we note that cone order-preserving operations with $\lambda_{rq}^{(ij)}\leq 1$, and $\lambda_{ij}^{(ij)} = 1$ consist in stochastic processes over $p_C$ projecting it to distributions in $\mathcal{F}'_{Ai},\mathcal{F}'_{Bj}$ along a cross section like those shown in Fig.~\ref{fig:TC_1}.  Indeed, the operations described previously represent the possibility of degrading Alice or Bob's resources for a given output, forming a set of RNG operations with respect to $\mathcal{F}$ in the exemplary case (See  Appendix \ref{app:total_correlations_HCO} for details). This fact reflects that the choice of $\mathcal{F}'$ not only has practical benefits but also is physically meaningful.

\begin{figure}[H]
	\centering
	\includegraphics[width=\linewidth]{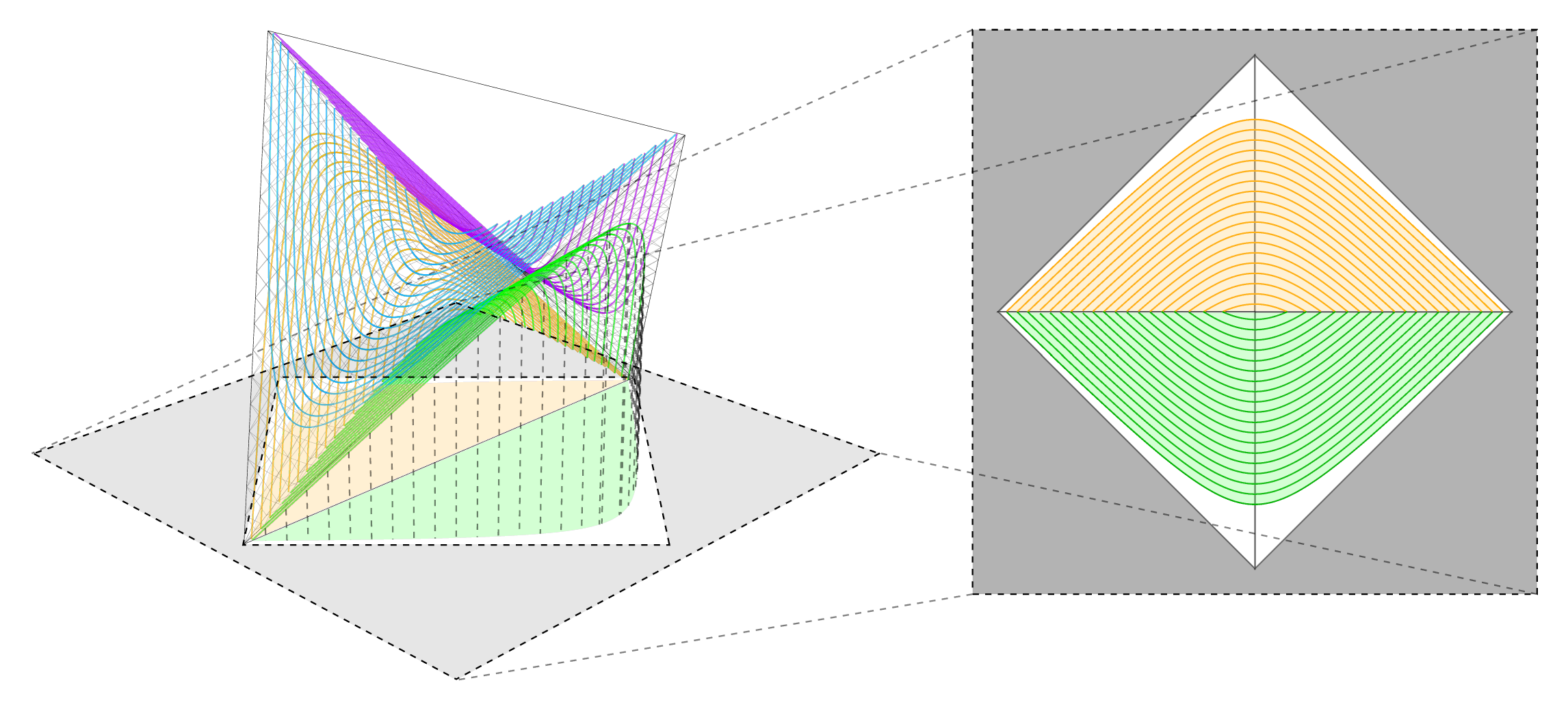}
	\caption{Four separating hyperbolic surfaces, based on the witnesses $\mathcal{W}_{TC}$. The surfaces separate the set $\mathcal{F}$ of uncorrelated distributions from a significant portion of those with total correlations.
    }
	\label{fig:TC_4}
    \end{figure}

    Now we have full construction of the theory based on the auxiliary free set $\mathcal{F}'\subset\mathcal{F}$ and we are ready to formulate a hyperbolic witness of total correlation steam from the monotones defined for $\mathcal{F}'$.
    In order to obtain the tightest witness, we consider the maximal value of monotone $\mathcal{G}_{\mathcal{L}}$ achievable by any uncorrelated probability distribution $p(a,b)\in\mathcal{F}$, 
    \begin{equation}
        \mathcal{G}_{\text{crit}} = \max_{p(a,b)\in\mathcal{F}} \mathcal{G}_{\mathcal{L}}\left(p(a,b)\mid\mathcal{F}'\right).
    \end{equation}
    In Appendix \ref{app:criticalcorrelations} we provide explicit calculations of the critical value, finding $\mathcal{G}_{\text{crit}} = \frac{1}{16}$.  Using the above we can define the hyperbolic witness of total correlations as
    \begin{equation}
        \mathcal{W}_{TC}\qty[p(a,b)] = \mathcal{G}_{\mathcal{L}}\left(p(a,b)\mid\mathcal{F}'\right) - \frac{1}{16}.
    \end{equation}
        We can tell that if $\mathcal{W}_{TC}\qty[p(a,b)] >0$, then $p(a,b)\notin\mathcal{F}$; It is not, however, a tight witness, i.e. $\mathcal{W}_{TC}\qty[p(a,b)] \leq 0$ does not certify absence of total correlations. Thus, it is the first resource theory up to date that allows to distinguish potentially local distributions from ones that must have used either classical or quantum shared randomness. For an extension of this approach to general networks' total correlations, see Appendix \ref{app:exttonetworks}.

\subsection{Witness of non-unistochasticity}

Classical stochastic processes require conservation of probability, which, on the level of transition matrices, translates into the normalization of columns -- the sum of (non-negative) elements in each of them is equal to one. 
When, additionally, the rows of a matrix are normalized such matrix $B$ is called  \emph{bistochastic}; for all $i$ we have $\sum_j B_{ij} = \sum_j B_{ji} = 1$. 

A ubiquitous notion in the quantum domain is a \emph{unitary} matrix $U$. 
Such matrices have orthonormal rows and columns; thus, the sum of squared absolute values of elements in rows/columns are equal to one, for all $i$ we have $\sum_j |U_{ij}|^2 = \sum_j |U_{ji}|^2 = 1$. 
Therefore, it is straightforward to construct a bistochastic matrix out of a unitary one
\begin{equation}\label{eq:bistochastic_and_unitary}
    B_{ij} \coloneqq |U_{ij}|^2.
\end{equation}

Alternatively, one may ask if it is possible to create a unitary matrix through some reverse operation for unspecified phases $\phi_{ij}$, defined as $U_{ij} \coloneqq e^{i\phi_{ij}} \sqrt{B_{ij}}$. 
However, since the unitarity requires not only the normalization but the orthogonality of the rows/columns, then there are additional constraints on the phases. 
More specifically, in every dimension $d>2$ there are bistochastic matrices without unitary counterparts. 
The problem of determining whether a given bistochastic matrix is \emph{unistochastic} (has a unitary counterpart) is solved fully for $d=3$~\cite{Au-Yeung_1979,Jarlskog_1988}, for $d=4$  several important sets were characterized~\cite{Bengtsson_2005,Rajchel_2022} together with a numerical algorithm determining the unistochasticity of a given matrix~\cite{Kruszynski_1987,Rajchel_2018}, while for $d>5$ only the unistochasticity of certain small-dimensional subsets was classified~\cite{Au-Yeung_1991,Rajchel_2018}. 

The unistochasticity problem is important from the perspective of particle physics~\cite{BS09,Di06,jarlskog}, quantum walks~\cite{aharonov1993quantum}, and classical to quantum transition~\cite{PKZ01,PTZ03}, to mention only a few of the applications.
The geometry of the set of bistochastic matrices is given by a Birkhoff polytope, whose vertices are permutation matrices. 
It is conjectured that the set of unistochastic matrices is star-shaped -- in fact, its superset was shown to be star-shaped with respect to the central, van der Waerden matrix~\cite{Rajchel_2022}.

\begin{figure}[h]
\centering
\begin{tikzpicture}
        \node at (0,0) {\includegraphics[width=\linewidth]{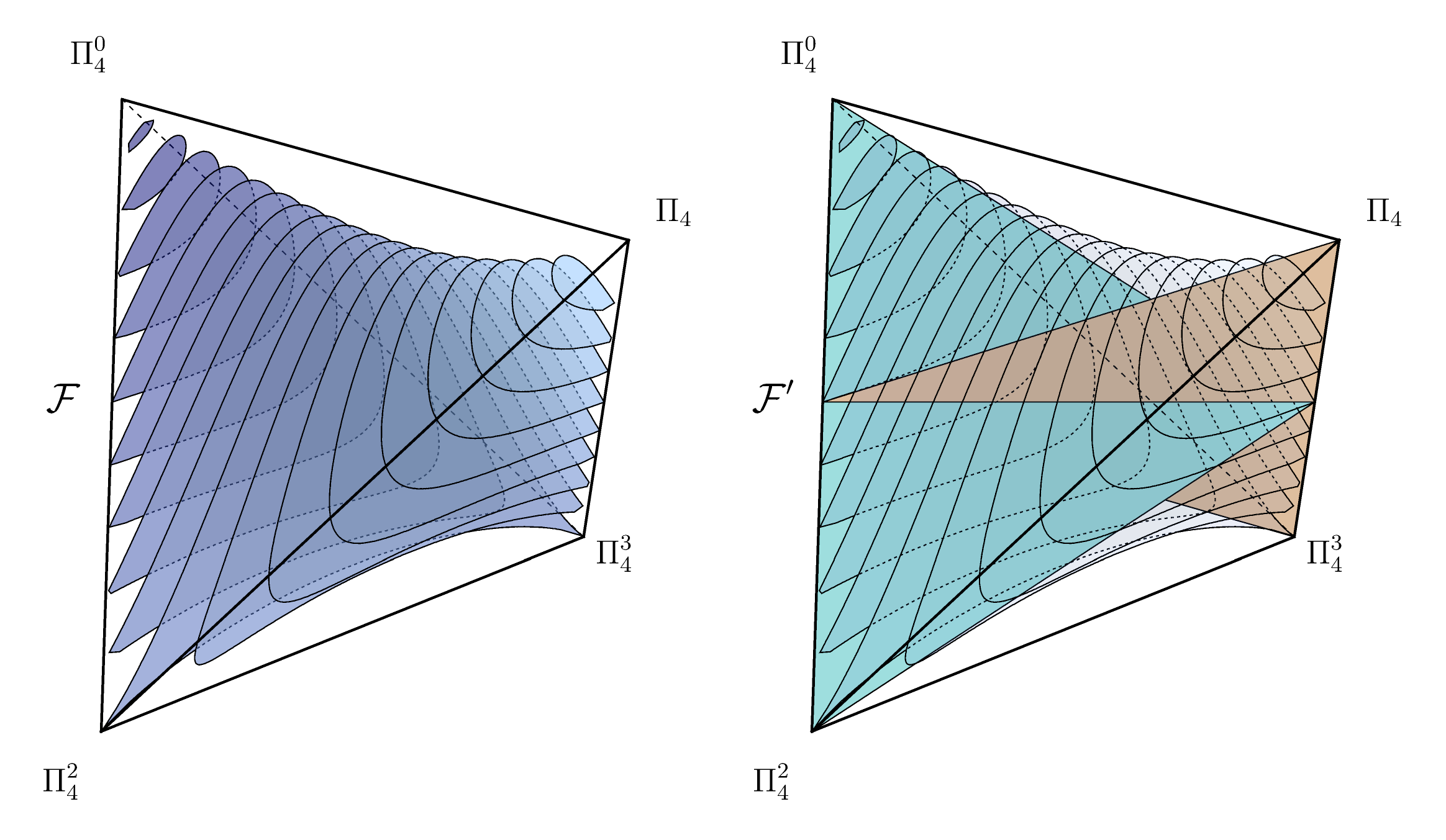}};
        
        \node at (-4.2,2.5) {(a)};
        \node at (0.,2.5) {(b)};
\end{tikzpicture}
    \centering
    \caption{The set of bistochastic circulant matrices of dimension 4 forms a regular tetrahedron. (a) Inside the tetrahedron, one can distinguish a non-convex set $\mathcal{F}$ of unistochastic matrices, depicted as a blue set. (b) The auxiliary free set $\mathcal{F}'$ of the advocated approach is given by a union of two equilateral triangles contained inside the 
    set of unistochastic matrices.}
    
    \label{fig:enter-label}
\end{figure}

Our rationale can be understood as a complementary effort to the certification of the unistochasticity of a given matrix. 
The validity of this approach lies in the disproving of unitary-evolution theories.
Although we believe that the microscopic description of the world is given by quantum mechanics, the actual experiments conducted in a laboratory result in a set of real numbers, not complex as would be suggested by the unitary evolution.
For example, any unitary transition between a set of quantum states, such as CKM (Cabibbo-Kobayashi-Maskawa) and PMNS (Pontecorvo-Maki-Nakagawa-Sakata) matrices ~\cite{Cabibbo_1963,KM_1973,Pontecorvo_1957, MNS_1962}, is probed by transition probabilities, which are squared absolute values of complex entries of the unitary matrix, forming a bistochastic one via Eq.~(\ref{eq:bistochastic_and_unitary}). However, bistochastic matrices without unitary counterparts cannot be modeled by an underlying unitary, quantum evolution.
Thus, suppose that the bistochastic matrix of neutrinos mixing, obtained in the experiment, is not unistochastic.  
This would show that there does not exist a PMNS matrix explaining their evolution, indicating a necessity of a new theory.

We illustrate how to apply the techniques introduced to address the problem of unistochasticity by focusing on the  circulant bistochastic
matrices of dimension $d = 4$ studied in depth in~\cite{Rajchel_2022} and shown in Fig.~\ref{fig:enter-label}. In this case, a natural free set consists of a set of unistochastic circulating matrices, which forms a star domain in its real representation within the tetrahedron of circulant bistochastic matrices. Using the operational task associated with a support cone, we can use the canonical fortress related to such a free set to detect each non-unistochastic circulating matrix. However, due to the presence of locally non-convex points along the boundary of unistochastic matrices, employing an auxiliary free set, similar to the strategy applied in the problem of 
total correlations, proves computationally more efficient.

In this example, as an auxiliary free set $\mathcal{F}'$, we choose the union of two triangles, $\mathcal{F}_{1} $ and $\mathcal{F}_{2}$, entirely included in the unistochastic set, described  in terms of the identity $\Pi_4^0$ and permutations $\Pi_4$, $\Pi_4^2$, $\Pi_4^3$~\cite{Rajchel_2022}, yielding:
\begin{equation}\label{eq:unistochastic_free_set}
    \begin{aligned}
        \mathcal{F}_{1} & : \qty{B=a \,\Pi_4^0 + b\,\Pi_4^2 + (1-a-b)\,\frac{\Pi_4+\Pi_4^3}{2}}, \\
        \mathcal{F}_{2} & : \qty{B=a \,\Pi_4 + b\,\Pi_4^3 + (1-a-b)\,\frac{\Pi_4^0+\Pi_4^2}{2}},
    \end{aligned}        
\end{equation}
with $0 \leq a,b \leq 1$. Joining them we compose the auxiliary free set,
\begin{equation}
    \mathcal{F}' = \mathcal{F}_{1}\cup  \mathcal{F}_{2}.
\end{equation}

Evidently $\mathcal{F}'$ is a star domain with kernel $\text{Ker}(\mathcal{F}'):\{ B = W_4 + \alpha \big(\frac{\Pi_4+\Pi_4^3}{2} - W_4\big)\}$, where $\alpha\in[-1,1]$ and $W_4$ is the van der Waerden matrix at the center of the tetrahedron. Also,  $\mathcal{F}_{1} $ and $\mathcal{F}_{2}$ constitute the only convex components of $\mathcal{F}'$, while the canonical fortress include the four cones: 

\begin{eqnarray*}
\mathcal{C}_{ij} & \! :\! & \left\{ B=W_{4}+\alpha_{+}\!\!\left(\frac{\Pi_{4}+\Pi_{4}^{3}}{2}-W_{4}\right)\!+\beta_{i}\!\left(\Pi_{4}^{i}-W_{4}\right)\right.\nonumber \\
 &  & \left.+\alpha_{-}\!\left(\frac{\Pi_{4}^{0}+\Pi_{4}^{2}}{2}\!-W_{4}\right)\!+\beta_{j}\!\left(\Pi_{4}^{j}\!-W_{4}\right)\!\right\}\!, 
\end{eqnarray*}
where $\alpha_{+},\alpha_{-},\beta_{i},\beta_{j}\geq0$, $i\in \{0,2\}$ and $j\in \{1,3\}$. 
Crucially, all four cones $\mathcal{C}_{ij} $ determine as free sections the sets  $\mathcal{F}_{1} $ and $\mathcal{F}_{2}$, making irrelevant to determine in which cone the target bistochastic matrix $B$ lies. The above fact paves the way for Bob certifying non-unistochasticity of $B$ by means of a sequence of operational tests, without complete knowledge of $B$.

\begin{figure}[H]
    \centering
    \includegraphics[width=.75\columnwidth]{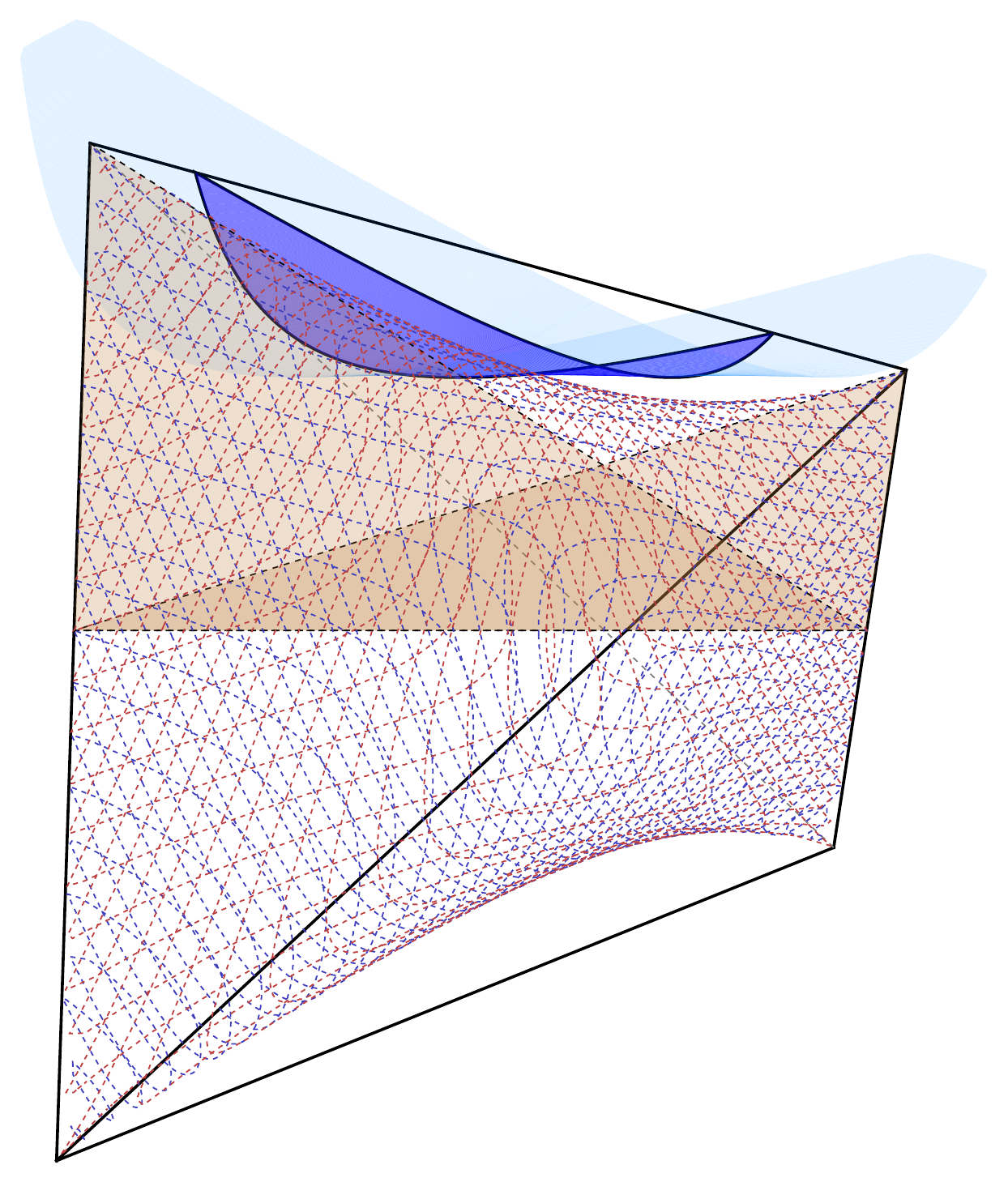}
    \caption{The blue surface divides the circulant bistochastic matrices of dimension $d=4$ into a subset containing only the non-unistochastic matrices.
    The figure emphasizes one quadrant of the full tetrahedron, while the behaviour in the others is analogous. 
    The maximal geometric mean of distances to the yellow planes for unistochastic matrices reads $\frac{1}{2\sqrt{2}} \approx 0.354$, and this value was chosen in order to construct the optimal dividing hyperbolic surface.    } 
    \label{fig:tetra}
\end{figure}

The certification protocol requires two steps. Firstly, we exploit the convexity bistochastic circulant  processes, to check that $B$ belongs to that set, by means of the discrimination protocols like those in \cite{TakagiRegula2019}. Later, if $B$ pass the previous test, we proceed with the operational task of Theorem 1, to distinguish $B$ from both $\mathcal{F}_{1} $,  $\mathcal{F}_{2}$, and compute the quantifier using  \eqref{operquant}. Finally  if the quantifier is greater than $\frac{1}{2\sqrt{2}}$ then the matrix is not unistochastic (for a proof of the critical value see Appendix \ref{Unisto}). 
Therefore, we have the following witness of non-unistochasticity:
\begin{equation}\label{eq:witness_non-unistochasticity}
        \mathcal{W}_{NU}\qty[B] = \mathcal{G}_{\mathcal{L}}\left(B\mid\mathcal{F}'\right) - \frac{1}{2\sqrt{2}}.
    \end{equation}
If the analyzed matrix $B$ satisfies $\mathcal{W}_{NU}\qty[B]>0 $, then it  does not correspond to any unitary transformation. The separation surface is depicted in Fig.~\ref{fig:tetra}.

Such reasoning can be extended to various setups in which we consider a transition between quantum  and classical evolution, as well as to classical walks without quantum counterparts. 
To sum up, proposed setup allow us to disprove \emph{operationally} the validity of a theory attempting to explain  quantum to classical transition, demonstrating the need for additional degrees of freedom.

\subsection{Witness of non-Markovianity}

When considering evolution with certain degree of randomness, it is common to resort to what is called a stochastic process. Moreover, oftentimes it is assumed that the process depends only on the most recent state of the system as opposed to the entire prior trajectory. These short-memory processes are called Markovian, forming an important non-convex subset of all possible stochastic processes as they find use in wide variety of settings, spanning statistical physics \cite{breuer2002theory}, chemistry \cite{anderson2011continuous}, biology \cite{zhang2019markovian}, and economics \cite{mirman2008qualitative}, to name a few.

In quantum mechanics one way of defining the distinction between Markovian and non-Markovian processes is through the CP divisibility as described in \cite{wolf2008dividing}.
The analogy between the classical and quantum Markovianity is understood immediately when we consider that at every infinitesimal time interval $\tau$ the only input to the channel is the prior state of the system, $\rho_{t+\tau} = \Lambda_\tau(\rho_t)$.

It was recently discovered that mixing two Markovian processes gives rise to a non-Markovian one~\cite{Megier_2017}.
In this application, we show a practical and effective approach to construct a witness of non-Markovianity for the set of channels obtained as convex combinations of three Pauli channels \cite{Vinayak-free}. Within this context, Pauli channels and their mixtures serve as a concrete demonstration of the non-convex structure of the space of Markovian quantum operations
\cite{DZP19,PRZ19}. A Pauli channel is defined as:

\begin{equation}\label{Paulichannel}
    \Gamma_i^p(\rho) = p \rho + \qty(1-p)\sigma_i\rho\sigma_i,
\end{equation}
 with $\sigma_i \in \qty{\sigma_1, \sigma_2, \sigma_3}$ being the standard Pauli matrices and $p$ being the probability of finding the input state $\rho$ unperturbed. A previous study \cite{Vinayak-free} demonstrated that a nontrivial convex combination of two distinct Pauli channels exhibits non-Markovian behavior despite the individual channels being Markovian. Precisely, for a convex combination of three Pauli channels,

\begin{equation}\label{Paulicomb}
    \Theta = a\Gamma^p_1 + b\Gamma_2^p +c\Gamma_3^p,
\end{equation}
 with non-negative $a,b,c$ such that $a+b+c=1$, the subset of Markovian channels is  non-convex. Indeed, \cite{Vinayak-free} describes analytically the form of the set of Markovian Pauli-channels, which forms a star domain with a kernel composed solely of the maximally depolarizing channel, $p=0,a=b=c=1/3$. Furthermore, the boundary of Markovian Pauli-channels contains only locally non-convex points, leading to a canonical fortress  of infinite support cones. Although, in this case, we could directly apply advocated methods, it is more practical  to define an auxiliary free set to certify resources, as in the previous applications. Furthermore, in this section, we propose to delve into the benefits of this technique by reducing the study space (\ref{Paulicomb}) to the cross-section  $p=1/2$, a vital case analyzed in \cite{Vinayak-free}.

\begin{figure}[h]
    \centering
    \includegraphics[width=\columnwidth]{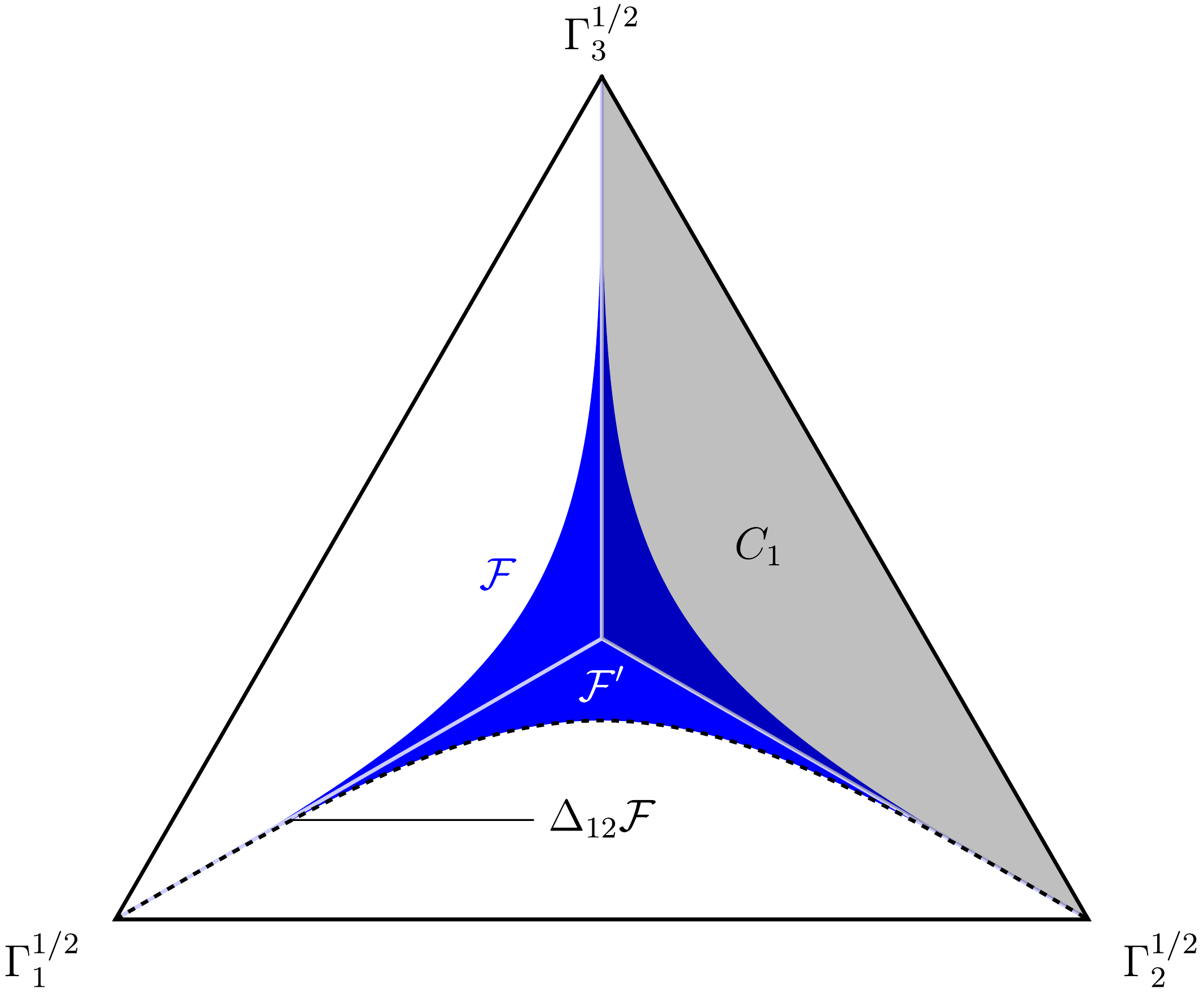}
    \caption{The theory of non-Markovian channels provides a rich  area of phenomena in which one can 
    apply the framework introduced. A concrete example 
 are convex combinations of Pauli channels $\Gamma_{i}^{p}$, with a particular section portrayed in  the figure (\mbox{$p=1/2$}). The blue region corresponds to  the Markovian  subset of channels $\mathcal{F}$, while the external regions correspond to the non-Markovian channels. The figure also highlights some relevant geometric elements such as the auxiliary free set $\mathcal{F}'\subset\mathcal{F}$ (gray lines), support cone $\mathcal{C}_1$ (opaque region) and the boundary arc 
$\varDelta_{12}\mathcal{F}$ (dotted curve).   }
    \label{vinayak}
\end{figure}

  Precisely, we can visualize all sets involved at the cross-section $p = 1/2$ within the triangle depicted in Fig.\ref{vinayak}. The triangle represents the channels of the cross-section employing
$(a,b,c)$ as barycentric coordinates, where the blue region $\mathcal{F}$
determines all Markovian channels with $p=1/2$. The boundary $\partial\mathcal{F}$
of the blue region form a triangle with parabolic arc edges, usually
denoted as parabolic horn triangle.
In barycentric coordinates we
obtain one of such arc edges by selecting a given $\bar{b}\leq\sqrt{5}-2$,
an $a$ coordinate:

\begin{equation}
\bar{a}_{\pm}=\frac{1-\bar{b}}{2}\left(1\pm\sqrt{\frac{5-\left(\bar{b}+2\right)^{2}}{1-\bar{b}^{2}}}\right),
\end{equation}
and then $\bar{c}_{\pm}=1-\bar{a}_{\pm}-\bar{b}=\bar{a}_{\mp}$, leading to points
$(\bar{a}_{\pm},\bar{b},\bar{a}_{\mp})$. Similarly, we obtain the
other two arc edges by replacing the starting coordinate $b$ either by $a$, or $c$ in the above
construction. Inside $\mathcal{F}$ we distinguish three segments:
\begin{equation}
    \mathcal{F}_j: \{\Phi_j\mid \Phi_j=q\Gamma_j^{1/2}+(1-q)\Gamma_d^{1/2} \},
\end{equation}
with $j\in\{1,2,3\}$, $0\leq q \leq 1$, and $\Gamma_d^{1/2}$ stands for the channel with barycentric coordinates $(1/3,1/3,1/3)$. Since segments $\mathcal{F}_j$ are tangent to $\partial\mathcal{F}$ at the vertices of the triangle, the kernel of $\mathcal{F}$ has a single element $\Gamma_d^{1/2}$, at the intersection of segments $\mathcal{F}_j$.

To better understand the relevance of the sets $\mathcal{F}_{j}$, let us observe that we could construct each parabolic arc $\varDelta_{ij}\mathcal{F}\in\partial\mathcal{F}$
that joins $\Gamma_{i}^{1/2}$ with $\Gamma_{j}^{1/2}$ using De Casteljau's
algorithm \cite{farin2000}:
\begin{equation}
\varDelta_{ij}\mathcal{F}\!:\!\{\Theta_{s}\!\mid\!\Theta_{s}\!=\!\left(1\!-\!s\right)^{2}\Gamma_{i}^{1/2}\!\!+\!2s\left(1-s\right)\Gamma_{d}^{1/2}\!\!+\!s^{2}\Gamma_{j}^{1/2}\},\label{eq:Bezier}
\end{equation}
with parameter $s\in\left[0,1\right]$, forming a B\'{e}zier curve
with control points $\Gamma_{i}^{1/2}$, $\Gamma_{d}^{1/2}$, $\Gamma_{j}^{1/2}$.
A direct consequence of (\ref{eq:Bezier}) is that every support cone
 of $\mathcal{F}$ intersects the boundary at most at three points: an apex $\Theta_{s}\in\varDelta_{ij}\mathcal{F}$
with the corresponding vertices $\Gamma_{i}^{1/2}$ and $\Gamma_{j}^{1/2}$. Hence, by computing proposed quantifier using an auxiliary
free set $\mathcal{F}^{\prime}$ containing the segments $\mathcal{F}_{j}$,
the loss of information will be negligible while the complexity of
the calculations is reduced. The simplest of such auxiliary free sets
corresponds to:
\begin{equation}
  \mathcal{F}'=\mathcal{F}_1 \cup \mathcal{F}_2 \cup \mathcal{F}_3.  
\end{equation}

It is clear that the set $\mathcal{F}'$ forms a star domain with its kernel being the depolarizing channel $\Gamma_d^{1/2}$ introduced before.
Moreover, the convex components coincide with segments $\mathcal{F}_{j}$, and the canonical fortress $\mathfrak{T}_{\mathcal{F}'}$ has a simple finite set of support cones:

\begin{equation}
   \mathcal{C}_k : \{\Gamma_{d}^{1/2} + \sum_{i\neq k}\alpha_i(\Gamma_{i}^{1/2}-\Gamma_{d}^{1/2})  |\quad \alpha_i \geq 0\}, 
\end{equation}
It is direct to verify that every cone has two free sections forming its boundary inside the study region, e.g. $\mathfrak{C}_{\mathcal{F}}^{(1)}=\{ \mathcal{F}_{2},\mathcal{F}_3 \}$. In consequence the domains $\mathcal{D}^{(k)}$ coincide with the support cones $\mathcal{C}_k$, lack overlapping interiors, and allow simple computation of (\ref{eq:Conemeasure}) and (\ref{eq:monotoneF}).

\begin{figure}[H]
    \centering
    \includegraphics[width=\columnwidth]{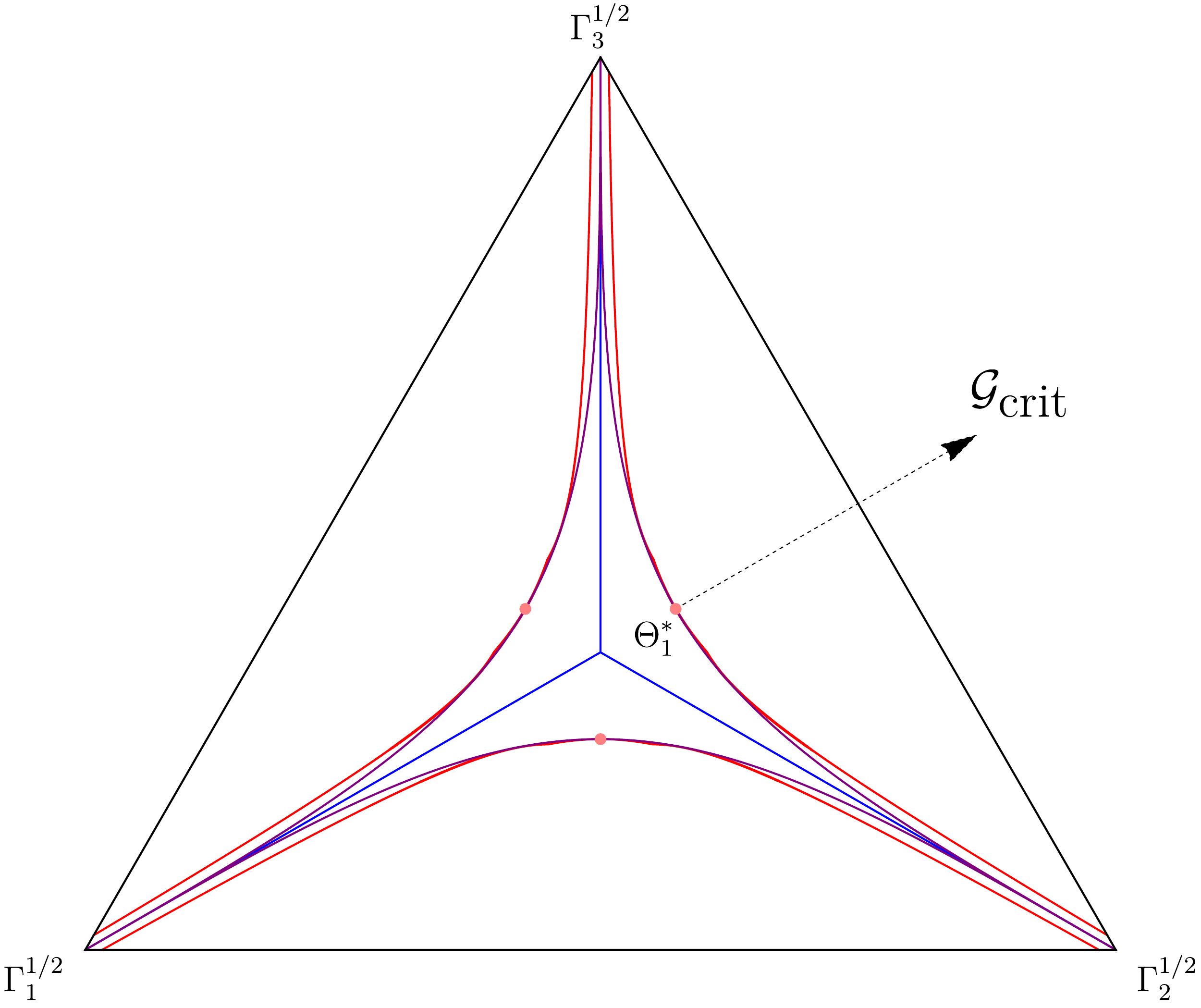}
    \caption{ The red hyperbolas indicate  the channels for which the $\mathcal{F}'$-based monotone provides the maximal value for Markovian channels  $\mathcal{G}_{\text{crit}}=\frac{1}{8} (7-3\sqrt{5})$ 
    Every channel beyond the red curve is detected as non-Markovian. Notably, the set of non-Markovian channels  undetected is confined to the small region between the gray parabolic arcs and the red curve.} 
    \label{critical-points}
\end{figure}

As in the previous applications, the goal is to find a tight upper bound in the value of the quantifier (\ref{eq:Conemeasure}) that Markovian channels can achieve. Geometrically, the tightest upper bound happens when the hyperbolic surfaces of equally resourceful channels are tangent to parabolic arcs $\varDelta_{ij}\mathcal{F}$. The parabolic arcs include, at their extremes, Pauli channels to which the hyperbolas of equal resources are asymptotic; in consequence both curves must coincide at a different point between the vertex and the parabola's focus. Because the above applies to both ends of the parabola, the tangent hyperbola can only include the focus. The channel at the focus of the  parabolic arc inside $\mathcal{C}_k$ is given by,

\begin{equation}
\Theta^{*}_{k}= (\sqrt{5}-2)\Gamma_{k}^{1/2} + \frac{1}{2} (3-\sqrt{5}  )\sum_{i\neq k}\Gamma_{i}^{1/2}.  
\end{equation}
Since we wish to illustrate clearly exemplary applications
of the techniques introduced, we simplify the channel discrimination by setting the input state to a part of the maximally entangled state. This allows us to exploit the Choi-Jamiołkowski isomorphism \cite{jamiolkowski1972linear,choi1975completely} to express all the combinations of Pauli channels as states. Thus, for a channel satisfying (\ref{Paulicomb}), with $p=1/2$ we have:

\begin{equation}\label{choi-matrix}
    J_\Theta=\begin{pmatrix}
\frac{1+c}{4} & 0 & 0 & \frac{1-c}{4} \\
0 & \frac{a+b}{4} & \frac{a-b}{4} & 0\\
0 &  \frac{a-b}{4}& \frac{a+b}{4}  & 0\\
 \frac{1-c}{4} & 0 & 0 &  \frac{1+c}{4}\\
\end{pmatrix},
\end{equation}
and the explicit Jamiołkowski states of $J_{\Phi_j} $ for an arbitrary $\Phi_j\in\mathcal{F}_j$,  presented in Appendix \eqref{markovian-calc}. Now, since computing the bound induced by a channel $\Theta^{*}_{k}$ in $\mathcal{D}^{(k)}=\mathcal{C}_k$ is symmetrical for every domain, it is enough to compute it for $\Theta^{*}_{1}\in\mathcal{D}^{(1)}$. In this case the distance quantifier gave us (see Appendix \eqref{markovian-calc}):

\begin{align}\label{markovian-bound} 
 \mathcal{L}\left(J_{\Theta^{*}_{1}}\mid\mathcal{F}_1\right) &=\min_{\Phi_1 \in \mathcal{F}_1} \frac{1}{2}||J_{\Theta^{*}_{1}}-J_{\Phi_1} ||_1 \nonumber \\ 
    &= \frac{1}{8} \Big (7-3\sqrt{5} \Big )\approx 0.0364745,
\end{align}
with the same value for $\mathcal{L}\left(J_{\Theta^{*}_{1}}\mid\mathcal{F}_2\right) $, since from the definition of focus, the point is equidistant to $\mathcal{F}_1$ and $\mathcal{F}_2$. Then, the corresponding geometric mean  (\ref{eq:Conemeasure}):

\begin{align}
    \mathcal{G}_{\mathcal{L}}\left(J_{\Theta^{*}_{1}}\mid\mathfrak{C}_{\mathcal{F}}^{(1)}\right)&=\mathbb{G}_{j}\left[\mathcal{L}\left(J_{\Theta^{*}_{1}}\mid\mathcal{F}_{j}^{(1)}\right)\right]\nonumber\\
    &=\sqrt{\mathcal{L}\left(J_{\Theta^{*}_{1}}\mid\mathcal{F}_1\right) \mathcal{L}\left(J_{\Theta^{*}_{1}}\mid\mathcal{F}_2\right)} \nonumber\\ 
    &=\mathcal{L}\left(J_{\Theta^{*}_{1}}\mid\mathcal{F}_1\right)  = \frac{1}{8} \Big (7-3\sqrt{5} \Big ), \label{boundmarkovian}
\end{align}
because the only domain that enters in the maximization is $\mathcal{D}^{(1)}$, (\ref{boundmarkovian}) also provides the value of $\mathcal{G}_{\mathcal{L}}\left(J_{\Theta^{*}_{1}}\mid\mathcal{F}\right)$.
As we pointed out before, this is an upper bound to the set of Markovian channels, since due to the symmetry of the problem, every $\mathcal{G}_{\text{crit}}=\mathcal{G}_{\mathcal{L}}\left(J_{\Theta^{*}_{k}}\mid{\mathcal{F}}\right)$ has the same value.  In conclusion for any $J_{\Theta}$ satisfying  $\mathcal{G}_{\mathcal{L}}\left(J_{\Theta}\mid{\mathcal{F}}\right)>\mathcal{G}_{\text{crit}}$ we can assure that the convex combination of Pauli channels  $\Theta$ is  non-Markovian, as portrayed in Fig.  \ref{critical-points}.

The section preserving operations include any mixing with $\Gamma_d^{1/2}$, representing a depolarizing noise. Also, every domain $\mathcal{D}^{(k)}$ admits all types of hyperbolic contracting operations (\ref{eq:HCO}) with respect to the auxiliary free set $\mathcal{F}'$, but only the conic non-increasing order  are RNG with respect to $\mathcal{F}$ (See Appendix \ref{ApxConicRNG} for details). Additionally, we remark that for every domain the  conic non-increasing operations have different physical meaning, for instance in $\mathcal{D}^{(3)}$ they essentially involve an additional dephasing, while for $\mathcal{D}^{(1)}$, and $\mathcal{D}^{(2)}$, the dephasing is restricted to imaginary or real parts, together with an extra stochastic classical action.

\section{Discussion} \label{sec:dis}

The standard approach to resource theories, particularly applying convex analysis, has been instrumental in understanding and exploiting quantum resources like entanglement. However, these frameworks need to improve in their capacity to fully capture the scope of quantum phenomena that are naturally non-convex. Our work addresses this limitation by introducing Star Resource Theories (SRTs), which rely on the unique geometric properties of star-shaped sets. This foundational shift motivates the development of a broader mathematical landscape: a \emph{Star-analysis}, capable of encompassing the full spectrum of known quantum properties.

Our results further justify the development of the above new mathematical field by demonstrating its ability to generalize and enhance existing quantum resource theories, including convex and crucial non-convex ones. This generalization opens up new possibilities for assessing and harnessing quantum resources, which were previously inaccessible under the constraints of convexity.

While recent research \cite{Schluck_2022,adesso2023} offers quantifiers for operationally assessing non-convex resources, we demonstrate that key experimentally valuable theories, such as quantum-state texture \cite{Parisio2024}, extend beyond their methodological capabilities but are effectively captured within our framework. Moreover, our approach makes it possible to define universal nontrivial free operations for SRTs. As a result, we introduce additional operationally testable examples where only our method currently enables the rigorous development of faithful witnesses and free operations, providing a complete resource theory.

Furthermore, the quantifiers designed explicitly for star resource theories, introduce concrete improvements. Firstly, they have desirable properties, such as faithfulness and being a convex function when appropriate. Secondly, they are also better at suppressing relative error (computational or from measurements) than previous methodologies. Furthermore,  the operational interpretations of the quantifiers introduced are unprecedented, evaluating new practical advantages, which simultaneously combine discrimination tasks and correlations, opening the possibility of novel applications.

We also describe the application of the formalism introduced and its techniques in four different research areas, grouped into two by their prior attainability. Firstly, we analyze the
quantum discord and non-Markovianity to show new analytical results with operational relevance. Secondly, the framework proposed is used to 
certify total correlations in a quantum system
and non-unistochasticity of certain bistochastic matrices.
In this way the resource theory is extended
to hitherto unattainable fields. We envision generalization of the results obtained for total correlations for classical and quantum networks as a promising area with multiple repercussions. 

Some potential future research directions
can consist in
adapting of the methods proposed
to computation of covariance matrices and their subsequent application to Bell nonlocality in networks \cite{Chaves2020,Tavakoli2022} or extending resource analysis of Buscemi nonlocality \cite{Buscemi2012,PatrykPRX2021} 
to networks of more than two parties. In classical networks, the approach studied can contribute to the classification of weighted networks \cite{horvath2011} or to determine the reducibility of multilayer networks \cite{DeDomenico2015}, with consequences ranging from biology to economics.

On top of that, extending these results to particle physics, verification of the unistochasticity of a transition matrix between neutrino eigenbases is a test of both the correctness of the measurement as well as the theory of the PMNS (Pontecorvo-Maki-Nakagawa-Sakata) mixing matrix~\cite{Pontecorvo_1957, MNS_1962}. In this process, the quantum unitary evolution underlays the real, bistochastic matrices observed in experiments. Thus, a very similar setup could test the quark families' mixing while testing the unistochasticity of the CKM (Cabibbo-Kobayashi-Maskawa) matrix~\cite{Cabibbo_1963,KM_1973}.

The broader implications of this work extend beyond the realm of quantum information theory. The capacity of the setup proposed to exclude a particular point from a specific set finds relevance in various settings outside quantum information, providing a valuable tool for validating models or experimental tests. This mathematical leap aligns with recent developments in non-linear witnesses \cite{Otfried2006,Shen2020,Sen2023}, and supports further applications of non-convex optimization,  for instance, in support vector machine learning \cite{ma2014support,Prateek2017}.

In summary, this article establishes a novel paradigm in resource theory, introducing new methodological capabilities that outperform previous quantum information methods. Moreover, the implications of SRTs reach far beyond quantum properties, offering the physical motivation to forge ahead into a new era of non-convex analysis. The wide-ranging nature of this work holds promise for diverse applications, pushing forward the frontiers of quantum information and mathematical analysis.
\vspace{0.5 cm}

\section*{Acknowledgements}

We are grateful to Carsten Voelkmann for studying the manuscript in detail and pointing out numerous typographical errors and further remarks. We thank Andrés Ducuara for providing fruitful comments and suggestions, as well as Nicolas Brunner and Francesco Buscemi for their insightful discussions and interest in this project.
This work realized within the  QuantERA II DQquant project  
 was supported by the National Science Centre, Poland, under the contract number 2021/03/Y/ST2/00193,
which received funding from the European Union’s Horizon 2020 research and innovation programme under Grant Agreement No 101017733.
RS acknowledges the financial support of the Foundation for Polish Science through the TEAM-NET project (contract no. POIR.04.04.00-00-17C1/18-00). JCz acknowledges financial support by NCN PRELUDIUM BIS no. DEC-2019/35/O/ST2/01049. RRR is financially supported by the Ministry for Digital Transformation and of Civil Service of the Spanish Government through the QUANTUM ENIA project call - Quantum Spain project, by the EU through the RTRP - NextGenerationEU within the framework of the Digital Spain 2026 Agenda, and by Conselleria d’Educació i Universitats del Govern de les Illes Balears and Fons Social Europeu+ through the contract POSTDOC2024\_17. 
GR-M acknowledges funding from the European Innovation Council accelerator grant COMFTQUA, no. 190183782, as well as ERC AdG NOQIA; MICIN/AEI (PGC2018-0910.13039/501100011033, CEX2019-000910-S/10.13039/501100011033, Plan National FIDEUA PID2019-106901GB-I00, FPI; MICIIN with NextGenerationEU (PRTR-C17.I1): QUANTERA MAQS PCI2019-111828-2); MCIN/AEI/ 10.13039/501100011033 and by QUANTERA DYNAMITE PCI2022-132919 within the QuantERA II Programme under Grant Agreement No 101017733 -- “Retos Colaboración” QUSPIN RTC2019-007196-7).
PH acknowledges partial support by the
Foundation for Polish Science (IRAP project, ICTQT, contract no. MAB/2018/5, co-financed by the EU within the Smart Growth Operational Programme).
The authors also acknowledge funding from the European Union’s Horizon Europe research and innovation programme under the project "Quantum Secure Networks Partnership" (QSNP, grant agreement No 101114043).

\appendix

\begin{widetext}
\section*{Appendix}

\begin{figure}[H]
    \centering
    \includegraphics[width=0.9\columnwidth]{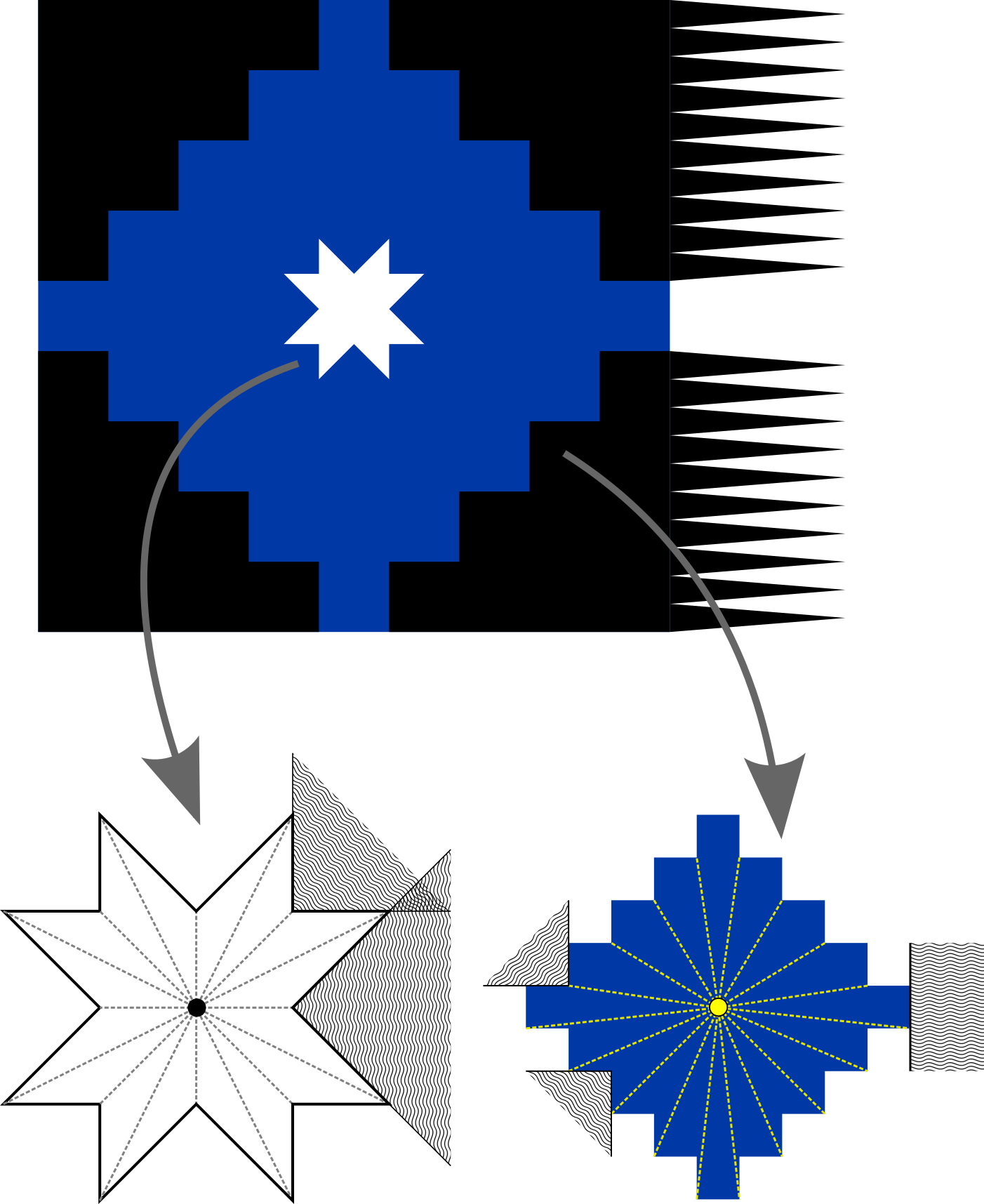}
    \caption{The study and artistic exploration of star-shaped figures highlight the intersection of mathematics and aesthetics, demonstrating how geometric concepts are expressed creatively in various cultural and artistic contexts. For instance, some ancient symbols from natives of Latin America, such as Lautaro's flag, provide several nontrivial shapes, which are good examples of admissible free sets $\F$ of an SRT, being star domains with a natural finite polyhedral fortress.}
    \label{fig:my_label}
\end{figure}

\section{Geometrical relations}
\subsection{Redundancy deletion of support cones satisfies the covering condition}
\label{app:covering}

Here we demonstrate that the fortress condition i) is still satisfied for the support cones, after redundancy deletion relative to $\mathcal{F}$. 

We start by recalling that for any point $z\notin\mathcal{F}$, we can select a $w\in\textrm{Ker\ensuremath{(\mathcal{F})}}$
to obtain $z_{w}=\lambda z+(1-\lambda)w$ at the boundary $\partial\mathcal{F}$,
leading to a support cone $\mathcal{C}_{z_{w}}^{\prime}\ni z$. Next, for every $\mathcal{C}_{z_{w}}^{\prime}\ni z$,
two scenarios unfold: either $\mathcal{C}_{z_{w}}^{\prime}=\mathcal{C}_{x^{*}}$
or there exists a $\mathcal{C}_{x^{*}}$ such that $$\mathcal{C}_{x^{*}}\cap\mathcal{F}\supset\mathcal{C}_{z_{w}}^{\prime}\cap\mathcal{F}.$$
In both instances, $z_{w}\in\mathcal{C}_{z_{w}}^{\prime}\cap\mathcal{F}$
holds, showing that $z_{w}\in\mathcal{C}_{x^{*}}$. 
Next, we define a cone $$\mathcal{Q}_{w}:\{w+\alpha_{1}\left(z-w\right)+\alpha_{2}\left(x^{*}-w\right)\mid\alpha_{1},\alpha_{2}\in\mathbb{R}_{\geq0}\},$$ then
by construction we have $\left(\mathcal{Q}_{w}+z_{w}-w\right)\ni z$,
simply taking $\alpha_{1}=1-\lambda,\alpha_{2}=0$.  However, this
means $$\left(\mathcal{Q}_{w}-w+x^{*}\right)\ni z-z_{w}+x^{*},$$ but
$\left(\mathcal{Q}_{w}-w+x^{*}\right)$ shifts the apex of $\mathcal{Q}_{w}$
to $x^{*}$ where the support cone is $\mathcal{C}_{x^{*}}$, implying
that $$\mathcal{C}_{x^{*}}\ni z-z_{w}+x^{*}$$ or equivalently $\left(\mathcal{C}_{x^{*}}+z_{w}-x^{*}\right)\ni z$. Thus, since $z_{w}\in\mathcal{C}_{x^{*}}$, shifting the apex of $\mathcal{C}_{x^{*}}$
to $z_{w}$ generates a subset of $\mathcal{C}_{x^{*}}$, and in consequence
$$\mathcal{C}_{x^{*}}\supseteq\left(\mathcal{C}_{x^{*}}+z_{w}-x^{*}\right)\ni z,$$ demonstrating the satisfaction of the fortress condition i) after redundancy deletion. \qed

Lastly, we would like to remark an interesting particular case. Whenever $\textrm{Int\ensuremath{(\textrm{Ker\ensuremath{(\mathcal{F})}})}}\neq\emptyset$,
we could choose $w\in\textrm{Int\ensuremath{(\textrm{Ker\ensuremath{(\mathcal{F})}})}}$
 then we must have $\textrm{Int}(\mathcal{C}_{z_{w}}^{\prime})\ni z$,
and following a reasoning among the same lines as before we would
conclude that there exists a $\mathcal{C}_{x^{*}}$ such that $z\in\textrm{Int}(\mathcal{C}_{x^{*}})$.
In the last case, we have: 
\begin{equation*}
\mathcal{F}\supseteq\left[\bigcup_{\ensuremath{x}}\textrm{Int}(\mathcal{C}_{x})\right]^{\textbf{c}}\equiv\bigcap_{x}\textrm{Ext}(\mathcal{C}_{x})
\end{equation*}
and because the converse inclusion is straightforward from ii),
we obtain: 
\begin{equation}\label{singlefortress}
\mathcal{F}=\bigcap_{x}\textrm{Ext}(\mathcal{C}_{x}).
\end{equation}

The above shows, that all fortress conditions could be reduced to the single condition (\ref{singlefortress}) for the special case $\textrm{Int\ensuremath{(\textrm{Ker\ensuremath{(\mathcal{F})}})}}\neq\emptyset$. However, to keep the theory general and include several applications where the kernel has no interior, we must keep the definition of fortress based on conditions i)-iv). 

\section{Proofs for general theory}\label{Genproofs}

\subsection{Proof of relative error suppression} \label{relerrosupr}

To estimate a relative error we must assume $\mathcal{M}\left(\Theta\mid\mathcal{F}\right)>0$
, which also implies $\mathcal{G}_{\mathcal{M}}\left(\Theta\mid\mathcal{F}\right)>0$
due to faithfulness of both quantifiers. Then, for $\mathcal{F}$
a star domain the monotones $\mathcal{M}\left(\Theta\mid\mathcal{F}\right)$
based on decomposition into sub-convex sets can be written as \cite{adesso2023,Kuroiwa2023}:
\begin{equation}
\mathcal{M}\left(\Theta\mid\mathcal{F}\right)=\inf_{\mathcal{F}_{\eta}^{(x)}\subseteq\mathcal{F}}\mathcal{M}\left(\Theta\mid\mathcal{F}_{\eta}^{(x)}\right).\label{eq:req1}
\end{equation}
Moreover,  we assume only that each observed quantity $\mathcal{M}(\Theta\mid\mathcal{F}_{\eta}^{(x)})$  deviates from its ideal value by at most $\delta\mathcal{M}(\Theta\mid\mathcal{F}_{\eta}^{(x)}) $, without making assumptions on the statistical distribution of these deviations. Then, we remark that any multivariate error-propagation formula is used here purely as a deterministic first-order sensitivity bound (valid for arbitrary perturbations within these intervals), not as a probabilistic variance-propagation formula. The resulting bounds, therefore, quantify the maximum possible deviation of the functional under worst-case perturbations.

Consequently, when assigning the error $\delta\mathcal{M}\left(\Theta\mid\mathcal{F}\right)$
to $\mathcal{M}\left(\Theta\mid\mathcal{F}\right)$ we should assign the largest error found in the evaluation of \eqref{eq:req1}, thus:

\begin{equation}
\delta\mathcal{M}\left(\Theta\mid\mathcal{F}\right) =\max_{\mathcal{F}_{\eta}^{(x)}\subseteq\mathcal{F}}\delta\mathcal{M}\left(\Theta\mid\mathcal{F}_{\eta}^{(x)}\right).\label{eq:req2}
\end{equation}
 Now, we obtain the error $\delta\mathcal{G}_{\mathcal{M}}(\Theta\mid\mathfrak{C}_{\mathcal{F}}^{(x)})$
for every $\mathcal{G}_{\mathcal{M}}(\Theta\mid\mathfrak{C}_{\mathcal{F}}^{(x)})$
with $\Theta\in\mathcal{D}^{(x)}$, applying the standard error propagation
formula \cite{clifford1973}:
\begin{eqnarray}
    \delta\mathcal{G}_{\mathcal{M}}\left(\Theta\mid\mathfrak{C}_{\mathcal{F}}^{(x)}\right) & = & \sqrt{\int_{\Sigma_{x}} d\xi\left(\delta\mathcal{M}\left(\Theta\mid\mathcal{F}_{\xi}^{(x)}\right)\frac{\partial\mathcal{G}_{\mathcal{M}}\left(\Theta\mid\mathfrak{C}_{\mathcal{F}}^{(x)}\right)}{\partial\mathcal{M}\left(\Theta\mid\mathcal{F}_{\xi}^{(x)}\right)}\right)^{2}}\nonumber \\
 & \overset{(a)}{=} & \frac{1}{\left|\Sigma_{x}\right|}\sqrt{\int_{\Sigma_{x}} d\xi\left(\mathcal{G}_{\mathcal{M}}\left(\Theta\mid\mathfrak{C}_{\mathcal{F}}^{(x)}\right)\frac{\delta\mathcal{M}\left(\Theta\mid\mathcal{F}_{\xi}^{(x)}\right)}{\mathcal{M}\left(\Theta\mid\mathcal{F}_{\xi}^{(x)}\right)}\right)^{2}}\nonumber \\
 & = & \frac{\mathcal{G}_{\mathcal{M}}\left(\Theta\mid\mathfrak{C}_{\mathcal{F}}^{(x)}\right)}{\left|\Sigma_{x}\right|}\sqrt{\int_{\Sigma_{x}} d\xi\left(\frac{\delta\mathcal{M}\left(\Theta\mid\mathcal{F}_{\xi}^{(x)}\right)}{\mathcal{M}\left(\Theta\mid\mathcal{F}_{\xi}^{(x)}\right)}\right)^{2}},
\end{eqnarray}
where in $(a)$ we used the most general form of the geometric mean:
$$\mathcal{G}_{\mathcal{M}}\left(\Theta\mid\mathfrak{C}_{\mathcal{F}}^{(x)}\right)=\exp\left\{\frac{1}{\left|\Sigma_{x}\right|} \int_{\Sigma_{x}} d\eta\ln\left(\mathcal{M}\left(\Theta\mid\mathcal{F}_{\eta}^{(x)}\right)\right)\right\}, $$
and $\left|\Sigma_{x}\right|=\int d\xi$ indicates the integration
over the whole collection of free sections $\mathfrak{C}_{\mathcal{F}}^{(x)}$.
Using the above we can bound the relative error:
\begin{eqnarray}
\frac{\delta\mathcal{G}_{\mathcal{M}}\left(\Theta\mid\mathfrak{C}_{\mathcal{F}}^{(x)}\right)}{\mathcal{G}_{\mathcal{M}}\left(\Theta\mid\mathfrak{C}_{\mathcal{F}}^{(x)}\right)} & = & \frac{1}{\left|\Sigma_{x}\right|}\sqrt{\int_{\Sigma_{x}} d\xi\left(\frac{\delta\mathcal{M}\left(\Theta\mid\mathcal{F}_{\xi}^{(x)}\right)}{\mathcal{M}\left(\Theta\mid\mathcal{F}_{\xi}^{(x)}\right)}\right)^{2}}\nonumber \\
 & \overset{(b)}{\leq} & \frac{1}{\left|\Sigma_{x}\right|}\sqrt{\int_{\Sigma_{x}} d\xi\left(\frac{\delta\mathcal{M}\left(\Theta\mid\mathcal{F}\right)}{\mathcal{M}\left(\Theta\mid\mathcal{F}\right)}\right)^{2}}\nonumber \\
 & = & \frac{1}{\sqrt{\left|\Sigma_{x}\right|}}\frac{\delta\mathcal{M}\left(\Theta\mid\mathcal{F}\right)}{\mathcal{M}\left(\Theta\mid\mathcal{F}\right)},\label{eq:supprelerr}
\end{eqnarray}
where in $(b)$ we combined both (\ref{eq:req1}) and (\ref{eq:req2})
methodological requirements to bound the relative error of each free
section $\mathcal{F}_{\eta}^{(x)}\subseteq\mathcal{F}.$ Since, (\ref{eq:supprelerr})
holds for every $\mathcal{D}^{(x)}\ni\Theta$, and defining $\Sigma=\left|\Sigma_{x}\right| \geq 1$ we arrive at the corresponding
relative error bound: 
\begin{equation}
\frac{\delta\mathcal{G}_{\mathcal{M}}\left(\Theta\mid\mathcal{F}\right)}{\mathcal{G}_{\mathcal{M}}\left(\Theta\mid\mathcal{F}\right)}\leq\frac{1}{\sqrt{\Sigma}}\frac{\delta\mathcal{M}\left(\Theta\mid\mathcal{F}\right)}{\mathcal{M}\left(\Theta\mid\mathcal{F}\right)},
\end{equation}
which shows the suppression of relative errors, when computing $\mathcal{G}_{\mathcal{M}}\left(\Theta\mid\mathcal{F}\right)$. \qed

\subsection{Proof of monotonicity}
\label{app:Monproofs}
We consider a monotone $\mathcal{M}(\cdot\!\!\mid\!\!\mathcal{X}):\mathcal{S}\rightarrow\mathbb{R}_{\geq0}$
faithful for a convex free set $\mathcal{X}$ and which satisfies
the identity: 
\begin{equation}
\mathcal{M}\left(\lambda\Theta+\left(1-\lambda\right)\Lambda\mid\!\mathcal{X}\right)\leq\lambda\mathcal{M}\left(\Theta\mid\!\mathcal{X}\right)+\left(1-\lambda\right)\mathcal{M}\left(\Lambda\mid\!\mathcal{X}\right)\quad\forall\lambda\in\left[0,1\right]\label{eq:triconv}
\end{equation}
which is true for typical monotones, such as trace distance or generalised
robustness. In such case the monotone $\mathcal{G}_{\mathcal{M}}(\cdot\mid\mathfrak{C}_{\mathcal{F}}^{(x)})$
applied on a domain $\mathcal{D}^{(x)}$ is given by the geometric
average over the monotones $\mathcal{M}(\Theta\mid\mathcal{F}_{j}^{(x)})$,
determined by each free section $\mathcal{F}_{j}^{(x)}\in\mathfrak{C}_{\mathcal{F}}^{(x)}$ :
\begin{equation}
\mathcal{G}_{\mathcal{M}}\left(\Theta\mid\mathfrak{C}_{\mathcal{F}}^{(x)}\right)=\mathbb{G}_{j}\left[\mathcal{M}\left(\Theta\mid\mathcal{F}_{j}^{(x)}\right)\right],
\end{equation}
where $\mathbb{G}_{j}\left[x_{j}\right]:=\sqrt[N]{\prod_{j=1}^{N}x_{j}}$
is the geometric mean of variables $x_{j}$, here understood as
every $\mathcal{M}(\Theta\mid\mathcal{F}_{j}^{(x)})$. Before continuing we need to show the following lemma: 

\begin{description}
\item [{Lemma 1}]  \emph{Given the free set $\mathcal{F}$ of an }SRT
\emph{with center $\Delta\in \mathrm{Ker} (\mathcal{F})$, and a fortress $\mathfrak{T}_{\mathcal{F}}=\left\{ \mathcal{C}_{x}\right\} _{x\in X}$, we have that if $\widehat{\Delta}_{\lambda}\left[\Theta\right]\in\mathcal{C}^{(x)}$
with $\lambda > 0$, then also $\Theta\in\mathcal{C}^{(x)}$. }
\end{description}
\emph{Proof. }If $\lambda=1$ the statement is trivial, then if for
$1>\lambda>0$ the object $\widehat{\Delta}_{\lambda}\left[\Theta\right]$ belongs
to the convex cone $\mathcal{C}^{(x)}$ with frame $\mathcal{T}^{(x)}$, we have: 
\begin{equation}
\widehat{\Delta}_{\lambda}\left[\Theta\right]=v_{0}^{(x)}+\sum_{y\in Y_x}\alpha_{y}\left(v_{y}^{(x)}-v_{0}^{(x)}\right),\label{eq:conelamb}
\end{equation}
 for some apex $v_{0}^{(x)}$,  vectors $v_{y}^{(x)}-v_{0}^{(x)}$ in the frame $\mathcal{T}^{(x)}$ with label set $Y_x$,
 and constants $\alpha_{y}\geq0$. Equation
(\ref{eq:conelamb}) directly leads to: 
\begin{eqnarray}
\Theta & = & \frac{1}{\lambda}v_{0}^{(x)}-\left(\frac{1-\lambda}{\lambda}\right)\Delta+\sum_{y\in Y_x}\frac{\alpha_{y}}{\lambda}\left(v_{y}^{(x)}-v_{0}^{(x)}\right)\nonumber \\
 & \overset{(a)}{=} & v_{0}^{(x)}+\left(\frac{1-\lambda}{\lambda}\right)\sum_{y\in Y_x}\gamma_{y}\left(v_{y}^{(x)}-v_{0}^{(x)}\right)+\sum_{y\in Y_x}\frac{\alpha_{y}}{\lambda}\left(v_{y}^{(x)}-v_{0}^{(x)}\right)\nonumber \\
 & \overset{(b)}{=} & v_{0}^{(x)}+\sum_{y\in Y_x}\beta_{y}\left(v_{y}^{(x)}-v_{0}^{(x)}\right),\label{eq:conetrans}
\end{eqnarray}
where in $(a)$ we used the fortress condition iii) $\textrm{Refl}(\mathcal{C}_{x})\supseteq \textrm{Ker}(\mathcal{F})$, and in $(b)$ we defined $\beta_{y}=\left(\frac{1-\lambda}{\lambda}\right)\gamma_{y}+\frac{\alpha_{y}}{\lambda}\geq0$.
By the definition of cone, equation (\ref{eq:conetrans}) shows that
$\Theta\in\mathcal{C}^{(x)}$. \qed

Now, let's study the monotonicity of $\mathcal{G}_{\mathcal{M}}(\Theta\mid\mathfrak{C}_{\mathcal{F}}^{(x)})$
under $\widehat{\Delta}_{\lambda}$. If $\widehat{\Delta}_{\lambda}\left[\Theta\right]\in\mathcal{F}$
then, by construction $\mathcal{G}_{\mathcal{M}}(\widehat{\Delta}_{\lambda}\left[\Theta\right]\mid\mathfrak{C}_{\mathcal{F}}^{(x)})=0\leq\mathcal{G}_{\mathcal{M}}(\Theta\mid\mathfrak{C}_{\mathcal{F}}^{(x)})$.
On the contrary, if $\widehat{\Delta}_{\lambda}\left[\Theta\right]\notin\mathcal{F}$,
the object $\widehat{\Delta}_{\lambda}\left[\Theta\right]$ must belong to one
or more cones $\mathcal{C}^{(x)}$, and from the definition of $\mathcal{G}_{\mathcal{M}}(\Theta\mid\mathfrak{C}_{\mathcal{F}}^{(x)})$
we have: 
\begin{eqnarray}
\mathcal{G}_{\mathcal{M}}(\widehat{\Delta}_{\lambda}\left[\Theta\right]\mid\mathcal{F}) & = & \sup_{\mathcal{D}^{(x)}:\widehat{\Delta}_{\lambda}\left[\Theta\right]\in\mathcal{D}^{(x)}}\mathcal{G}_{\mathcal{M}}(\widehat{\Delta}_{\lambda}\left[\Theta\right]\mid\mathfrak{C}_{\mathcal{F}}^{(x)})\nonumber \\
 & = & \sup_{\mathcal{D}^{(x)}:\widehat{\Delta}_{\lambda}\left[\Theta\right]\in\mathcal{D}^{(x)}}\mathbb{G}_{j}\left[\mathcal{M}\left(\widehat{\Delta}_{\lambda}\left[\Theta\right]\mid\mathcal{F}_{j}^{(x)}\right)\right]\nonumber \\
 & \leq & \sup_{\mathcal{D}^{(x)}:\widehat{\Delta}_{\lambda}\left[\Theta\right]\in\mathcal{D}^{(x)}}\mathbb{G}_{j}\left[\lambda\mathcal{M}\left(\Theta\mid\mathcal{F}_{j}^{(x)}\right)+\left(1-\lambda\right)\mathcal{M}\left(\Delta\mid\mathcal{F}_{j}^{(x)}\right)\right]\nonumber \\
 & \overset{(a)}{=} & \sup_{\mathcal{D}^{(x)}:\widehat{\Delta}_{\lambda}\left[\Theta\right]\in\mathcal{D}^{(x)}}\mathbb{G}_{j}\left[\lambda\mathcal{M}\left(\Theta\mid\mathcal{F}_{j}^{(x)}\right)\right]\nonumber \\
 & = & \lambda\sup_{\mathcal{D}^{(x)}:\widehat{\Delta}_{\lambda}\left[\Theta\right]\in\mathcal{D}^{(x)}}\mathcal{G}_{\mathcal{M}}(\Theta\mid\mathfrak{C}_{\mathcal{F}}^{(x)})\nonumber \\
 & \overset{(b)}{=} & \lambda\mathcal{G}_{\mathcal{M}}(\Theta\mid\mathfrak{C}_{\mathcal{F}}^{(x^{*})})\nonumber \\
 & \leq & \mathcal{G}_{\mathcal{M}}(\Theta\mid\mathfrak{C}_{\mathcal{F}}^{(x^{*})})\nonumber \\
 & \overset{(c)}{\leq} & \sup_{\mathcal{D}^{(x)}:\Theta\in\mathcal{D}^{(x)}}\mathcal{G}_{\mathcal{M}}(\Theta\mid\mathfrak{C}_{\mathcal{F}}^{(x)})\nonumber \\
 & = & \mathcal{G}_{\mathcal{M}}(\Theta\mid\mathcal{F}),\label{eq:monfin}
\end{eqnarray}
where in $(a)$ we used (\ref{eq:triconv}), in $(b)$ we evaluate
the domain $\mathcal{D}^{(x^{*})}$ with the maximal value such that cone
$\widehat{\Delta}_{\lambda}\left[\Theta\right]\in\mathcal{C}^{(x^{*})}$ and in
$(c)$ we use lemma 1 which ensures that $\Theta\in\mathcal{C}^{(x^{*})}$
and thus $\Theta\in\mathcal{D}^{(x^{*})}$. \qed

\subsection{Proof of operational interpretation of distance base monotone}\label{OperTask1}
We proceed to demonstrate the operational meaning of the monotone
$\mathcal{G}_{\mathcal{L}}$ when the underlying monotone $\mathcal{M}$
is a distance $\mathcal{L}$ which discriminates the resource from
the free set. In the proof we will use the symbol $\Theta$ for channels,
but the same proof holds for states and measurements under an appropriate
choice of the distance $\mathcal{L}$. In particular when $\Theta$
belongs to the domain $\mathcal{D}$ which contains the cone $\mathcal{C}$
and free sections $\left\{ \mathcal{F}_{1},...,\mathcal{F}_{M}\right\} $
we would like to prove that:
\begin{equation}
P\left(\bigoplus_{j=1}^{M}\left[a_{j}\oplus1\right]=0\mid\Theta\right)=\frac{1}{2}\left[1+\prod_{j=1}^{M}\mathcal{L}\left(\Theta\mid\mathcal{F}_{j}\right)\right],\label{eq:optmean1}
\end{equation}
with $\oplus$ the sum mod(2) and every $a_{j}$ is the binary truth
value of discrimination between $\Theta$ and $\mathcal{F}_{j}$ i.e.
$a_{j}=1$ iff $\Theta\notin\mathcal{F}_{j}$ and $a_{j}=0$ iff $\Theta\in\mathcal{F}_{j}$,
\begin{equation}
P\left(a_{j}=1\mid\Theta\right)=\frac{1}{2}\left[1+\mathcal{L}\left(\Theta\mid\mathcal{F}_{j}\right)\right].\label{eq:discrim}
\end{equation}
Note, that we write (\ref{eq:optmean1}) using the relation $e_j=a_{j}\oplus1$.

\emph{Proof.} Since, for $M=1$ the result is trivial, let's check
the $M=2$ case: 
\begin{eqnarray}
P\left(\bigoplus_{j=1}^{2}\left[a_{j}\oplus1\right]=0\mid\Theta\right) & = & P\left(\bigoplus_{j=1}^{2}a_{j}=0\mid\Theta\right)\nonumber \\
 & = & P\left(a_{1}=0\mid\Theta\right)P\left(a_{2}=0\mid\Theta\right)+P\left(a_{1}=1\mid\Theta\right)P\left(a_{2}=1\mid\Theta\right)\nonumber \\
 & \overset{(a)}{=} & \frac{1}{4}\left[1-\mathcal{L}\left(\Theta\mid\mathcal{F}_{1}\right)\right]\left[1-\mathcal{L}\left(\Theta\mid\mathcal{F}_{2}\right)\right]+\frac{1}{4}\left[1+\mathcal{L}\left(\Theta\mid\mathcal{F}_{1}\right)\right]\left[1+\mathcal{L}\left(\Theta\mid\mathcal{F}_{2}\right)\right]\nonumber \\
 & = & \frac{1}{2}\left[1+\prod_{j=1}^{2}\mathcal{L}\left(\Theta\mid\mathcal{F}_{j}\right)\right].\label{eq:discrpair}
\end{eqnarray}
Let's assume now that if $M$ is odd, by inductive hypothesis the
equation (\ref{eq:optmean1}) is true for all $M^{\prime}<M$. Then,
\begin{eqnarray}
P\left(\bigoplus_{j=1}^{M}\left[a_{j}\oplus1\right]=0\mid\Theta\right) & \overset{(a)}{=} & P\left(\left[\bigoplus_{j=1}^{M}a_{j}\right]\oplus1=0\mid\Theta\right)\nonumber \\
 & = & P\left(\left[\bigoplus_{j=1}^{M-1}a_{j}\right]\oplus\left[a_{M}\oplus1\right]=0\mid\Theta\right)\nonumber \\
 & = & P\!\left(\left[\bigoplus_{j=1}^{M-1}a_{j}\right]\!\!=0\!\mid\!\Theta\right)P\left(a_{M}\!\oplus\!1=\!0\!\mid\!\Theta\right)\!+\!P\!\left(\left[\bigoplus_{j=1}^{M-1}a_{j}\right]\!\!=1\!\mid\!\Theta\right)P\left(a_{M}\!\oplus\!1=\!1\!\mid\!\Theta\right)\nonumber \\
 & \overset{(b)}{=} & P\!\left(\bigoplus_{j=1}^{M-1}\left[a_{j}\!\oplus\!1\right]\!=\!0\!\mid\!\Theta\right)P\!\left(a_{M}=1\!\mid\!\Theta\right)\!+\!P\!\left(\left[\bigoplus_{j=1}^{M-1}\left[a_{j}\!\oplus\!1\right]\right]\!=\!1\!\mid\!\Theta\right)P\!\left(a_{M}=1\!\mid\!\Theta\right)\nonumber \\
 & \overset{(c)}{=} & \frac{1}{4}\left[1+\prod_{j=1}^{M-1}\mathcal{L}\left(\Theta\mid\mathcal{F}_{j}\right)\right]\left[1+\mathcal{L}\left(\Theta\mid\mathcal{F}_{M}\right)\right]+\frac{1}{4}\left[1-\prod_{j=1}^{M-1}\mathcal{L}\left(\Theta\mid\mathcal{F}_{j}\right)\right]\left[1-\mathcal{L}\left(\Theta\mid\mathcal{F}_{M}\right)\right]\nonumber \\
 & = & \frac{1}{4}\left[1+\prod_{j=1}^{M}\mathcal{L}\left(\Theta\mid\mathcal{F}_{j}\right)\right],\label{eq:indc1}
\end{eqnarray}
where in $(a)$ we use the fact that $M$ is odd and $\oplus$ is
the sum mod(2), in $(b)$ we take advantage of $M-1$ being even and
in $(c)$ we use the inductive hypothesis. Let's assume now that if
$M$ is even, by inductive hypothesis the equation (\ref{eq:optmean1})
is true for all $M^{\prime}<M$. Then, 
\begin{eqnarray}
P\left(\bigoplus_{j=1}^{M}\left[a_{j}\oplus1\right]=0\mid\Theta\right) & \overset{(a)}{=} & P\left(\left[\bigoplus_{j=1}^{M}a_{j}\right]=0\mid\Theta\right)\nonumber \\
 & \overset{(b)}{=} & P\left(\left[\bigoplus_{j=1}^{M-1}\left[a_{j}\oplus1\right]\right]\oplus\left[a_{M}\oplus1\right]=0\mid\Theta\right)\nonumber \\
 & = & P\!\left(\bigoplus_{j=1}^{M-1}\left[a_{j}\!\oplus\!1\right]\!=\!0\!\mid\!\Theta\right)P\!\left(a_{M}\!\oplus\!1\!=\!0\!\mid\!\Theta\right)\!+\!P\!\left(\bigoplus_{j=1}^{M-1}\left[a_{j}\!\oplus\!1\right]\!=\!1\!\mid\!\Theta\right)P\!\left(a_{M}\!\oplus\!1\!=\!1\!\mid\!\Theta\right)\nonumber \\
 & = & P\!\left(\bigoplus_{j=1}^{M-1}\left[a_{j}\!\oplus\!1\right]\!=\!0\!\mid\!\Theta\right)P\!\left(a_{M}=1\!\mid\!\Theta\right)\!+\!P\!\left(\bigoplus_{j=1}^{M-1}\left[a_{j}\!\oplus\!1\right]\!=\!\!1\!\mid\!\Theta\right)P\!\left(a_{M}=1\!\mid\!\Theta\right)\nonumber \\
 & \overset{(c)}{=} & \frac{1}{4}\left[1+\prod_{j=1}^{M-1}\mathcal{L}\left(\Theta\mid\mathcal{F}_{j}\right)\right]\left[1+\mathcal{L}\left(\Theta\mid\mathcal{F}_{M}\right)\right]+\frac{1}{4}\left[1-\prod_{j=1}^{M-1}\mathcal{L}\left(\Theta\mid\mathcal{F}_{j}\right)\right]\left[1-\mathcal{L}\left(\Theta\mid\mathcal{F}_{M}\right)\right]\nonumber \\
 & = & \frac{1}{4}\left[1+\prod_{j=1}^{M}\mathcal{L}\left(\Theta\mid\mathcal{F}_{j}\right)\right],\label{eq:indc1-1}
\end{eqnarray}
where in $(a)$ we use the fact that $M$ is even and $\oplus$ is
the sum mod(2), in $(b)$ we take advantage of $M-1$ being odd and
in $(c)$ we use the inductive hypothesis. Now, since we show (\ref{eq:optmean1})
to be true for $M=1,2$ the induction in (\ref{eq:indc1}) shows (\ref{eq:optmean1})
for $M=3$, then (\ref{eq:optmean1}) is true for $M=1,2,3$ and (\ref{eq:indc1-1})
shows (\ref{eq:optmean1}) for $M=4$, and then we continue alternating
(\ref{eq:indc1}) with (\ref{eq:indc1-1}) to show by induction (\ref{eq:optmean1})
for any natural $M$. From (\ref{eq:optmean1}) follows that:
\begin{equation}
P\left(\bigoplus_{j=1}^{M}\left[a_{j}\oplus1\right]=0\mid\Theta\right)=\frac{1}{2}\left[1+\left[\mathcal{G}_{\mathcal{L}}\left(\Theta\mid\mathfrak{C}_{\mathcal{F}}^{(x)}\right)\right]^{M}\right],
\end{equation}
which shows the operational interpretation in terms of correlations
$\bigoplus_{j=1}^{M}\left[a_{j}\oplus1\right]=0$ induced by $\Theta$. \qed

\subsection{Proof of operational interpretation of robustness base monotone}\label{OperTask2}
Here we show the operational meaning of the robustness based quantifier in terms of the discrimination of an ensemble of quantum combs \cite{Chiribella2008}. We start proving a more general statement than Theorem 2, by considering
an arbitrary collection of convex free sets $\left\{ \mathcal{F}_{1},...,\mathcal{F}_{M}\right\} $,
with an equally arbitrary, but ordered, set of resources $\left\{ \Theta_{1},...,\Theta_{M}\right\} $, and then show that:
\begin{equation}
\max_{\left\{ p_{j},\Xi_{j}\right\} }\frac{P_{\textrm{succ}}\left(\left\{ p_{j},\Xi_{j}\right\} ,\Theta_{1},...,\Theta_{M}\right)}{\max_{\left\{ \Phi_{k}\in\mathcal{F}_{k}\right\} _{k=1}^{M}}P_{\textrm{succ}}\left(\left\{ p_{j},\Xi_{j}\right\} ,\Phi_{1},\ldots,\Phi_{M}\right)}=\prod_{k=1}^{M}\left(1+\mathcal{R}\left(\Theta_{k}\mid\mathcal{F}_k\right)\right),\label{eq:robproductdisc}
\end{equation}
where $\left\{ p_{j},\Xi_{j}\right\} _{j=1}^{N}$is an ensemble of a quantum comb's evolution branches
 $\Xi_{j}:\textrm{CPTP}^{M}\longrightarrow q_{j}$, for
all $j$ with a probability $q_{j}\in\left[0,1\right]$, $\sum_{j}q_{j}=1$, as described in subsection \ref{robmonotenesection}.
$P_{\textrm{succ}}\left(\left\{ p_{j},\Xi_{j}\right\} ,\left(\cdot\right),\ldots,\left(\cdot\right)\right)$
is the success probability in the discrimination of the branches
by means of the corresponding input channels and $\mathcal{R}\left(\Theta_{k}\mid\mathcal{F}_k\right)$
is the generalised robustness of the channel $\Theta_{k}$ with respect
to the free set $\mathcal{F}_{k}$. 

\emph{Proof.} From the definition of generalised robustness we know
that for every $\mathcal{F}_{k}$ there exists a $\Phi_{k}^{*}\in\mathcal{F}_{k}$
such that: 
\begin{equation}
\left(1+\mathcal{R}\left(\Theta_{k}\mid\mathcal{F}_k\right)\right)\Phi_{k}^{*}\succeq\Theta,\label{eq:robup}
\end{equation}
where $\succeq$ means majorization in the semidefinite positive sense.
Then by definition, 
\begin{eqnarray*}
P_{\textrm{succ}}\left(\left\{ p_{j},\Xi_{j}\right\} ,\Theta_{1},...,\Theta_{M}\right) & = & \sum_{j}p_{j}\Xi_{j}\left[\Theta_{1},...,\Theta_{M}\right]\\
 & \overset{(a)}{\leq} & \prod_{k=1}^{M}\left(1+\mathcal{R}\left(\Theta_{k}\mid\mathcal{F}_k\right)\right)\sum_{j}p_{j}\Xi_{j}\left[\Phi_{1}^{*},\ldots,\Phi_{M}^{*}\right]\\
 & \leq & \prod_{k=1}^{M}\left(1+\mathcal{R}\left(\Theta_{k}\mid\mathcal{F}_k\right)\right)\max_{\left\{ \Phi_{k}\in\mathcal{F}_{k}\right\} _{k=1}^{M}}P_{\textrm{succ}}\left(\left\{ p_{j},\Xi_{j}\right\} ,\Phi_{1},\ldots,\Phi_{M}\right),
\end{eqnarray*}
where in $(a)$ we used the linearity of quantum combs \cite{Chiribella2008,Chiribella2009} and (\ref{eq:robup}).
For the lower bound we remember that every $\mathcal{R}\left(\Theta_{k}\mid\mathcal{F}_k\right)$
is the result of an semidefinite program (SDP):
\begin{eqnarray*}
\max & \,\textrm{Tr}\left[Y_{k}J_{\Theta_{k}}\right]-1,\\
\textrm{subject to:} & Y_{k}\succeq0\\
 & \,\textrm{Tr}\left[Y_{k}J_{\Phi}\right]\leq1 & \forall J_{\Phi}\in\mathbb{O}_{\mathcal{F}_{k}}
\end{eqnarray*}
where $J_{\Theta_{k}}$ is the Jamiołkowski state of $\Theta_{k}$ and $\mathbb{O}_{\mathcal{F}_{k}}$
is the set of all Jamiołkowski states with a $\Phi\in\mathcal{F}_{k}$.
Then, we construct an ensemble with $p_{1}^{*}=1$ and $p_{j}^{*}=0$
for all $j>1$ with $\Xi_{1}$ is given by:
\begin{equation}
\Xi_{1}^{*}\left[\Lambda_{1},\ldots,\Lambda_{M}\right]=\prod_{k=1}^{M}\textrm{Tr}\left[\mathbb{I}\otimes\Lambda_{k}\left(\left|\Psi_{+}\right\rangle \!\left\langle \Psi_{+}\right|\right)\frac{Y_{k}}{\left\Vert Y_{k}\right\Vert _{\infty}}\right],\label{eq:combsp}
\end{equation}
where each $Y_{k}$ is a solution of the previous SDP program associated
to $\mathcal{R}\left(\Theta_{k}\mid\mathcal{F}_k\right)$ and $\left\Vert \cdot\right\Vert _{\infty}$
is the operator norm. The other $\Xi_{j}^{*}$ for $j>1$ are all
$2^{M}-1$ combs resulting from replacing the effects $\frac{Y_{k}}{\left\Vert Y_{k}\right\Vert _{\infty}}$
for $I-\frac{Y_{k}}{\left\Vert Y_{k}\right\Vert _{\infty}}$ in (\ref{eq:combsp})
in all possible combinations, representing all possible outputs from
measurements $\left\{ \frac{Y_{k}}{\left\Vert Y_{k}\right\Vert _{\infty}},I-\frac{Y_{k}}{\left\Vert Y_{k}\right\Vert _{\infty}}\right\} $.
Now, by direct computation we have:
\begin{eqnarray*}
\frac{P_{\textrm{succ}}\left(\left\{ p_{j}^{*},\Xi_{j}^{*}\right\} ,\Theta_{1},\ldots,\Theta_{M}\right)}{\max_{\left\{ \Phi_{k}\in\mathcal{F}_{k}\right\} _{k=1}^{M}}P_{\textrm{succ}}\left(\left\{ p_{j}^{*},\Xi_{j}^{*}\right\} ,\Phi_{1},\ldots,\Phi_{M}\right)} & = & \frac{\prod_{k=1}^{M}\textrm{Tr}\left[\mathbb{I}\otimes\Theta_{k}\left(\left|\Psi_{+}\right\rangle \!\left\langle \Psi_{+}\right|\right)\frac{Y_{k}}{\left\Vert Y_{k}\right\Vert _{\infty}}\right]}{\max_{\left\{ \Phi_{k}\in\mathcal{F}_{k}\right\} _{k=1}^{M}}\prod_{k=1}^{M}\textrm{Tr}\left[\mathbb{I}\otimes\Phi_{k}\left(\left|\Psi_{+}\right\rangle \!\left\langle \Psi_{+}\right|\right)\frac{Y_{k}}{\left\Vert Y_{k}\right\Vert _{\infty}}\right]}\\
 & \overset{(a)}{=} & \frac{\prod_{k=1}^{M}\textrm{Tr}\left[Y_{k}J_{\Theta_{k}}\right]}{\max_{\left\{ \Phi_{k}\in\mathcal{F}_{k}\right\} _{k=1}^{M}}\prod_{k=1}^{M}\textrm{Tr}\left[Y_{k}J_{\Phi}\right]}\\
 & \overset{(b)}{\geq} & \prod_{k=1}^{M}\left(1+\mathcal{R}\left(\Theta_{k}\mid\mathcal{F}_k\right)\right),
\end{eqnarray*}
where in $(a)$ we cancel the products of $\left\Vert Y_{k}\right\Vert _{\infty}$
and replaced by the definitions of Jamiołkowski state, while finally
in $(b)$ the inequality follows from the conditions of the SDP programs
defining every $\mathcal{R}\left(\Theta_{k}\mid\mathcal{F}_k\right)$. \qed

Then, in the above result, we consider the specific case in which
all resources are the same, i.e. $\Theta_{1}=\ldots=\Theta_{M}=\Theta$
belongs to the domain $\mathcal{D}$ including the free set sections $\left\{ \mathcal{F}_{1},...,\mathcal{F}_{M}\right\} $.
In this case, by replacing the arbitrary collection with the free
sections in the domain, we obtain Theorem 2 from the main text.

Now, as explained in the main text we use as utility function $\mathbf{u}$
for a coalition $T$ of agents as the maximum relative advantage they
obtain in quantum comb discrimination tasks by using $\Theta$. From
(\ref{eq:robproductdisc}) it follows directly the following relation
between $\mathbf{u}$ and the robustness measures:
\begin{equation}
\mathbf{u}\left(\{\Theta_{k}\}_{k\in T}\mid T\right)=\prod_{k\in T}\left(1+\mathcal{R}\left(\Theta_{k}\mid\mathcal{F}_k\right)\right)-1.\label{eq:utility}
\end{equation}
Thus, the dividend $\mathbf{d}_{\mathbf{u}}\left(\{\Theta_{k}\}_{k\in S}\mid S\right)$
determines the surplus that is created by a coalition $S$ of agents
when using the respective resources $\{\Theta_{k}\}_{k\in S}$ in a task with utility $\mathbf{u}$.
In cooperative game theory~\cite{harsanyi2010} the surplus is specified as the difference
between the the coalition's utility and the surplus created by all
potential subcoalitions:
\begin{equation}
\mathbf{d}_{\mathbf{u}}\left(\{\Theta_{k}\}_{k\in S}\mid S\right)=\mathbf{u}\left(\{\Theta_{k}\}_{k\in S}\mid S\right)-\sum_{T\subset S}\mathbf{d}_{\mathbf{u}}\left(\{\Theta_{k}\}_{k\in T}\mid T\right),
\end{equation}
and $\mathbf{d}_{\mathbf{u}}\left(\Theta_i\mid\left\{ i\right\} \right)=\mathbf{u}\left(\Theta_i\mid\left\{ i\right\} \right)$
for a single agent $\left\{ i\right\} $. Applying the previous recursive
definition, we will show that for a coalition of the agents $\left\{ A_{k}\right\} _{k=1}^{M}$
 respectively holding resources $\left\{ \Theta_{1},...,\Theta_{M}\right\}$,
their dividend in a quantum comb discrimination is:
\begin{equation}
\mathbf{d}_{\mathbf{u}}\left(\left\{ \Theta_{k}\right\} _{k=1}^{M}\mid\left\{ A_{k}\right\} _{k=1}^{M}\right)=\prod_{k=1}^{M}\mathcal{R}\left(\Theta_{k}\mid\mathcal{F}_k\right).\label{eq:product}
\end{equation}
Since $M=1$ is trivial, we begin the induction with the case $M=2$:
\begin{eqnarray}
\mathbf{d}_{\mathbf{u}}\left(\left\{ \Theta_{k}\right\} _{k=1}^{2}\mid\left\{ A_{k}\right\} _{k=1}^{2}\right) & = & \mathbf{u}\left(\left\{ \Theta_{k}\right\} _{k=1}^{2}\mid\left\{ A_{k}\right\} _{k=1}^{2}\right)-\mathbf{d}_{\mathbf{u}}\left(\Theta_1\mid\left\{ A_{1}\right\} \right)-\mathbf{d}_{\mathbf{u}}\left(\Theta_2\mid\left\{ A_{2}\right\} \right)\nonumber \\
 & \overset{(a)}{=} & \mathbf{u}\left(\left\{ \Theta_{k}\right\} _{k=1}^{2}\mid\left\{ A_{k}\right\} _{k=1}^{2}\right)-\mathbf{u}\left(\Theta_1\mid\left\{ A_{1}\right\} \right)-\mathbf{u}\left(\Theta_2\mid\left\{ A_{2}\right\} \right)\nonumber \\
 & \overset{(b)}{=} & \prod_{k=1}^{2}\left(1+\mathcal{R}\left(\Theta_k\mid\mathcal{F}_k\right)\right)-\sum_{k=1}^{2}\left(1+\mathcal{R}\left(\Theta_k\mid\mathcal{F}_k\right)\right)+1\nonumber \\
 & = & \prod_{k=1}^{2}\mathcal{R}\left(\Theta_k\mid\mathcal{F}_k\right),\label{eq:step1ind}
\end{eqnarray}
where in $(a)$ we use the fact that dividend for single parties is
equal to their utility and in $(b)$ the relation (\ref{eq:utility}).
Now, we will assume as induction hypothesis that relation (\ref{eq:product})
is valid for all subcoalitions $T$ such that $\left|T\right|<M$.
Now from the definition we have: 
\begin{eqnarray}
\mathbf{d}_{\mathbf{u}}\left(\left\{ \Theta_{k}\right\} _{k=1}^{M}\mid\left\{ A_{k}\right\} _{k=1}^{M}\right) & = & \mathbf{u}\left(\left\{ \Theta_{k}\right\} _{k=1}^{M}\mid\left\{ A_{k}\right\} _{k=1}^{M}\right)-\sum_{T\subset\left\{ A_{k}\right\} _{k=1}^{M}}\mathbf{d}_{\mathbf{u}}\left(\left\{ \Theta_{k}\right\} _{k\in T}\mid T\right)\nonumber \\
 & \overset{(a)}{=} & \prod_{k=1}^{M}\left(1+\mathcal{R}\left(\Theta_k\mid\mathcal{F}_k\right)\right)-1-\sum_{T\subset\left\{ A_{l}\right\} _{l=1}^{M}}\mathbf{d}_{\mathbf{u}}\left(\left\{ \Theta_{k}\right\} _{k\in T}\mid T\right)\nonumber \\
 & \overset{(b)}{=} & \sum_{T\subseteq\left\{ A_{l}\right\} _{l=1}^{M}}\prod_{A_{k}\in T}\mathcal{R}\left(\Theta_k\mid\mathcal{F}_k\right)-\sum_{T\subset\left\{ A_{k}\right\} _{k=1}^{M}}\prod_{A_{k}\in T}\mathcal{R}\left(\Theta_k\mid\mathcal{F}_k\right)\nonumber \\
 & \overset{(c)}{=} & \prod_{A_{k}\in\left\{ A_{l}\right\} _{l=1}^{M}}\mathcal{R}\left(\Theta_k\mid\mathcal{F}_k\right)\nonumber \\
 & = & \prod_{k=1}^{M}\mathcal{R}\left(\Theta_k\mid\mathcal{F}_k\right),\label{eq:step2ind}
\end{eqnarray}
where in $(a)$ we use relation (\ref{eq:utility}), in $(b)$ we
expand the first term and used the induction hypothesis on the second
term. Finally, in $(c)$ we note that the only term which doesn't
cancel out in the previous step is $T=\left\{ A_{l}\right\} _{l=1}^{M}$
on the first sum. Since, the hypothesis is true for $M=1,2$ we can
use (\ref{eq:step2ind}) to show the hypothesis is true if it so for
all subcoalitions $T$ such that $\left|T\right|<M$, the result follows
by induction. \qed  

Analogously to the case (\ref{eq:robproductdisc}), if all resources are the same, and equal to $\Theta \in \mathcal{D}$, again by replacing the arbitrary collection with the free
sections in the domain $\mathcal{D}$, we recover Theorem 3 from the main text.  Remarkably, the particular case in which the resource used at every slot of the comb is the same, but the reference collection of convex free sets is arbitrary allows us to apply (\ref{eq:robproductdisc}) in the multi-resource \cite{Sparaciari2020} context, relevant for quantum thermodynamics  \cite{Guryanova2015} or even in the most general non-convex scenario described in \cite{adesso2023,Kuroiwa2023}.

Moreover, the general results (\ref{eq:robproductdisc}) and (\ref{eq:step2ind}) include another notable particular case: when we leave the set of resources $\left\{ \Theta_{1},...,\Theta_{M}\right\} $ arbitrary, and all convex free sets reduce to one, i.e., $ \mathcal{F}_{1}=...=\mathcal{F}_{M}=\mathcal{F} $, which allows us to cover the multi-object scenarios \cite{Brunner2021,DucuaraMO2020,salazar2022}. Among these scenarios, the similarity between the lower bound for the success probability of Shor's Algorithm obtained in \cite{Ahnefeld2022}, and the advantage quantified by the right-hand side of (\ref{eq:robproductdisc}) is striking.

Lastly, as mentioned in the main text, we could use the Harsanyi dividend to compute the Shapley value $\varphi_{i}\left(\mathbf{u}\right)$
for each player $\left\{ i\right\}$, \cite{peleg2003,chakravarty2015}:
\begin{equation}
\varphi_{i}\left(\mathbf{u}\right)=\sum_{T^{(i)}\subseteq S}\frac{\mathbf{d}_{\mathbf{u}}\left(\left\{ \Theta_{k}\right\} _{k\in T^{(i)}}\mid T^{(i)}\right)}{\left|T^{(i)}\right|},
\end{equation}
where $T^{(i)}$ is any coalition including the player $\left\{ i\right\} $.
For the comb discrimination games studied here, this reads: 
\begin{equation}
\varphi_{i}\left(\mathbf{u}\right)=\sum_{T^{(i)}\subseteq S}\frac{1}{\left|T^{(i)}\right|}\prod_{k\in T^{(i)}}\mathcal{R}(\Theta_{k}\mid\mathcal{F}_{k}).\label{eq:solutionconcept}
\end{equation}
By definition $i\in T^{(i)}$, and consequently every term of (\ref{eq:solutionconcept})
includes the factor $\mathcal{R}(\Theta_{i}\mid\mathcal{F}_{i})$,
following that the Shapley value $\varphi_{i}\left(\mathbf{u}\right)\propto\mathcal{R}(\Theta_{i}\mid\mathcal{F}_{i})$,
as it would be expected. Our results on the Harsanyi dividend translate
directly to the study of players' individual profits. This connection
can facilitate the analysis of more complex scenarios of competing
coalitions in comb discrimination games, which we will study in future
research.

 \section{Quantum discord }\label{Qdiscord}
Here, to demonstrate Theorem 6 we consider the set $\Omega$ of two-qubit
states: 
\begin{equation}
\rho\left(\mathbf{x},\mathbf{y},\mathbf{t}\right)=\frac{1}{4}\left\{ I\otimes I+\sum_{i}x_{i}\sigma_{i}\otimes I+\sum_{i}y_{i}I\otimes\sigma_{i}+\sum_{i}t_{i}\sigma_{i}\otimes\sigma_{i}\right\} .\label{eq:second case}
\end{equation}
Then, following \cite{Borivoje2010} the zero-discord states (\ref{eq:zero-discord})
(i.e. on Alice side) of form (\ref{eq:second case}) must satisfy
the condition: 
\begin{equation}
\left\Vert \mathbf{x}\right\Vert ^{2}+\left\Vert \mathbf{t}\right\Vert ^{2}-\kappa_{\max}=0,\label{eq:zerodiscordAlice}
\end{equation}
where $\kappa_{\max}$ is the maximum eigenvalue of a matrix $K$
with entries $K_{ij}=x_{i}x_{j}+t_{i}t_{j}\delta_{ij}$. To determine
the set of states with zero-discord we proceed to compute the characteristic
polynomial of $K$:
\begin{equation}
q_{K}\left(\kappa\right)=\kappa^{3}-\kappa^{2}\left(\left\Vert \mathbf{x}\right\Vert ^{2}+\left\Vert \mathbf{t}\right\Vert ^{2}\right)+\kappa\left(\sum_{i\neq j}t_{i}^{2}t_{j}^{2}+\sum_{\left(i,j,k\right)\in\pi_{C}}t_{i}^{2}\left(x_{j}^{2}+x_{k}^{2}\right)\right)-\left(t_{1}^{2}t_{2}^{2}t_{3}^{2}+\sum_{\left(i,j,k\right)\in\pi_{C}}x_{i}^{2}t_{j}^{2}t_{k}^{2}\right),\label{eq:polK}
\end{equation}
where $\pi_{C}$ stand for the cyclic permutations of $(1,2,3)$.
Applying the Descartes' rule of signs we see that $q_{K}\left(\kappa\right)$
has no negative roots and 3 or 1 positive roots whenever $\left\Vert \mathbf{x}\right\Vert ^{2}+\left\Vert \mathbf{t}\right\Vert ^{2}>0$.
From the above it turns out that $\kappa_{\max}>0$ exist for $\left\Vert \mathbf{x}\right\Vert ^{2}+\left\Vert \mathbf{t}\right\Vert ^{2}>0$,
while for $\left\Vert \mathbf{x}\right\Vert ^{2}+\left\Vert \mathbf{t}\right\Vert ^{2}=0$
the result $\kappa_{\max}=0$ is trivial. Now, for zero-discord states,
from (\ref{eq:zerodiscordAlice}) we have $\kappa_{\max}=\left\Vert \mathbf{x}\right\Vert ^{2}+\left\Vert \mathbf{t}\right\Vert ^{2}$
and replacing this condition on $q_{K}\left(\kappa_{\max}\right)=0$
we have:
\begin{equation}
\left(\left\Vert \mathbf{x}\right\Vert ^{2}+\left\Vert \mathbf{t}\right\Vert ^{2}\right)\left(\sum_{i\neq j}t_{i}^{2}t_{j}^{2}+\sum_{\left(i,j,k\right)\in\pi_{C}}t_{i}^{2}\left(x_{j}^{2}+x_{k}^{2}\right)\right)-\left(t_{1}^{2}t_{2}^{2}t_{3}^{2}+\sum_{\left(i,j,k\right)\in\pi_{C}}x_{i}^{2}t_{j}^{2}t_{k}^{2}\right)=0,
\end{equation}
which after some algebraic transformations reads:
\begin{equation}
2t_{1}^{2}t_{2}^{2}t_{3}^{2}+\sum_{i\neq j}x_{i}^{2}t_{i}^{2}t_{j}^{2}+\sum_{\left(i,j,k\right)\in\pi_{C}}t_{i}^{4}t_{j}^{2}t_{k}^{2}+\left(\left\Vert \mathbf{x}\right\Vert ^{2}+\left\Vert \mathbf{t}\right\Vert ^{2}\right)\sum_{\left(i,j,k\right)\in\pi_{C}}t_{i}^{2}\left(x_{j}^{2}+x_{k}^{2}\right)=0.\label{eq:condD0}
\end{equation}
Since every term on the right hand side of (\ref{eq:condD0}) is non
negative, it must be that all of them are zero, from that follows
that only one of the coordinate pairs $\left\{ \left(t_{i},x_{i}\right)\right\} _{i=1}^{3}$
could be non zero. Concluding the demonstration of Theorem 6. \qed

To demonstrate Theorem 7 consider $\mathbf{x}=\left(x_{1},x_{2},x_{3}\right)$ and $\mathbf{y}=\left(y_{1},y_{2},y_{3}\right)$
with the additional constraint that $\mathbf{y}$ must have the same
support as $\mathbf{x}$, i.e. if a coordinate $x_{i}$ is zero this
implies that a coordinate $y_{i}$ is also zero, but the converse
is not necessarily true. Since $\mathbf{x}$ and $\mathbf{y}$
have the same support the free set considers only states $\chi\left(\mathbf{x},\mathbf{y},\mathbf{t}\right)$
with two triples $\left\{ x_{i},y_{i},t_{i}\right\} $
out of the three labeled by $i\in\left\{ 1,2,3\right\} $ necessarily equal to  $\left\{ 0,0,0\right\}$. Then, from the
Bloch matrix representation, the conditions for $\chi\left(\mathbf{x},\mathbf{y},\mathbf{t}\right)\in\Omega_{0}$
to be a trace one positive matrix are (See section IV.D in \cite{OmarGamel2016}): 
\begin{eqnarray}
3-\left(x^{2}+y^{2}+t^{2}\right) & \geq & 0,\label{eq:cond1}\\
1+2xty-\left(x^{2}+y^{2}+t^{2}\right) & \geq & 0,\label{eq:cond2}\\
8xty+\left(x^{2}+y^{2}+t^{2}-1\right)^{2}-4y^{2}\left(x^{2}+t^{2}\right)-4x^{2}t^{2} & \geq & 0,\label{eq:cond3}
\end{eqnarray}
where $x,y,t$ stand for the single triple $\left\{ x_{i},y_{i},t_{i}\right\} $
which of potentially non zero terms in $\mathbf{x},\mathbf{y},\mathbf{t}$,
because $\chi\left(\mathbf{x},\mathbf{y},\mathbf{t}\right)\in\Omega_{0}$.
Now, for every fixed value of $y$ we  describe the set in coordinates
$x,t$. The first condition (\ref{eq:cond1}) requires each point $\left(x,t\right)$
to lie inside a circle of radius $3-y^{2}$, while to see the geometric
requirement of the second condition (\ref{eq:cond2}) we rotate coordinates
by angle of $\pi/4$:
\begin{equation*} 
x  =  \frac{u+v}{\sqrt{2}} \quad
t  =  \frac{u-v}{\sqrt{2}}.
\end{equation*}
After such a replacement  condition (\ref{eq:cond2}) turns into: 
\begin{equation*}
\frac{u^{2}}{1+y}+\frac{v^{2}}{1-y}\leq1,
\end{equation*}
which determines an ellipse with principal axes $a=\sqrt{1+y}$ and
$b=\sqrt{1-y}$ which on coordinates $x,t$ is always inside the unit
circle, because $\left|y\right|\leq1$. For the third condition (\ref{eq:cond3})
we need to use some algebra:
\begin{eqnarray}
8xty+\left(x^{2}+y^{2}+t^{2}-1\right)^{2}-4y^{2}\left(x^{2}+t^{2}\right)-4x^{2}t^{2} & \geq & 0\nonumber \\
8xty+\left(x^{2}+t^{2}-R^{2}\right)^{2}-4y^{2}\left(x^{2}+t^{2}\right)-4x^{2}t^{2} & \overset{(a)}{\geq} & 0\nonumber \\
4y\left(u^{2}-v^{2}\right)+\left(u^{2}+v^{2}-R^{2}\right)^{2}-4y^{2}\left(u^{2}+v^{2}\right)-\left(u^{2}-v^{2}\right)^{2} & \overset{(b)}{\geq} & 0\nonumber \\
4yL+\left(K-R^{2}\right)^{2}-4y^{2}K-L^{2} & \overset{(c)}{\geq} & 0\nonumber \\
K^{2}-L^{2}-2\left(R^{2}+2y^{2}\right)K+4yL+R^{4} & \geq & 0\nonumber \\
\left(K-h\right)^{2}-\left(L-g\right)^{2}+R^{4}+\left(g^{2}-h^{2}\right) & \geq & 0\nonumber \\
\left(K-h\right)^{2}-\left(L-g\right)^{2}+2y^{2}\left(1-y^{2}-R^{2}\right) & \geq & 0\nonumber \\
\left(K-h\right)^{2}-\left(L-g\right)^{2} & \geq & 0\nonumber \\
\left(K-h\right)^{2} & \geq & \left(L-g\right)^{2}\nonumber \\
\left|K-h\right| & \geq & \left|L-g\right|,\label{eq:cond33}
\end{eqnarray}
where in $(a)$ we define $R=\sqrt{1-y^{2}}$, in $(b)$ we use the
same change of coordinates to $u,v$ as for (\ref{eq:cond2}) and
in $(c)$ we define the auxiliary variables:
\begin{eqnarray*}
K & = & u^{2}+v^{2},\\
L & = & u^{2}-v^{2},\\
h & = & R^{2}+2y^{2},\\
g & = & 2y.
\end{eqnarray*}
But, $K=u^{2}+v^{2}=x^{2}+t^{2}\leq1\leq1+y^{2}=R^{2}+2y^{2}=h$,
where we use that $\left(x,t\right)$ is inside the unit circle due
to (\ref{eq:cond2}). Then, inequality (\ref{eq:cond33}) leads to only to two possibilities i) $h-K\geq L-g$ or ii) $h-K\geq g-L$.
If $L\geq g$ we have:
\begin{eqnarray}
L-g & \leq & h-K\nonumber \\
L+K & \leq & h+g\nonumber \\
2u^{2} & \leq & 1+y^{2}+2y\nonumber \\
2u^{2} & \leq & \left(1+y\right)^{2}\nonumber \\
\left|u\right| & \leq & \left(1+y\right)/\sqrt{2},\label{eq:cond33i}
\end{eqnarray}
and if $L\leq g$ we have: 
\begin{eqnarray}
g-L & \leq & h-K\nonumber \\
K-L & \leq & h-g\nonumber \\
2v^{2} & \leq & 1+y^{2}-2y\nonumber \\
2v^{2} & \leq & \left(1-y\right)^{2}\nonumber \\
\left|v\right| & \leq & \left(1-y\right)/\sqrt{2}.\label{eq:cond33ii}
\end{eqnarray}
Thus, conditions (\ref{eq:cond33i}) and (\ref{eq:cond33ii}) imply
that $\left(x,t\right)$ is inside a rectangle of sides $2\left(1-y\right)/\sqrt{2}$
and $2\left(1+y\right)/\sqrt{2}$. On the $u,v$ coordinates we can
see that such rectangle is inscribed in the ellipse of (\ref{eq:cond2}), thus concluding the proof.\qed

Additionally, we will describe the star domain structure of Alice's zero-discord states.
When the set of states is of the form (\ref{eq:second case}), their description correspond to a six dimensional
space $\Omega\subset\mathbb{R}^{6}$, defined by positivity conditions
\cite{OmarGamel2016}. The natural orthonormal basis of $\Omega$
are the unit vectors $\left\{ \hat{\mathbf{x}}_{i},\hat{\mathbf{t}}_{i}\right\} _{i=1}^{3}$, however we will use a rotated basis of vectors $\left\{ \hat{\mathbf{u}}_{i},\hat{\mathbf{v}}_{i}\right\} _{i=1}^{3}$:
\begin{equation*}
\hat{\mathbf{u}}_{i}  =  \frac{\hat{\mathbf{x}}_{i}+\hat{\mathbf{t}}_{i}}{\sqrt{2}}\quad
\hat{\mathbf{v}}_{i}  =  \frac{\hat{\mathbf{x}}_{i}-\hat{\mathbf{t}}_{i}}{\sqrt{2}}.
\end{equation*}
The reason to choose the basis $\left\{ \hat{\mathbf{u}}_{i},\hat{\mathbf{v}}_{i}\right\} _{i=1}^{3}$
is that it aligns the unit vectors with the symmetry axis of the rectangles
forming the the free set $\Omega_{0}$.
Indeed,  the previous analysis implies that 
the free set is given by: 
\begin{equation}
\Omega_{0}:\left\{ \mathbf{r}\mid\mathbf{r}=\alpha\hat{\mathbf{u}}_{i}+\beta\hat{\mathbf{v}}_{i},\left|\alpha\right|\leq\left(1+y_{i}\right)/\sqrt{2},\left|\beta\right|\leq\left(1-y_{i}\right)/\sqrt{2}\textrm{ for some }i\in\left\{ 1,2,3\right\} \right\}, \label{eq:freediscdef}
\end{equation}
where  $\mathrm{Ker}\left(\Omega_{0}\right)=\left\{\mathbf{0}\right\}$ representing $I\otimes I/4$, and there exist only three rectangular convex components $\Omega_{0,i}$
described by an individual $i\in\left\{ 1,2,3\right\} $ in definition (\ref{eq:freediscdef}).

\section{Total correlations }
\label{app:total_correlations}

\subsection{Hyperbolic contraction operations for the total correlations restricted free set} \label{app:total_correlations_HCO}

Here we provide a survey of  hyperbolic contraction operations for the theory of total correlations. The frames $\mathcal{T}^{(ij)}$ as described in \eqref{frameTotCorr} contain vectors     $\bold{l}^{ab}_{ij} \equiv \delta^{ab}_{ij} - \eta$, to which we associate the constants $\lambda_{rq}^{(ij)}$ defining the action of the operations as in \eqref{eq:HCO}:
    \begin{equation}
        \eta + \alpha_{ij} \bold{l}^{ab}_{ij} + \sum_{k\neq i,\,j\neq l}\qty(\alpha_{kj}\bold{l}^{ab}_{kj} + \alpha_{jl}\bold{l}^{ab}_{jl})\mapsto
        \eta + \alpha_{ij} \lambda_{ij}^{(ij)} \bold{l}^{ab}_{ij} + \sum_{k\neq i,\,j\neq l}\qty(\alpha_{kj} \lambda_{kj}^{(ij)} \bold{l}^{ab}_{kj} + \alpha_{jl} \lambda_{il}^{(ij)} \bold{l}^{ab}_{il}).
    \end{equation}
    When we consider cone order-preserving operations, i.e. all $\lambda_{rq}^{(ij)}\leq 1$, the physical interpretation of the operation is a stochastic process over $p_C$ taking it closer to distributions in $\mathcal{F}'$. Notably, in general cone order-preserving operations are also RNG operations of the star resource theory based on $\mathcal{F}$. Concretely, we can write the uncorrelated probability distributions in the following
way:

\begin{eqnarray*}
\left(1-\epsilon\right)p_{A}(a)p_{B}(b)+\epsilon\eta & = & \eta+\left(1-\epsilon\right)\left(p_{A}(a)p_{B}(b)-\eta\right)\\
 & = & \eta+\left(1-\epsilon\right)\left(\sum_{k,l}p_{A}^{k}p_{B}^{l}\delta_{kl}^{ab}-\eta\right)\\
 & \overset{(t)}{=} & \eta+\left(1-\epsilon\right)\sum_{k,l}p_{A}^{k}p_{B}^{l}\left[\delta_{kl}^{ab}-\eta\right]\\
 & = & \eta+\sum_{k,l}\left(1-\epsilon\right)p_{A}^{k}p_{B}^{l}\left[\delta_{kl}^{ab}-\eta\right],
\end{eqnarray*}
where in $(t)$ we used the normalization of distributions $p_{A}^{k}$
and $p_{B}^{l}$. Consequently it follows the condition $\alpha_{kl}=(1-\epsilon)p_{A}^{k}p_{B}^{l}$
for the conic span. Thus, after a conic non-increasing operation with
parameters $\lambda_{kl}$ we have: 
\begin{equation}
\alpha_{kl}^{\prime}=(1-\epsilon)\lambda_{kl}p_{A}^{k}p_{B}^{l}.
\end{equation}
Then, a sufficient condition for the new distribution to be uncorrelated
is: 
\begin{equation}
\lambda_{kl}=\bar{\lambda}_{k}\tilde{\lambda}_{l},\label{eq:suffcond}
\end{equation}
with $\bar{\lambda}_{k},\tilde{\lambda}_{l}\in[0,1]$ for every $k,l$.
Indeed, if the operation satisfies (\ref{eq:suffcond}) we have:
\begin{eqnarray}
p_{C}^{\prime}(a,b) & = & \eta+\sum_{k,l}\left(1-\epsilon\right)\lambda_{kl}p_{A}^{k}p_{B}^{l}\left[\delta_{kl}^{ab}-\eta\right]\nonumber \\
 & = & \eta+\sum_{k,l}\left(1-\epsilon\right)\bar{\lambda}_{k}\tilde{\lambda}_{l}p_{A}^{k}p_{B}^{l}\left[\delta_{kl}^{ab}-\eta\right]\nonumber \\
 & \overset{(r)}{=} & \eta+\sum_{k,l}\left(1-\epsilon\right)\bar{\lambda}_{A}\tilde{\lambda}_{B}\bar{p}_{A}^{k}\tilde{p}_{B}^{l}\left[\delta_{kl}^{ab}-\eta\right]\nonumber \\
 & \overset{(s)}{=} & \eta+\sum_{k,l}\left(1-\epsilon^{\prime}\right)\bar{p}_{A}^{k}\tilde{p}_{B}^{l}\left[\delta_{kl}^{ab}-\eta\right],\label{eq:productform}
\end{eqnarray}
where in $(r)$ we defined the probability distributions:
\begin{equation*}
\bar{p}_{A}^{k}=\frac{\bar{\lambda}_{k}p_{A}^{k}}{\bar{\lambda}_{A}}\quad\tilde{p}_{B}^{l}=\frac{\tilde{\lambda}_{l}p_{B}^{l}}{\tilde{\lambda}_{B}},
\end{equation*}
with $\bar{\lambda}_{A}$, $\tilde{\lambda}_{B}$ the corresponding
normalization factors. Lastly, in $(s)$ we define $\epsilon^{\prime}$
such that:
\begin{equation}
\left(1-\epsilon^{\prime}\right)=\left(1-\epsilon\right)\bar{\lambda}_{A}\tilde{\lambda}_{B}.
\end{equation}
It is straightforward to note that $\epsilon\leq\epsilon^{\prime}\leq1$,
and then (\ref{eq:productform}) shows that the operation is resource
non-generating. Finally, in the exemplary case $n_{A}=n_{B}=2$, when
$\lambda_{ij}=1$ in $\mathcal{D}^{(ij)}$, condition (\ref{eq:suffcond})
is automatically satisfied by constants $\bar{\lambda}_{i}=\tilde{\lambda}_{j}=1$,
$\bar{\lambda}_{i\oplus1}=\lambda_{i\oplus1,j}$, and $\tilde{\lambda}_{j\oplus1}=\lambda_{i,j\oplus1}$.
Note that $\lambda_{i\oplus1,j\oplus1}$ could take any value because
in $\mathcal{D}^{(ij)}$ the component $\alpha_{ij}$ is always zero. \qed

    Other free operations in this class arise when $n_A=n_B=n$ by taking every pair $\lambda_{ik}^{(ij)}, \lambda_{lj}^{(ij)}$ with $k\neq j, l\neq i$  to define a hyperbolic rotation $\lambda_{ik}^{(ij)}=e^{\varphi}$,  $\lambda_{lj}^{(ij)}=e^{-\varphi}$.  The above is also a squeeze map where we consider the product of all $\lambda_{rq}^{(ij)}$, but with $\lambda_{ij}^{(ij)}=1$. When $n_A\neq n_B$, the squeeze map is not simulable as hyperbolic rotations, nevertheless we still can construct one:  without loss of generality consider $n_A < n_B$, first make all $\lambda_{ik}^{(ij)}$ with $k\neq j$ equal to a constant $\lambda_A \leq 1$, second take all $ \lambda_{lj}^{(ij)}$ equal to $\lambda_B= \lambda_A^{-n_A/n_B} $, and here also $\lambda_{ij}^{(ij)}=1$, obtaining a squeeze map which generalize the previous map. However, while the hyperbolic rotations, and squeeze maps also correspond to specific post-processing of outputs they are not RNG operations for the theory with the full free set $\mathcal{F}$.
    
\subsection{Computing critical value} 
\label{app:criticalcorrelations}

Here, we compute the critical value of the
quantifier based on
$\mathcal{F}'$, beyond which there exist only correlated distributions. To achieve this objective, we parameterize the free set $\mathcal{F}$ writing explicitly any distribution $\v{f}\in\mathcal{F}$ as:

\begin{equation}
    \v{f} = (1-\epsilon)\mqty(xy,&(1-x)y,&x(1-y),&(1-x)(1-y)) + \frac{\epsilon}{4},
\end{equation}
in particular any $\v{f}\in\mathcal{C}_{00}$ satisfies $x,y\geq\frac{1}{2}$. It is also straightforward to project arbitrary $\v{f}$ onto the faces as follows:

\begin{equation}
    \begin{aligned}
        \mathcal{F}'_{A0} \ni \v{f}_A = (1-\epsilon)\mqty(xy,&(1-x)y,&\frac{1-y}{2}, & \frac{1-y}{2}) + \frac{\epsilon}{4}, \\
        \mathcal{F}'_{B0} \ni \v{f}_B = (1-\epsilon)\mqty(xy,&\frac{1-x}{2},&x(1-y),&\frac{1-x}{2})
        + \frac{\epsilon}{4},
    \end{aligned}
\end{equation}
from which it is simple to calculate the product of the 1-norms, serving as a proxy to the monotone

\begin{equation}
    \sqrt{\norm{\v{f} - \v{f}_A}_1\norm{\v{f} - \v{f}_B}_1} = \sqrt{(1-\epsilon)^2\abs{(1-y)\qty(\frac{1}{2}-x)}\abs{(1-x)\qty(\frac{1}{2} - y)}}.
\end{equation}

A typical optimization provides maximum of this function is: $\frac{1}{16}$. This bound yields the maximum  resource achievable by uncorrelated distributions. Thus any larger value of of the quantifier based on $\mathcal{F}'$, 
witnesses total correlations in the evaluated distribution. \qed

\subsection{Extension to network total correlations} \label{app:exttonetworks}

A  straightforward extension of the SRT of total correlations considers an arbitrary number $N$ of parties $A_1,\hdots,A_N$ with an alphabet of $n_{A_i}$ letters for outputs $\bold{a}=(a_1,\hdots,a_N)$ with correlations:

\begin{equation}\label{multipartfree}
    \mathcal{F}\ni p_C(a_1,\hdots,a_N)= \qty{(1-\epsilon)\qty(\prod_{i=1}^N p_{A_i}(a_i)) + \epsilon\qty(\prod_{i=1}^N\frac{1}{n_{A_i}}):p_{A_i}(a_i)\in\Delta_{n_{A_i}}},
\end{equation}
which evidently form a star with the flat distribution at its kernel. Moreover, we can also extend the free sections of the auxiliary free set: 

\begin{equation}\label{multipartfreeAux}
    \mathcal{F}'_{A_i, \bold{k}}\ni p_C(\bold{a})= \qty{(1-\epsilon)p_{A_i}(a_i)\qty(\prod_{j\neq i}\delta_{a_j,k_j}) + \epsilon\qty(\prod_{i=1}^N\frac{1}{n_{A_i}}):p_{A_i}(a_i)\in\Delta_{n_{A_i}}},
\end{equation}
and then considering all possible values for the vector $\bold{k}$. Then, it is possible to carry on the analysis and construction of witnesses in complete analogy as with two parties. 

However, we can apply this simple extension to the important field of network correlations by identifying each party with a node of a weighted graph or hyper-graph and the weight of an edge/hyper-edge with the marginal distribution of $ p_C(a_1,\hdots,a_N)$ including only the parties connected by the edge/hyper-edge. 

The development of resource theories for weighted networks with concrete topologies it is a promising area due to its multiple applications, yet unexplored to date. Crucially, one can apply the techniques described here to any network topology. Since the weights can be continuously reduced regardless of the graph until obtaining a network with independent nodes, such a network must belong to the free set's kernel. Consequently, we can use star domains to characterize their properties and possible transformations for any network topology.

\section{Non-Markovianity }
\subsection{Explicit calculation of the minimization problem in Eq. (\ref{markovian-bound})}\label{markovian-calc}

In this appendix, we provide the necessary proofs to find the witness of non-markovianity  from the main text.

Now, inside the free set $\mathcal{F}$ of Markovian Pauli channels we distinguish the following three free sections:
\begin{equation}
    \mathcal{F}_j: \{\Phi_j\mid \Phi_j=q\Gamma_j^{1/2}+(1-q)\Gamma_d^{1/2} \},
\end{equation}
with $j\in\{1,2,3\}$, $0\leq q \leq 1$, $\Gamma_i^{1/2}(\rho) = \frac{1}{2} \rho + \frac{1}{2}\sigma_i\rho\sigma_i$ and $\Gamma_d^{1/2}(\rho)=\frac{1}{2} \rho + \frac{1}{4}I$ stands for the 1/2 mixture between the identity channel and the fully depolarizing channel. Then, for an arbitrary $\Phi_j\in\mathcal{F}_j$,  the respective Jamiołkowski states representing single--qubit operations read:

\begin{align}\label{choi-free0}
    J_{\Phi_1}=\frac{1}{3}\!\begin{pmatrix}
1+\frac{q}{2} & 0 & 0 & \frac{1-q}{2} \\
0 & \frac{1-q}{2} & 0 & 0\\
0 &  0 & \frac{1-q}{2}  & 0\\
 \frac{1-q}{2} & 0 & 0 &  1+\frac{q}{2}\\
\end{pmatrix}\!, 
J_{\Phi_2}=\frac{1}{3}\!\begin{pmatrix}
1-\frac{q}{4} & 0 & 0 & \frac{2+q}{4} \\
0 & \frac{2+q}{4} & -\frac{q}{4} & 0\\
0 &  -\frac{q}{4} & \frac{2+q}{4}  & 0\\
 \frac{2+q}{4} & 0 & 0 &  1-\frac{q}{4}\\
\end{pmatrix}\!, 
J_{\Phi_3}=\frac{1}{3}\!\begin{pmatrix}
1-\frac{q}{4} & 0 & 0 & \frac{2+q}{4} \\
0 & \frac{2+q}{4} & \frac{q}{4} & 0\\
0 &  \frac{q}{4} & \frac{2+q}{4}  & 0\\
 \frac{2+q}{4} & 0 & 0 &  1-\frac{q}{4}\\
\end{pmatrix}\!.
\end{align}

Employing the above expressions for $J_{\Phi_j}$, the minimization problem (\ref{markovian-bound}) takes the following form:

\begin{equation}
\begin{array}{rrclcl}
\displaystyle \min_{q} & \multicolumn{3}{l}{\frac{1}{12} \Big (\frac{1}{2} \lvert 7-3 \sqrt{5} - 4q \rvert + \lvert -7 +3\sqrt{5} +q \rvert + \frac{1}{2} \lvert 7-3\sqrt{5} +2q \rvert  \Big )}\\
\textrm{s.t.} & 0\leq q \leq 1. \\
\end{array}
\end{equation}

As we are dealing with inequality constraints, we will make use of slack variables $s^2$ and $t^2$ that will convert the problem to the form

\begin{equation}
\begin{array}{rrclcl}
\displaystyle \min_{q} & \multicolumn{3}{l}{\frac{1}{12} \Big (\frac{1}{2} \lvert 7-3 \sqrt{5} - 4q \rvert + \lvert -7 +3\sqrt{5} +q \rvert + \frac{1}{2} \lvert 7-3\sqrt{5} +2q \rvert  \Big )}\\
\textrm{s.t.} & q-s^2=0,\\
& q+t^2-1=0. \\
\end{array}
\end{equation}
Now, we are able to construct the Lagrangian
\begin{align}\label{Lagrangian}
\mathscr{L} &= \frac{1}{12} \Big (\frac{1}{2} \lvert 7-3 \sqrt{5} - 4q \rvert + \lvert -7 +3\sqrt{5} +q \rvert + \frac{1}{2} \lvert 7-3\sqrt{5} +2q \rvert  \Big ) - \mu(q-s^2) + \lambda(q+t^2-1),
\end{align}
and the equations that need to be fulfilled for this Lagrangian are the following
\begin{align}\label{min-equations}
    \frac{\partial \mathscr{L}}{\partial q}&= \mp \frac{1}{6} \pm \frac{1}{12} + \frac{1}{12} + \lambda - \mu =0 \\
    \frac{\partial \mathscr{L}}{\partial \lambda}&=q + t^2 - 1 =0\\
    \frac{\partial \mathscr{L}}{\partial \mu}&= -(q-s^2) =0,
\end{align} \\
where $\mp$ and $\pm$ denote the two possibilities when taking the derivative of each absolute value, the first one when the expression inside the absolute value is positive and the second one when it is negative. 

Now, using KKT conditions (see \cite{karush1939minima, kuhn19511951}) we know that at the optimal solution of this problem, $q^*$, either the Lagrange multiplier corresponding to the inequality constraint is zero or the corresponding inequality constraint is active, i.e. the corresponding slack variable is zero and we have an equality. Therefore there are four possibilities:

  \begin{enumerate}
    \centering
    \item $\mu=0$, $\lambda=0$, $s^2\geq 0$ and $t^2 \geq 0$.
    \item $\mu=0$, $\lambda \neq 0$, $s^2\geq 0$ and $t^2=0$.
    \item $\mu \neq 0$, $\lambda \neq 0$, $s^2 = 0$ and $t^2=0$.
    \item $\mu \neq 0$, $\lambda = 0$, $s^2= 0$ and $t^2 \geq 0$.
\end{enumerate}

If we investigate the equation (\ref{min-equations}), the only way for this equation to be zero the cases in which $\mu=0$ and $\lambda=0$ is when the expression inside the first absolute value is positive and also the one inside the second one. Inspecting at the expressions we see that this is never possible and therefore we must rule out the first possibility. 

Regarding the other three possibilities, we see that the result is the same regardless of the signs in (\ref{min-equations}) because (\ref{min-equations}) does not give us information regarding $q$. So, the second case give us $q^*=1$ and the corresponding objective function, $f=\frac{1}{4}(\sqrt{5}-1)$. The third one is not feasible, because it gives us that $q^*=0$ and also $q^*=1$ at the same time and the last one $q^*=0$ and $f=\frac{1}{6}(7-3\sqrt{5})$.

The remaining cases to try are those in which the expression inside the two first absolute values are equal to 0. This is the case for $q^*=\frac{7-3\sqrt{5}}{4}$ and $f=\frac{1}{8} \Big (7-3\sqrt{5} \Big )$, and $q^*=7-3\sqrt{5}$ and $f=\frac{1}{4} \Big (7-3\sqrt{5} \Big )$ respectively. 

Finally, comparing all the cases we find the mininum equal to $\frac{1}{8} \Big (7-3\sqrt{5} \Big )$, and achieved when $q=\frac{7-3\sqrt{5}}{4}$. \qed

\subsection{Conic non-increasing operations are RNG} \label{ApxConicRNG}
Using the De Casteljau's form of $\varDelta_{ij}\mathcal{F}$ we can
write elements of the section of $\mathcal{F}$ inside $\mathcal{D}^{(k)}$
with $k\neq i$, $k\neq j$ by,
\begin{eqnarray}
\Phi & = & \left(1-\epsilon\right)\left[\left(1-s\right)^{2}\Gamma_{i}^{1/2}\!+\!2s\left(1-s\right)\Gamma_{d}^{1/2}\!+\!s^{2}\Gamma_{j}^{1/2}\right]+\epsilon\Gamma_{d}^{1/2}\nonumber \\
 & = & \Gamma_{d}^{1/2}+\left(1-\epsilon\right)\left[\left(1-s\right)^{2}\Gamma_{i}^{1/2}\!+\!2s\left(1-s\right)\Gamma_{d}^{1/2}\!+\!s^{2}\Gamma_{j}^{1/2}-\Gamma_{d}^{1/2}\right]\nonumber \\
 & \overset{(r)}{=} & \Gamma_{d}^{1/2}+\left(1-\epsilon\right)\left[\left(1-s\right)^{2}\left(\Gamma_{i}^{1/2}-\Gamma_{d}^{1/2}\right)\!+\!s^{2}\left(\Gamma_{j}^{1/2}-\Gamma_{d}^{1/2}\right)\right],\label{eq:coneform}
\end{eqnarray}
where in $(r)$ we use the fact that $\left(1-s\right)^{2}+2s\left(1-s\right)+s^{2}=1$.
Now, from (\ref{eq:coneform}) follows that $\alpha_{i}=\left(1-\epsilon\right)\left(1-s\right)^{2}$
and $\alpha_{j}=\left(1-\epsilon\right)s^{2}$. In consequence, after
a conic non-increasing operation with parameters $\lambda_{i},\lambda_{j}\in[0,1]$
, we have a channel $\Phi^{\prime}$ with conic components: 
\begin{equation}
\alpha_{i}^{\prime}=\lambda_{i}\left(1-\epsilon\right)\left(1-s\right)^{2}\quad\alpha_{j}^{\prime}=\lambda_{j}\left(1-\epsilon\right)s^{2}.\label{eq:transformconicrng}
\end{equation}
Now, we define the parameters: 
\begin{equation}
\bar{s}=\frac{\sqrt{\lambda_{j}}s}{\sqrt{\lambda_{j}}s+\sqrt{\lambda_{i}}\left(1-s\right)}\quad\left(1-\epsilon^{\prime}\right)=\left(1-\epsilon\right)\left[\sqrt{\lambda_{j}}s+\sqrt{\lambda_{i}}\left(1-s\right)\right]^{2}
\end{equation}
Then, replacing $\bar{s}$ and $\epsilon^{\prime}$ in (\ref{eq:transformconicrng})
we get: 
\begin{equation}
\alpha_{i}^{\prime}=\left(1-\epsilon^{\prime}\right)\left(1-\bar{s}\right)^{2}\quad\alpha_{j}^{\prime}=\left(1-\epsilon^{\prime}\right)\bar{s}^{2},
\end{equation}
demonstrating that any conic non-increasing operation is RNG. \qed

 \subsection{Non-Unistochasticity}
 \label{Unisto}

A bistochastic matrix corresponding to 
a well-defined unitary transition matrix
is unistochastic: its entries
are equal to squared moduli of some underlying unitary mixing matrix.
Thus it is important to verify, whether
a given bistochastic matrix satisfies this property. 
In this Section we demonstrate how to obtain the optimal value of the witness of non-unistochasticity, Eq.~(\ref{eq:witness_non-unistochasticity}). 
We analyze the tetrahedron of circulant bistochastic matrices in dimension $N=4$
shown in Fig.  \ref{fig:enter-label}.
Without loss of generality, let us fix the analyzed quadrant to the one defined in Eq.~(\ref{eq:unistochastic_free_set}).
First observation comes from the symmetry of the problem -- the closest non-unistochastic matrix to the central van der Waerden matrix lies on the line $W_4 \leftrightarrow \frac{1}{2}(\Pi_4^0 + \Pi_4)$.

It is straightforward to find the matrix, given via $(1-\frac{1}{\sqrt{2}})W_4 + \frac{1}{\sqrt{2}}(\Pi_4^0+\Pi_4)$, together with the geometric mean of the distances to the triangles that define the quadrant, being $\frac{1}{2\sqrt{2}}$.
What is left is a simple algebraic check that this is indeed optimal; i.e., the hyperboloid with the above parameter has only one common point with the set of the unistochastic matrices, proving its suitability for the task of witnessing non-unistochasticity. \qed

\section{Illustrative examples of the advantages of our framework} \label{app:counterexample}

As noted in subsection \ref{subsec:mon}, we present resource theories that fall outside the scope of earlier approaches but are fully manageable within our framework. Prior works \cite{Schluck_2022,adesso2023} relied solely on generalized robustness, which fails in scenarios where the free set lies entirely on the boundary of the objects set, as with quantum-state texture. Thus, for resource theories where (a) the free set is entirely on the boundary and (b) forms a non-convex star domain, our distance-based monotone provides the only viable methodology to date.

A concrete example derives from our study of total correlations. If we further reduce the auxiliary free set $\mathcal{F}'$  in \ref{eq:auxfree}  to the distributions in the union of $\mathcal{F}'_{A0}$, and $\mathcal{F}'_{B0}$ both with $\epsilon=0$, we have: 
\begin{equation}\label{eq:auxauxfree}
        \mathcal{F}''_{00}:\qty{p_C(a, b) =  p_{B}(b)\delta_{a0} } \cup \qty{p_C(a, b) =  p_{A}(a)\delta_{b0} } .
    \end{equation}

Such a free set resides entirely on the boundary of the classical probability simplex and constitutes a star domain centered at the distribution $p_C(a, b) = \delta_{a0}\delta_{b0} $. The interpretation is simple: imagine a provider who hands out malfunctioning devices with two outputs: only one output can vary in probability, and the other always returns zero. Testers can request any desired probability for the variable output and rename the output labels, but they lack any internal source of randomness. Under these conditions, the resource theory evaluates how much performance an agent with a fully functioning device gains over testers who rely on such constrained resources. 

Another example present in quantum channel resource theories occurs if a provider delivers any of the Pauli channels, but only one at a time. Additionally, the tester can access a source of randomness, so he is free to apply any convex combination between the provided Pauli and the identity channel. In such cases, the free set of channels for the tester is: 

\begin{equation}\label{eq:auxauxchannfree}
        \mathcal{F}'': \{\Phi\mid \Phi=\Gamma_1^{p} \}\cup \{\Phi\mid \Phi=\Gamma_2^{p} \}\cup \{\Phi\mid \Phi=\Gamma_3^{p} \} ,
    \end{equation}
with the $\Gamma_j^{p} $ defined as in (\ref{Paulichannel}) and $p$ assuming any value between zero and one. The previous free set lies entirely on the boundary of the Pauli channel mixtures and forms a star domain with the identity channel as the kernel. Similarly, as in the previous case, the resource theory quantifies the power of an agent employing an arbitrary Pauli channel mixture compared to those testers with access to the above provider. 

One can construct several examples analogous to  these discussed in this work in both classical and quantum scenarios. Given the recent inception of quantum-state texture, we expect their proliferation in the future, underscoring the relevance of the techniques proposed here.
 
\end{widetext}

\bibliographystyle{quantum_abbr}
\bibliography{references}

\onecolumn

    \begin{figure}[h]
        \centering
        \includegraphics[height= 0.75\paperheight]{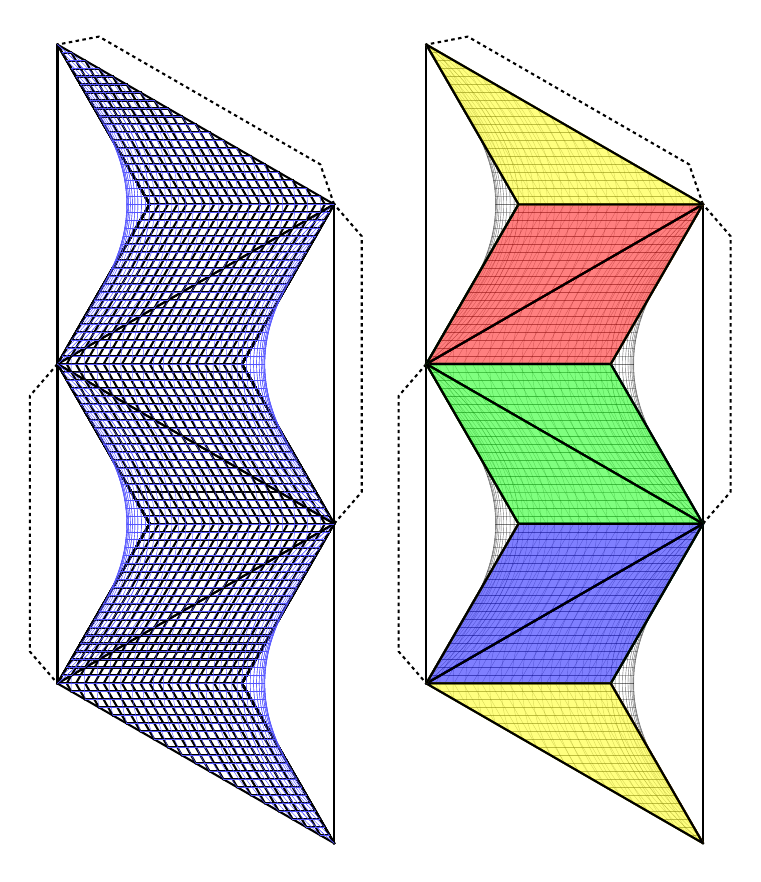}
        \caption{Paper folding versions of Figures \ref{fig:TC_1} and \ref{fig:TC_2}.}
        \label{fig:Fold12}
    \end{figure}

    
    \begin{figure}[h]
       \centering
        \includegraphics[angle = 90, height= 0.75\paperheight ]{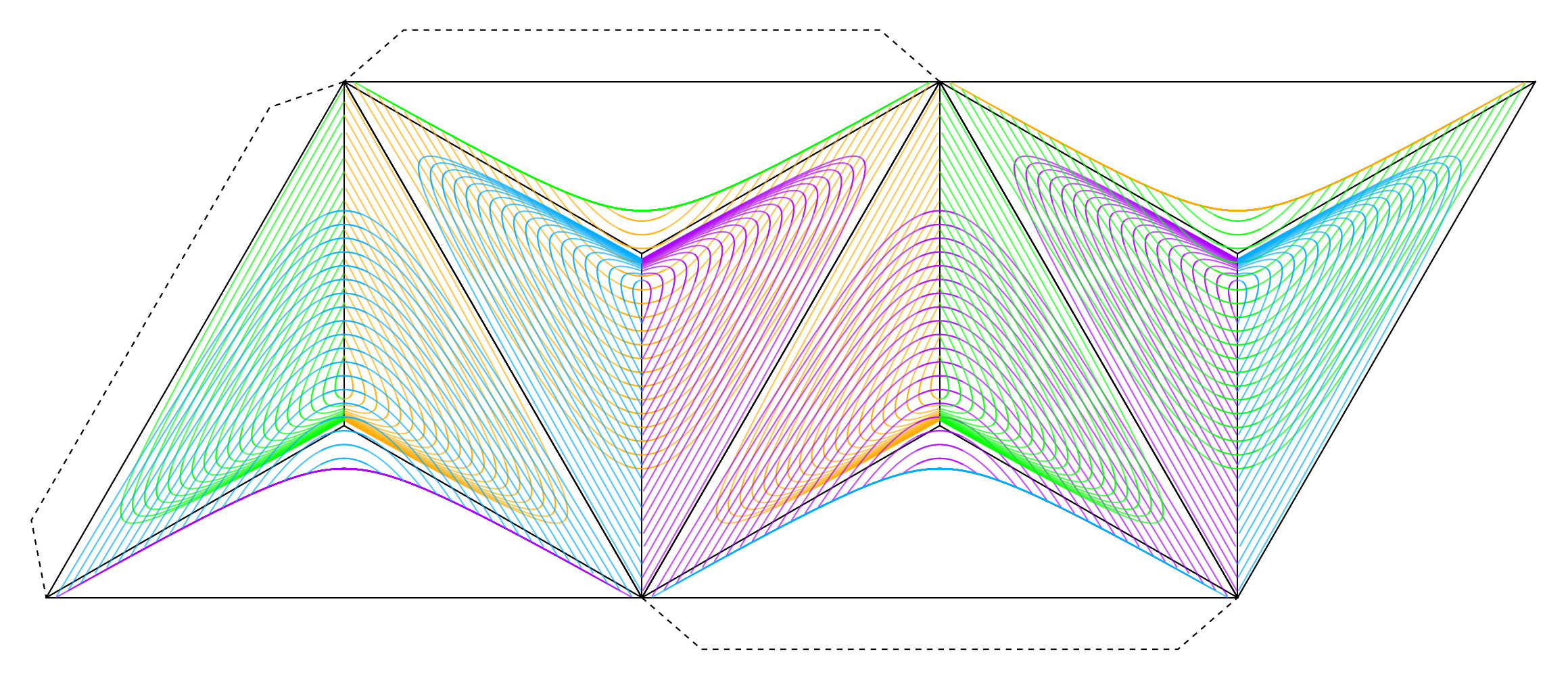}
        \label{fig:Fold34}
        \caption{Paper folding version of Figure \ref{fig:TC_4}.}
    \end{figure}
    



\end{document}